\documentclass[12pt,paper=letter,oneside]{book}

\usepackage{epsfig,graphics}
\input epsf
\usepackage{graphics}
\usepackage{myreportstyle}
\include{Shortcuts}
\usepackage{geometry}
\usepackage{pstcol} 
\usepackage{pst-node}

\makeatletter

\begin{document}


\newcommand{\captionfonts}{\small}

\makeatletter  
\long\def\@makecaption#1#2{%
  \begin{quotation}
  \vskip\abovecaptionskip
  \sbox\@tempboxa{{\captionfonts \noindent \textbf{#1}: #2}}%
  \ifdim \wd\@tempboxa >\hsize
    {\captionfonts \noindent \textbf{#1}: #2\par}
  \else
    \hbox to\hsize{\hfil\box\@tempboxa\hfil}%
  \fi
  \vskip\belowcaptionskip \end{quotation}}
\makeatother   

\renewcommand{\figurename}{Fig.}

\bibliographystyle{alpha}


\title{Jet Study in Ultra-Relativistic Heavy-Ion Collisions with the
ALICE Detectors at the LHC}

\author{Sarah-Louise Blyth}


\maketitle

\thispagestyle{empty}
\begin{center} \textbf{Abstract} \end{center}

In ultra-relativistic heavy-ion collisions at $\sqrt{s_{NN}}$ = 5.5
TeV at the ALICE experiment at the LHC, interactions between the
high-$p_{T}$ partons and the hot, dense medium produced in the
collisions, are expected to lead to jet energy loss (jet-quenching) resulting in
changes in the jet fragmentation functions as compared to the
unquenched case. In order to reconstruct jet fragmentation functions,
accurate information on the jet energy, direction and momentum
distribution of the jet particles is needed.
This thesis presents first results on jet reconstruction in simulated
Pb+Pb collisions using the ALICE detectors and a UA1-based cone jet
finding algorithm which has been modified and optimised to reconstruct
high-$p_{T}$ jets on an event-by-event basis. Optimisation of the
algorithm parameters and methods used to suppress the large background
energy contribution while maximising the algorithm efficiency, are
discussed and the resulting jet energy and direction resolutions are
presented. Accurate jet reconstruction will allow measurement of the
jet fragmentation functions and consequently the degree of quenching
and therefore provide insight on the 
properties of the hot and dense medium (for example the initial gluon
density) created in the collisions.

\vspace{5cm}
Publications and presentations on this work include:
\begin{enumerate}
  \item S-L Blyth, \emph{Jet study in ultra-relativistic heavy-ion
  collisions with the ALICE detector at the LHC}. Talk presented at Quark
  Matter 2004 Conference, Oakland, California, USA, January 2004
  \item S-L Blyth, \emph{Jet study in ultra-relativistic heavy-ion
  collisions with the ALICE detector at the LHC}. Paper to be published in
  Journal of Physics G: Nuclear and Particle Physics, QM2004
  proceedings
  \item S-L Blyth and M. J. Horner, \emph{Jet Study in
  Ultra-relativistic Heavy-ion Collisions at the LHC}. Poster
  presented at the Gordon Conference on Nuclear Chemistry, New
  Hampshire, USA, June 2004
\end{enumerate}



\thispagestyle{empty}
\begin{center}
\large{\textbf{Acknowledgements}}\end{center}

\normalsize
\noindent There are many people I'd like to thank that helped me in one way or
another (or in many ways!) in producing this thesis:\\

\noindent Thank you Prof. Cleymans and Grazyna for allowing me to
spend time at Lawrence Berkeley National Laboratory 
and work on this very interesting topic. It has been a very exciting
time and a great opportunity.\\

\noindent Thank you Grazyna, Jenn, Spencer and Mark for all your
patience, input and help. I greatly appreciate it. I'd also like to
thank Heather, Marco, Andreas, Tom, Alexei and the 
rest of the ALICE and ALICE-USA collaborations for your help.\\ 

\noindent Thank you Mark for teaching me ROOT, for all your
encouragement and for generally putting up with me!\\

\noindent Thank you Mom and Dad for making me believe I could do it.\\


\pagenumbering{roman} \setcounter{page}{1}
\tableofcontents
\listoffigures
\listoftables 


\pagebreak
\pagenumbering{arabic}

\addtolength{\parskip}{\baselineskip} 


\chapter{Introduction } 
 
\section{Heavy-ion Collisions and the QGP}
The ultimate goal of ultra-relativistic heavy-ion collision experiments is to
create and study the quark-gluon plasma (QGP), a phase of
matter in which quarks and gluons move freely in thermal
and chemical equilibrium \cite{Collins}. The QGP is predicted by
the theory of quantum chromodynamics (QCD) \cite{KarschQCD};
the theory of the strong colour force which is responsible for binding the 
constituent quarks and gluons (collectively called partons) inside
nucleons i.e. protons and neutrons \cite{Perkins}.

In nature, quarks are never found alone. Instead, they are always
found bound together in threes, inside baryons, or twos, as
quark-antiquark pairs inside mesons. This feature of QCD, that the effective
coupling constant $\alpha_{s}(Q^{2})$ between quarks increases at low
temperature and momentum transfer $Q$, or large
distances, is called confinement. However in QCD, at
high temperature and momentum transfer $Q$, or short distances ($\sim$1 fm),
$\alpha_{s}(Q^{2})$ decreases logarithmically
and the quarks become asymptotically free or
deconfined \cite{GWil, Polit}.

\subsection{The QCD Phase Diagram} 
According to the standard model,
the evolution of the early universe included a QGP phase at a very early stage 
(a few microseconds after the Big Bang) before
it cooled down and the quarks and gluons hadronised into ordinary
nuclear matter.
        
\begin{figure}[!hbt]
          \center
          \includegraphics[scale=0.35]{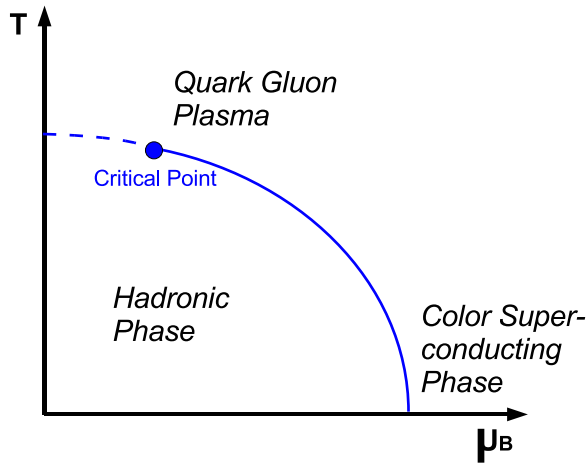}
          \protect\caption{Sketch of QCD phase diagram indicating the
          critical point for the case of two light quarks and 1 heavy
          quark. The solid line indicates a first order phase 
          transition and the dashed line indicates a crossover
          transition. Based on \cite{QM2004}.  
          }
          \protect\label{fig:CritPoint}
        \end{figure}

A sketch of the QCD phase diagram \cite{QM2004} of temperature versus baryon
chemical potential ($T$ vs. $\mu_{B}$) is shown in
Fig.~\ref{fig:CritPoint}.
As temperature increases, the coupling between quarks becomes weak
leading to deconfinement of the quarks and
chiral symmetry restoration (the quarks appear as nearly massless
particles compared to having masses of the order of more than 100 MeV when they are
confined inside hadrons). This state of freely moving quarks and
gluons, close to or at equilibrium, QGP, is predicted by QCD and indicated
in Fig.~\ref{fig:CritPoint} above the curved line. 

In QCD, analytical calculations can be performed using perturbative
methods (pQCD) above a scale set by $\lambda_{QCD}\approx$ 2
GeV. Below this limit, due to divergences, the equations
cannot be solved analytically.
Another approach to solving the equations of QCD, is to use
non-perturbative lattice gauge
theory techniques. This method attempts to discretise, on a lattice,
the continuous integral 
describing the QCD partition function and is
complicated due to the increase in calculation complexity as a
function of decreasing the step-size of the lattice and requires large
computing resources. 

Lattice gauge theory calculations
predict that at zero chemical potential ($\mu_{B} = 0$) the
critical temperature at which the phase transition 
from a hadron gas to a QGP occurs is $T_{c}=160-170$ MeV (e.g. $T_{c}
= 172 \pm 3.5$ MeV in \cite{Fodor})
which relates to an energy density of 1-2
GeV/$\rm{fm^{3}}$~\footnote{The MIT hadron bag model relates energy density
  to temperature for a QGP as $\epsilon = \frac{37\pi^{2}}{30}T^{4}$
  \cite{Wong}}.
The order of the phase transition (i.e. first order,  
second order or a crossover) at $\mu_{B} = 0$ is model dependent. 
For a first order phase transition, the Gibbs energies of the two phases
are equal at the transition temperature but the first derivatives with
respect to pressure and temperature are discontinuous 
at the point of transition.
A second order phase transition
is one where the entropy of the two phases on each side of the phase boundary is 
the same.
A cross-over means that there is no discontinuity in between the phases
of the matter involved in the transition, i.e the thermodynamic quantities
change smoothly. For cases with 2 light quarks
and 1 heavy quark such that $m_{s}\gg m_{u,d}\neq 0$, a smooth
crossover is predicted\cite{Fodor,Rajagopal}, see Fig.~\ref{fig:CritPoint}.
At large $\mu_{B}$, models predict that the phase transition is of the
first order. Thus the first order phase transition line may end at some
critical point ($\mu_{E}, T_{E}$) in the phase diagram
where for $\mu<\mu_{E}$ the transition is a crossover. The
position of this critical point is not yet established although some attempts
have been made to calculate its position using lattice calculations\cite{Fodor}.

At low temperature and very high baryon density (indicated in
Fig.~\ref{fig:CritPoint} by 
high baryon chemical potential, $\mu_{B}$), chiral symmetry is also
predicted to be restored. A state of deconfined quarks and gluons may
exist within neutron stars in the form of a color superconducting
state \cite{Rajagopal} where quarks may form Cooper pairs bound together
by gluons.

\subsection{Heavy-ion Collisions}
QCD predicts that the energy density $\epsilon$ at mid-rapidity grows
logarithmically with the centre-of-mass energy of a collision and
with $A^{2/3}$, where $A$ is the nuclear mass of the particles
involved in the collision
\cite{HarrisAndMuller}. Thus heavy-ion collisions at relativistic
energies are an ideal tool to create, in the laboratory, the
high energy densities to test the predictions of QCD and the expected formation
of the hypothetical state of deconfined matter, the QGP \cite{Chapline}.

In the centre-of-mass frame of a relativistic heavy-ion collision, the
colliding nuclei appear as two Lorentz-contracted discs as they fly
towards each other. When the nuclei collide, they deposit a large
amount of energy (dependent on the amount of energy carried by each
nucleus) into a very small volume. If the energy density created in
the collision is high enough to create a fireball of deconfined quarks
and gluons and the system lives long enough that the quarks and gluons have time
to interact and thermalise, then a QGP will be created. As the
fireball continues expanding rapidly, it cools down and hadronisation
occurs. The hadronisation mechanisms involved at different momentum ranges
are described phenomenologically.

As the
system continues expanding, inelastic and elastic scatterings 
occur between the particles. Chemical freeze-out occurs when the
inelastic scatterings
between the hadrons stop. Thermal freeze-out occurs once the elastic
scatterings between particles stop i.e. the system has expanded to
the point where the particles can no longer interact with each
other. Experiments measure the properties of the hadrons after thermal 
freeze-out.

        \begin{figure}[!hbt]
          \center
          \includegraphics[scale=0.8]{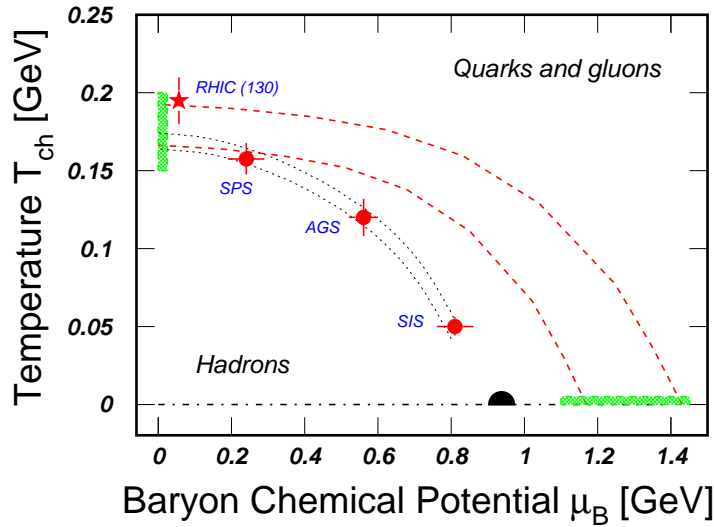}
          \protect\caption{QCD phase plot from
          \cite{XuKaneta}.
          The dashed lines indicate the boundary between interactions
          involving hadronic and partonic degrees of freedom. The black half
          circle represents the ground state of nuclei.  
          }
          \protect\label{fig:PhaseDiag1}
        \end{figure}

Heavy-ion experiments have covered a broad range of energies and 
masses of colliding nuclei, providing a large amount of data to
constrain theoretical calculations.  
Using a small number of particle ratios, (enough to constrain the model parameters),
the temperature $T$, and the baryon density $\mu_{B}$ at thermal freeze-out 
have been calculated using
statistical thermal models at various energies, see
Fig.~\ref{fig:PhaseDiag1} from \cite{XuKaneta}. The values of $T$ 
and $\mu_{B}$ obtained from the thermal model fits are found to describe well the 
majority of other particle ratios observed experimentally
\cite{XuKaneta,Cleymans,BraunM}. In Fig.~\ref{fig:PhaseDiag1}, the 
$T$ and $\mu_{B}$
of the created systems from different heavy-ion experiments, 
calculated using the thermal model, are plotted along the dotted lines. The
collision energy of the experiments increases towards the $T$-axis. 
Thus the experiments at RHIC measure
systems which have calculated values of $T$ and $\mu_{B}$ which 
are at the boundary of the phase transition at 
very low baryon density. The Large Hadron Collider (LHC) experiments, 
scheduled to go online in 2007, will measure systems produced with a
collision energy $\sim 30$ times that of RHIC and which are expected to have
energy densities about 20 times greater and thus initial temperatures
of $T_{LHC}\approx 2.1 T_{RHIC}$ \cite{Eskola}.

\section{Probes of the Hot, Dense, Partonic Medium/QGP}\label{sec:QGPSearch}
Since a hypothetical QGP, if created in a heavy-ion collision, will
have an extremely short life-time ($\sim 5-10$ fm/$c$) before
hadronisation occurs \cite{HarrisAndMuller}, its 
presence and characteristics can only be determined using indirect
observables. A variety of signatures has been proposed
although experimental measurements are complicated due to the
time/space evolution of the system and final state hadronic interactions.

The proposed experimental signatures may provide
information on different characteristics of the system produced
in the collision. For example, proposed signatures of \emph{deconfinement}
include suppression of quarkonium production \cite{MatsuiSatz} and enhancement of
multi-strange particle 
production \cite{RafelskiStrange}. \emph{Chiral symmetry restoration}
is expected 
to be observed through measuring modifications of vector meson masses
\cite{Pisarski}.
Hard probes \cite{WangProbes} such as high-$p_{T}$ jets, dileptons or
direct photons \cite{MStrickland},
heavy quarkonia or $W^{\pm}$ or $Z^{0}$ may also be used to study the
\emph{thermalisation} of the partons and \emph{early stage evolution} 
of the QGP.
Hard probes are created in high-$Q^{2}$ interactions and at early
times where $\tau_{hard}\sim1/Q\leq0.01$ fm/$c$. The bulk of the
secondary matter is formed later at around $\tau\sim 1/T_{0} \sim 0.2$
fm/$c$. Thus the hard probe becomes embedded in the secondary matter
and as it traverses the hot, dense medium it undergoes softer 
secondary interactions. Properties of the hard probe may therefore
be modified due to medium effects and provide information on
properties of the medium, most notably, the initial gluon density.
Properties of the system
such as energy density $\epsilon$, temperature $T$, and pressure $P$,
may be measured using the resulting hadron rapidity and
transverse energy distributions as well as the flow distributions
\cite{HarrisAndMuller}.

\subsection{High-$p_{T}$ Jets and Energy Loss in the Medium}
\begin{figure}[!hbt]
          \center
          \includegraphics[scale=0.5]{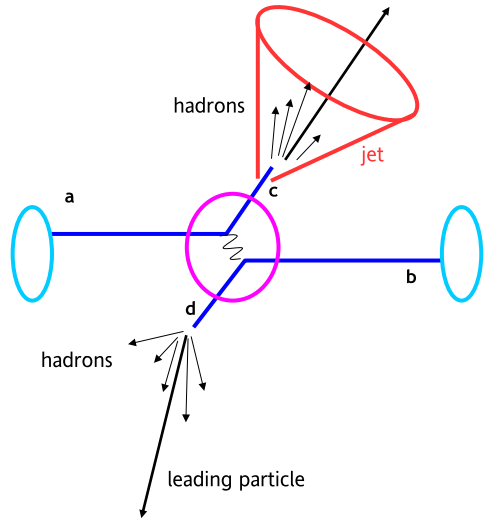}
          \protect\caption{The formation of a di-jet event from the hard
          scattering of partons (\textbf{a} and \textbf{b}) from
          nucleons within the incoming 
          nuclei. The hard partons (\textbf{c} and \textbf{d})
          traverse the medium formed in the 
          collision (purple circle) before fragmenting into jets of 
          hadrons. From
          \cite{JennKlay}.
          }
          \protect\label{fig:JetCartoon}
        \end{figure}

This thesis will focus on a specific hard-probe, namely high-$p_{T}$
jets. Jets, which result from hard parton-parton scatterings,
have been observed in $e^{+}$+$e^{-}$, p+p and 
p+$\overline{\rm{p}}$ experiments. A working definition of a jet is a
localised (in space) group of hadrons which 
originate from the fragmentation of an initial hard 
scattered parton, see Fig.~\ref{fig:JetCartoon} \cite{JennKlay}. The jet particles are
contained within a cone of radius
$R=\sqrt{(\Delta\eta)^2+(\Delta\phi)^2} = 1$. The total energy and 
direction of the jet is expected to be closely related to that
of the parent parton.

It has been predicted that a
high-$p_{T}$ parton traversing hot nuclear matter, for example a QGP, will
experience collisional energy loss (via scattering) and 
radiative energy loss (via induced gluon bremsstrahlung)
\cite{WangProbes, Wang1992, WangHuang}. Radiative energy loss is
expected to dominate over collisional energy loss. The amount of 
energy loss, $\Delta E$, is predicted to be proportional to the 
gluon/energy density of the medium and a function of the path length
of the parton through the medium. Thus jets, which
have lost energy while traversing the medium, can be used as tools to
probe the medium and provide information on its properties and
structure at an early stage \cite{WangProbes}.


\begin{figure}[!hbt]
          \center
          \includegraphics[scale=0.6]{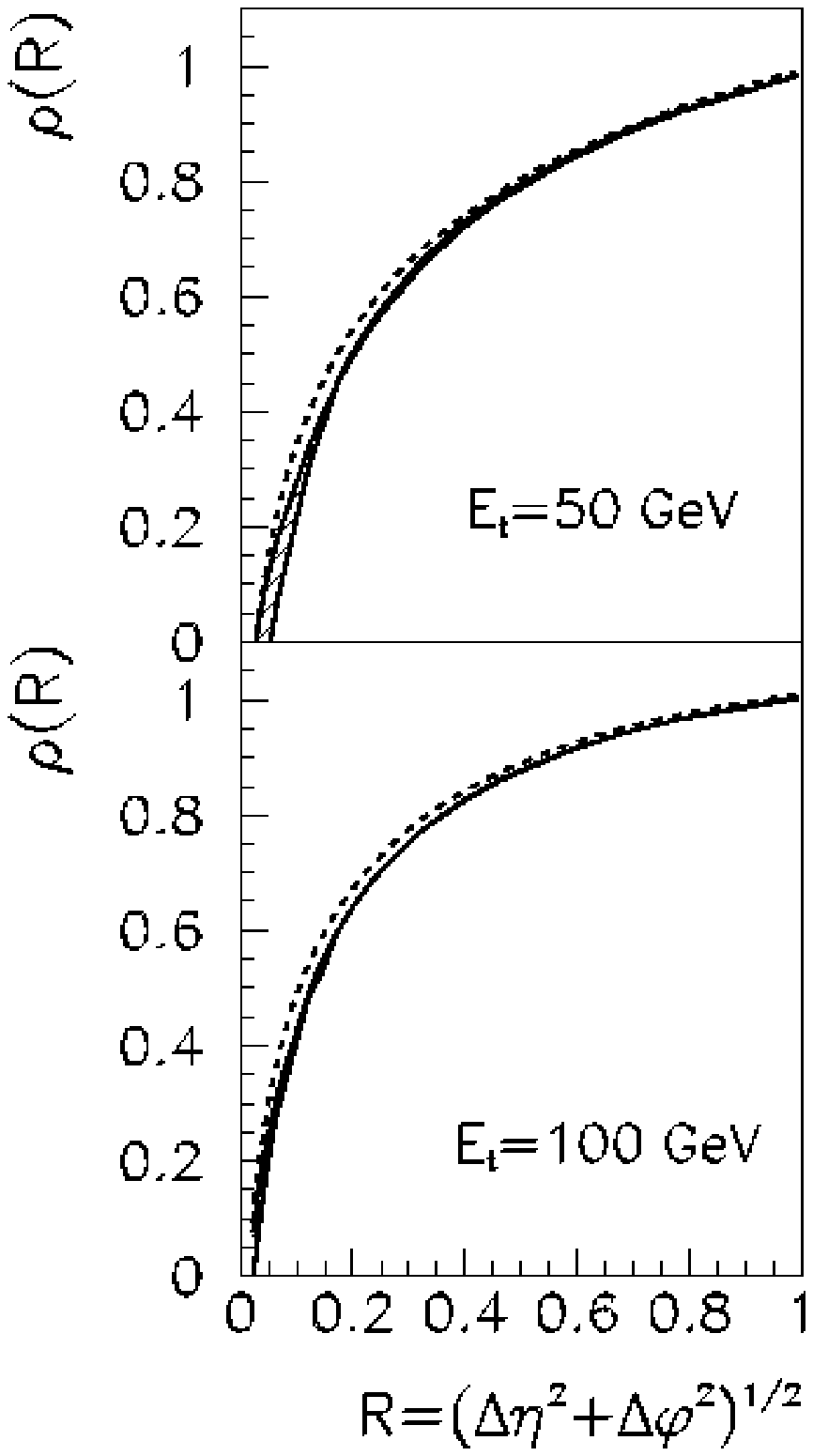}
          \protect\caption{Fraction of total jet $E_{T}$, $\rho(R)$,
          for quark-led jets fragmenting 
          in vacuum (dashed line) and fragmenting in a dense QCD
          medium (solid line) based on pQCD calculations from
          \cite{WiedemannShapes}
          . The upper panel represents the case for a 50GeV jet and
          the lower panel the case for a 100GeV jet. From 
          \cite{WiedemannShapes}.  
          }
          \protect\label{fig:JetEDist}
        \end{figure}

The pQCD-based calculations in
\cite{WiedemannShapes} predict that the angular energy distribution
for jets which 
have undergone energy-loss in a dense QCD medium is
very similar to the energy distribution of jets which fragment in vacuum, see
Fig.~\ref{fig:JetEDist} from \cite{WiedemannShapes}.

\begin{figure}[!hbt]
          \center
          \includegraphics[scale=0.25]{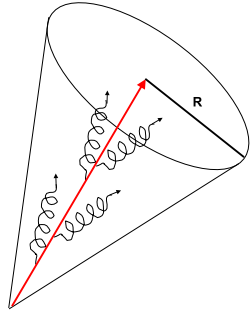}
          \protect\caption{During radiative energy loss, gluons
          (coiled lines) are
          radiated at angles close to the jet-axis (arrow) resulting in the
          total energy inside the cone remaining unchanged
          \cite{WiedemannShapes}. Picture from
          \cite{JennKlay}.  
          }
          \protect\label{fig:GluonRad}
        \end{figure}

The gluons are radiated at angles close to the jet axis \cite{WiedemannShapes} and therefore remain
iside the jet cone, see Fig.~\ref{fig:GluonRad}. Therefore,
measuring the total energy within a cone of radius $R$ (where $R =
\sqrt{{(\Delta\eta)}^2 + {(\Delta\phi)}^2}$ ) around the jet axis
will not provide information on jet energy loss since the total energy
inside the cone will include both the final jet energy plus the radiated
energy.

\begin{figure}[!hbt]
          \center
          \includegraphics[scale=0.5]{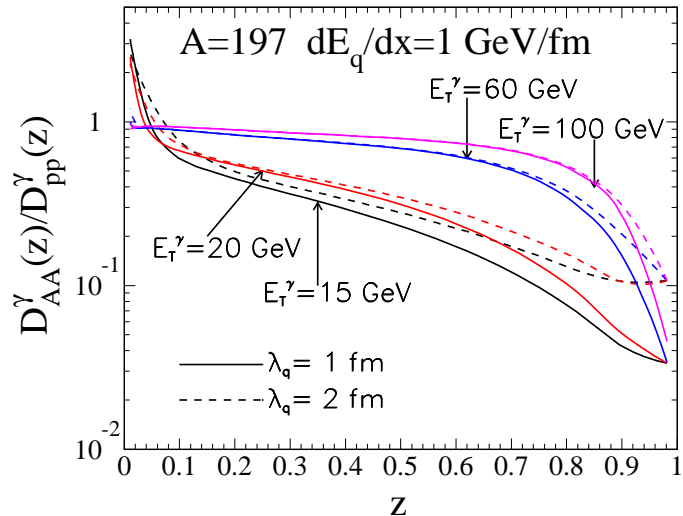}
          \protect\caption{Ratio of the inclusive jet fragmentation
          functions for $\gamma$-tagged jets with energy loss to the
          fragmentation 
          function of jets without energy loss in central Au+Au
          collisions. The solid lines indicate the calculation for the
          case of inelastic scattering mean free path $\lambda_{q} =$
          1 fm and the dashed line is the case for $\lambda_{q}$ = 2 fm. From 
          \cite{WangHuang}.  
          }
          \protect\label{fig:WangFF}
        \end{figure}

However, the momentum distribution of the particles within the jet
cone is expected be modified for jets which have lost energy due to
medium effects \cite{WangHuang}. This in turn causes a modification of
the jet fragmentation functions which are defined following
\cite{UA12} as
\begin{equation}\label{eqn:ffunc}
D(z) = \frac{1}{N_{jets}}\frac{dN_{ch}}{dz}
\end{equation}  
where, 
\begin{equation}\label{eqn:z}
  z = p_{L}/E_{T_{jet}},
\end{equation}
$p_{L}$ is the momentum component of a particle in the jet
in the direction parallel to the jet axis and $E_{T_{jet}}$ is the
transverse energy of the jet ($E_{T}=E\sin\theta$).
The jet energy loss can be quantified by measuring
the jet fragmentation functions and comparing to fragmentation
functions measured in p+p collisions or peripheral heavy-ion
collisions \cite{WangHuang}, see Fig.~\ref{fig:WangFF}.

Fig.~\ref{fig:WangFF} from \cite{WangHuang} shows the calculated ratio of the
inclusive jet fragmentation functions of jets with energy loss versus
jets with no energy loss as a function of the fractional momenta $z$
(where, in this case $z = p_{T}/E_{T_{jet}}$, differing from
the definition in equation~\ref{eqn:z}) of
the jet hadrons.
Jet energy loss is manifested by a shift of the fragmentation function
towards low $z$, as shown by the increase above unity of
low momentum (i.e. low-$z$) particles and the decrease in the number
of high-$z$ particles compared to the case with no energy loss i.e
there is a shift of particles from high-$p_{T}$ to low-$p_{T}$.
Lower energy jets are more affected than higher energy jets as shown
in Fig.~\ref{fig:WangFF}.  

\begin{figure}[!hbt]
          \center
          \includegraphics[scale=0.6]{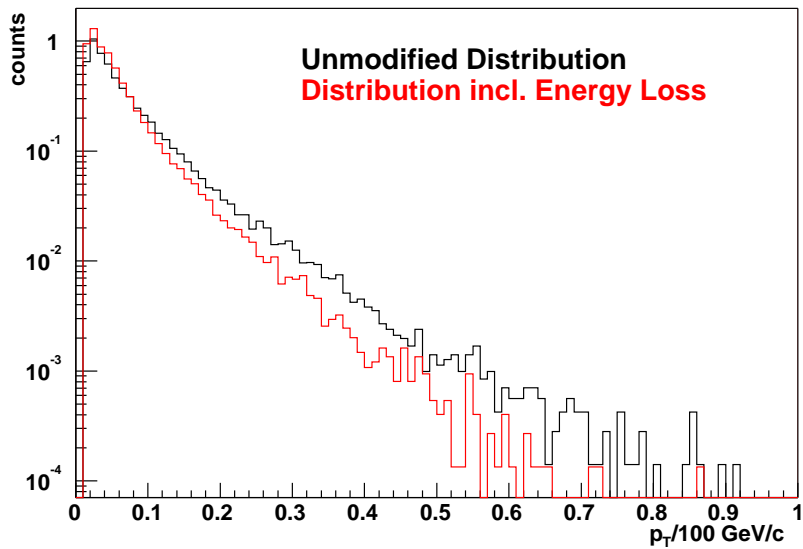}
          \protect\caption{Fragmentation functions of simulated 100
          GeV jets for 
          two different cases: the black histogram shows the
          fragmentation function for a sample of unmodified 100 GeV PYTHIA
          jets. The red histogram shows the fragmentation function of
          a sample of
          jets which have undergone energy-loss due to the medium as
          modeled by the superposition of a 20 GeV PYTHIA jet on a 80
          GeV PYTHIA jet after
          \cite{Andreas}
          from
          \cite{Mark}.
          }
          \protect\label{fig:FragF8020}
        \end{figure}

As yet, there is no Monte Carlo event generator which can be used to simulate
jet energy loss in a medium in heavy-ion collisions. 
However, a `toy' model of jet energy loss has been constructed \cite{Andreas}.
A jet in a heavy-ion collision is modeled by the superposition of a
p+p jet event 
from PYTHIA \cite{Pythia} on a high-multiplicity A+A event simulated
using HIJING  
\cite{Hijing}. In PYTHIA, jets were modified
according to the `toy' model prescription and, for example 20$\%$ 
energy loss from a 100 GeV jet, was mocked up by superimposing a
20 GeV PYTHIA jet on an 80 GeV PYTHIA jet on top of a HIJING background event.
Fig.~\ref{fig:FragF8020} shows the
resulting modified fragmentation function from a sample of
100 GeV jets with energy loss modeled in this way compared to the
fragmentation function from 100 GeV jets with no energy loss. The
fragmentation function for the 
energy-loss case is shifted to lower $p_{T}$ as expected. The
magnitude of the shift may provide information on the medium properties. Thus, in experiment, in
order to measure precisely the resulting jet fragmentation functions
according to equations~(\ref{eqn:ffunc}) and (\ref{eqn:z}), 
accurate measurement of the jet energy, jet axis and momentum distribution of
the constituent jet hadrons is needed. 

\subsection{Jets at RHIC}\label{RHICjets}
There are four heavy-ion experiments at RHIC which have studied Au+Au
collisions 
at three center-of-mass energies, $\sqrt{s_{NN}}=130$ GeV, $\sqrt{s_{NN}}=200$
GeV and $\sqrt{s_{NN}}=62.4$ GeV. The experiments, in increasing order of collaboration size,
are: BRAHMS (Broad Range Hadron Magnetic Spectrometers Experiment at
RHIC), PHOBOS\footnote{PHOBOS is a name, not an acronym},
PHENIX (Pioneering High Energy Nuclear Interaction 
Experiment) and STAR (Solenoidal Tracker at RHIC). Details of all
RHIC experiments can be found in \cite{YellowBook}.

\begin{figure}[!hbt]
          \center
          \includegraphics[scale=0.5]{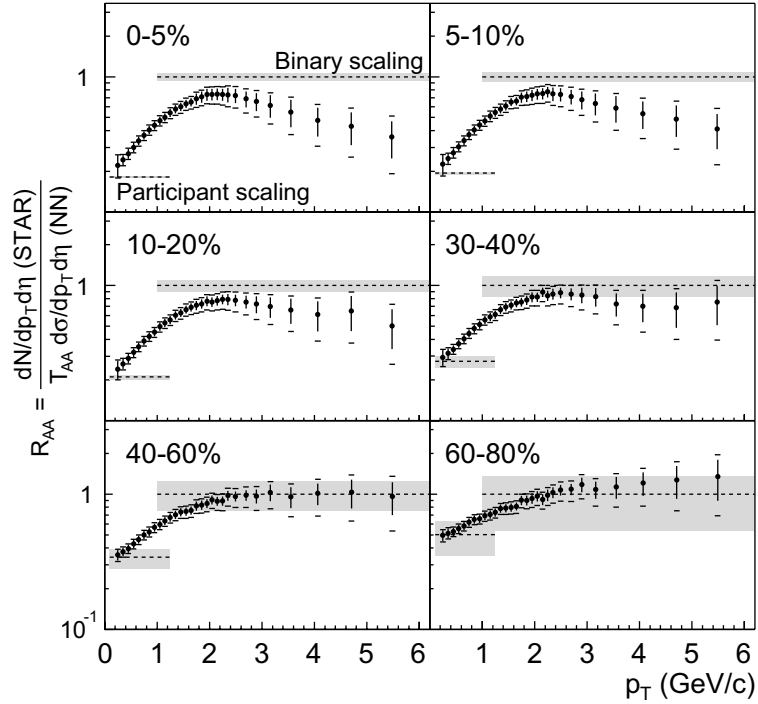}
          \protect\caption{$R_{AA}(p_{T})$ as a function of
          centrality. The dashed lines show scaling with
          $\langle N_{bin}\rangle$ and $\langle N_{part} \rangle$ and
          the gray bands show the systematic uncertainties. From 
          \cite{STAR130Raa}.
          }
          \protect\label{fig:JennRaa}
        \end{figure}

Evidence for jet energy loss has been reported by all four of the RHIC
experiments. Three main pieces of evidence exist to support this, namely, 
the increasing suppression of the number of
high-$p_{T}$ hadrons with 
increasing centrality in Au+Au events when compared to p+p events,
\cite{STAR130Raa,STAR200Raa, PHENIX200Raa}, see
Fig~\ref{fig:JennRaa}, the disappearance of away-side correlations in
central Au+Au collisions compared to p+p collisions \cite{PRL908}, and
the d+Au results \cite{STARdAu, PHENIXdAu} which allow discrimination
between initial and final state effects. 


Fig.~\ref{fig:JennRaa} from the STAR experiment shows the nuclear
modification factor, $R_{AA}$, defined as the ratio of
the inclusive momentum spectrum of charged hadrons from Au+Au
events to a p+p reference spectrum scaled to account for the
nuclear geometry. The different
panels show the evolution of $R_{AA}$ as a function of collision
centrality. It is expected that the number of hard processes in an
event scales
with the average number of binary collisions $\langle N_{bin} \rangle$,
in the absence of nuclear effects, which should result in $R_{AA}=1$
over a threshold defined by $p_{Thard} \approx 2$ GeV/$c$.
Fig.~\ref{fig:JennRaa} shows that there is a suppression from the expected 
value of unity which increases as a
function of centrality. This result is consistent 
with jet energy loss in a dense medium although from these results
alone, it cannot be determined whether the suppression is due to
initial or final state interactions. Similar results were obtained by
the PHENIX experiment \cite{PHENIX200Raa}.

        \begin{figure}[!hbt]
          \center
          \includegraphics[scale=0.6]{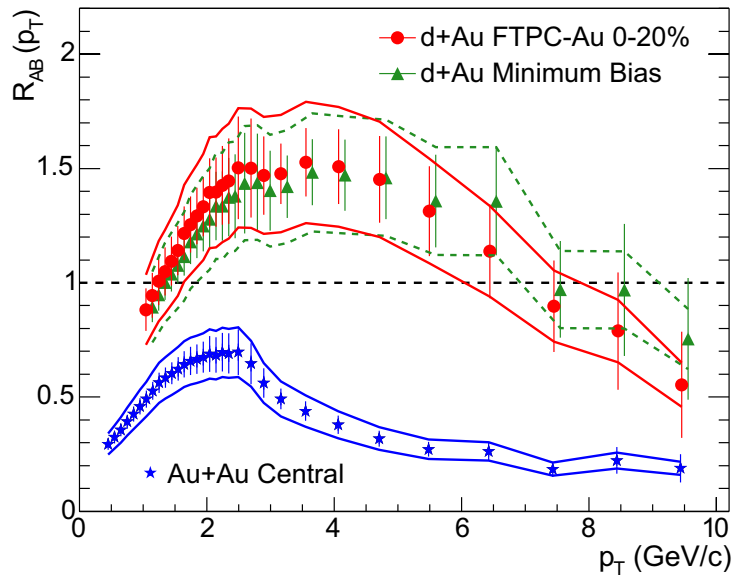}
          \protect\caption{$R_{AB}(p_{T})$ for central and minimum
          bias d+Au collisions and central Au+Au collisions at STAR. For
          clarity the d+Au minimum bias data are displaced to the
          right. From 
          \cite{STARdAu}. 
          }
          \protect\label{fig:RaadAu}
        \end{figure}

Fig.~\ref{fig:RaadAu} shows the nuclear modification factor for
central Au+Au events ($R_{AA}$) and d+Au events ($R_{AB}$) at
$\sqrt{s_{NN}}=200$ GeV as measured by the STAR experiment 
\cite{STARdAu}. It can be seen that the
inclusive hadron yield in d+Au events is enhanced relative to p+p
events which is the opposite behaviour to Au+Au collisions.
Due to the small size of the system created in a d+Au collision
(cannot be geometrically bigger than the size of the deuteron)
it is not expected that a dense medium is formed in these
collisions. Therefore, d+Au collisions can be used as a control
experiment to probe initial state effects. From the lack of
suppression of high-$p_{T}$ particles in d+Au collisions, 
it can be concluded that the suppression is due to final state
effects in Au+Au collisions. Consistent results were reported by
the PHENIX experiment 
in \cite{PHENIXdAu}.  

        \begin{figure}[!hbt]
          \center
          \includegraphics[scale=0.5]{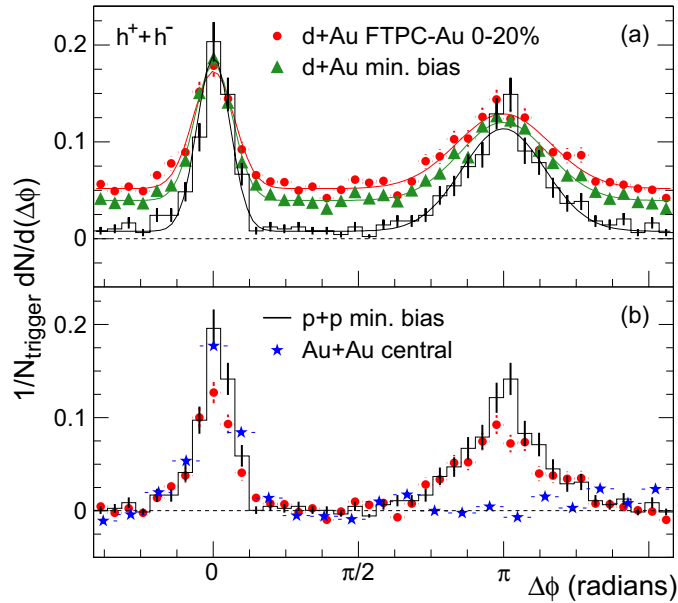}
          \protect\caption{Top panel: Two-particle azimuthal
          distributions for central (circles) and minimum bias
          (triangles) d+Au collisions 
          compared to p+p collisions (solid line histogram). Bottom
          panel: Two-particle 
          azimuthal distributions for central Au+Au collisions (stars)
          compared to central d+Au collisions (circles) and p+p
          collisions (solid line histogram). From 
          \cite{STARdAu}.  
          }
          \protect\label{fig:JetdAu}
        \end{figure}


The two-particle azimuthal distribution for central and minimum bias
d+Au collisions as measured by the STAR experiment is shown in the top
panel of Fig.~\ref{fig:JetdAu} taken from \cite{STARdAu}. The
distribution is obtained by measuring charged hadrons around a
high-$p_{T}$ trigger particle ($\Delta\phi\sim 0$) and at
$180^{\rm{o}}$ from the trigger particle ($\Delta\phi\sim\pi$).
In Fig.~\ref{fig:JetdAu}, the 
histogram indicates the case for p+p minimum bias collisions. The d+Au
data show a near-side peak at $\Delta\phi\sim 0$ from the leading
particle and an away-side peak
at $\Delta\phi\sim\pi$ from the leading particle which is
very similar to that of p+p collisions and consistent with di-jet
production. This is different to the azimuthal distributions for
central Au+Au collisions as shown in the lower panel of
Fig.~\ref{fig:JetdAu} where there is a clear peak in the Au+Au
distribution on the near-side and a large suppression on the away-side
resulting in a flat distribution.

Therefore the difference in the $R_{AB}$ and azimuthal distributions
between d+Au collisions and central Au+Au collisions rules out that
the suppression of high-$p_{T}$ hadrons be attributed to initial state
effects: the suppression is due to final state effects produced
in central Au+Au collisions. 

Jet-like correlations in central Au+Au collisions have been reconstructed on a
statistical basis by the STAR collaboration
\cite{Fuqiang}. This is done by correlating charged hadrons in a
certain energy range with a high-$p_{T}$ trigger particle. The
momentum distributions for the charged hadrons on the near-side
($\Delta\phi\sim$ 0), with
respect to the trigger particle, are reported to be similar to 
those from p+p collisions while the away-side ($\Delta\phi\sim$ $\pi$)
distributions 
(opposite in azimuth to the trigger particle) are shifted to lower $p_{T}$
compared to the p+p case, another indication of jet energy loss.
However, due to the small signal-to-background ratio and large
fluctuations in the 
high-multiplicity Au+Au  collisions, jets in heavy-ion collisions cannot be
reconstructed on an event-by-event basis at RHIC, only
statistically. Thus it is not yet possible at RHIC to measure jet
energies and directions in heavy-ion collisions on an event-by-event basis.

\subsection{Jets at LHC}
\begin{figure}[!hbt]
          \center
          \includegraphics[scale=0.6]{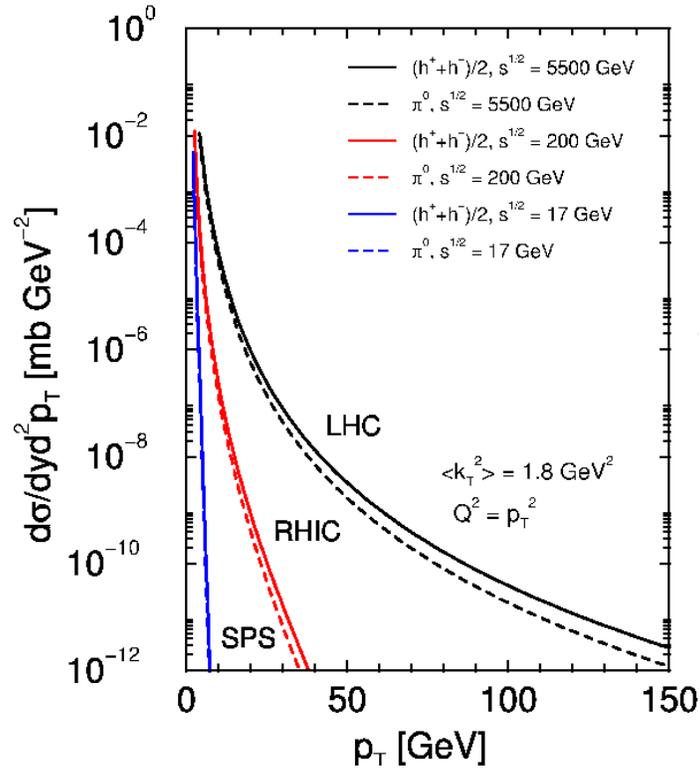}
          \protect\caption{Cross-sections for charged hadron and
          neutral pion production calculated using pQCD for different
          collision energies. From 
          \cite{GyulassyVitev}.  
          }
          \protect\label{fig:HardXSection}
        \end{figure}

The Large Hadron Collider (LHC) at CERN will be able to collide Pb ions at 
centre-of-mass energies of $\sqrt{s_{NN}}= 5.5$ TeV. The dedicated
heavy-ion experiment, ALICE (A Large Ion Collider Experiment) at the
LHC, has been designed to 
measure observables both in the soft and hard physics regimes.
The predicted cross-sections for hard processes at the LHC are
much higher than at RHIC energies as shown in
Fig.~\ref{fig:HardXSection}
from \cite{GyulassyVitev}.

\begin{figure}[!hbt]
          \center
          \includegraphics[scale=0.6]{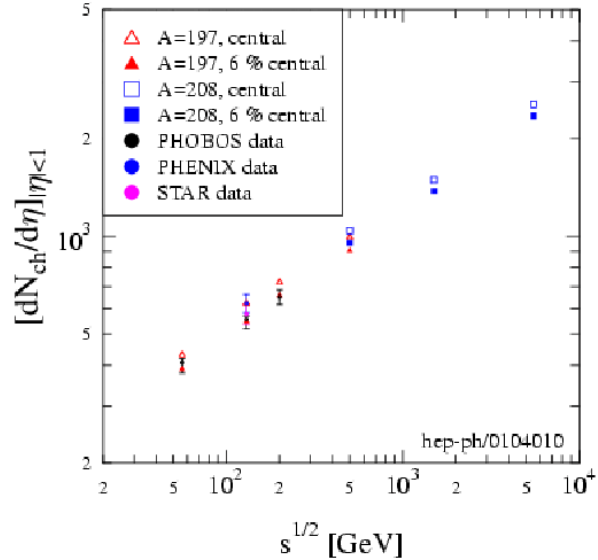}
          \protect\caption{Charged particle multiplicity averaged over
          an interval of $-1<\eta<1$ as a function of collision
          energy. Experimental data from RHIC experiments are also
          plotted. From 
          \cite{EskolaMult}.  
          }
          \protect\label{fig:Multiplicity}
        \end{figure}

The predicted particle multiplicities for LHC follow a slow
growth trend \cite{EskolaMult} from SPS and RHIC and amount to
a factor of $\sim$4 greater than at RHIC, leading to an expected
multiplicity of $dN_{ch}/dy=$ 2~500 as
shown in Fig.~\ref{fig:Multiplicity}. This is not the case for hard
probes, for example, the cross-section for 30 GeV jets is $\sim$7~000 times higher
at LHC than at RHIC \cite{GyulassyVitev}). Therefore, hard 
observables will be
able to be measured above the soft-background more easily at the LHC
than at RHIC. 

\subsection{Jet Studies at ALICE}\label{sec:ALICEJets}
With the predicted large cross-sections for jet production at LHC,
ALICE will be able to study jets over a very broad energy range 
(from $\sim$5 GeV to 250 GeV). Monte Carlo simulations have shown that Pb+Pb
events at top LHC energies 
contain a very large number of low energy ($E_{T}<20$ GeV) jets
which can be studied using particle correlation techniques \cite{AndreasLowPt}.
The number of produced jets
decreases with $p_{T}$ and fewer high energy jets are
expected ($\sim$100 000 jets with $E_{T}>$ 100 GeV within the experimental
acceptance per year
of ALICE Pb+Pb running). The high-$p_{T}$ jets will be reconstructed
on an event-by-event basis.

Information from both the high-resolution tracking detectors
(Inner Tracking System (ITS) and Time-Projection Chamber (TPC)) and the
Electromagnetic Calorimeter (EMCal) will be used to reconstruct
jets. The EMCal will function as a fast jet trigger
and will increase the yield of reconstructed jet events by a factor of $\sim$200
\cite{EMCalProposal}. (The TPC cannot be used as a jet trigger due to
its drift time of $\sim$90 $\mu$s \cite{ALICEppr}.) 

The energy and direction of the jets will be
measured using a combination of information from the ALICE tracking
detectors (TPC, ITS)  and the
EMCal. Thus both the charged and neutral energy components
of the jets will be recorded. The jet energy resolution, obtained using
the combined detector information, is expected to be
significantly improved compared to that obtained from using
information from only one detector.


The motivation behind jet studies at ALICE is to investigate
properties of the medium produced in Pb+Pb collisions at
$\sqrt{s_{NN}}=5.5$ TeV. This will be achieved through measuring
medium-induced modifications to the Pb+Pb jet fragmentation functions 
compared to the p+p case.

Precise measurement of the jet fragmentation
functions at ALICE will involve several steps. Accurate measurement of the jet
energy, jet axis and momentum distribution of the jet hadrons on an
event-by-event basis will be required. 
The raw jet energy spectrum will then be compiled from the measured
jet energies. However, this raw energy spectrum will be smeared in energy
due to detector effects and resolution effects caused by the
fluctuating background in Pb+Pb collisions. It will then be necessary
to deconvolute these effects (for example using the method in
\cite{CDFAkopian}) 
in order to extract the true jet energy spectrum and enable comparison with p+p
results, before analysing the fragmentation functions.

\subsection{Thesis Scope}
The focus of this thesis is 
the development of a jet-finding algorithm for
heavy-ion collisions, based on the UA1 approach \cite{STARNOTE196}, used in p+p
collisions, in order to reconstruct jet energies and 
directions accurately and efficiently on an event-by-event basis.
Modifications are included 
to take into account, and correct for, the 
fact that in heavy-ion collisions at LHC energies, the background particle 
multiplicities and energy fluctuations from the `underlying event' are
much greater than in p+p collisions. This is the first attempt at
event-by-event full jet reconstruction in heavy-ion collisions.

Deconvolution of the jet energy spectrum and measurement
of jet fragmentation functions are beyond the scope of this thesis.

\section{ALICE Experiment Overview}\label{sec:ALICEexp}
\begin{figure}[!hbt]
          \center
          \includegraphics[scale=0.5]{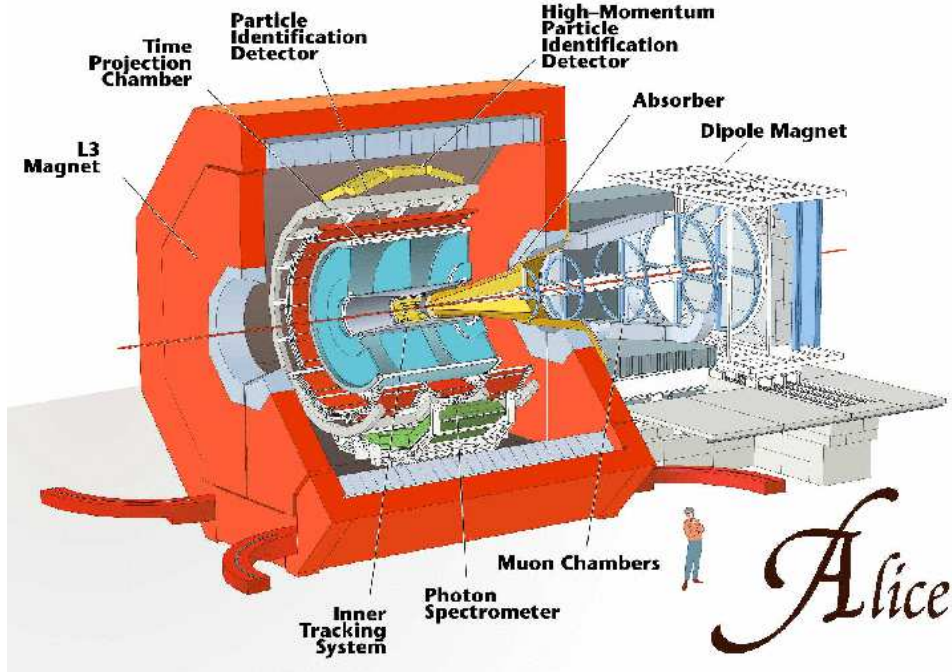}
          \protect\caption{Diagram showing ALICE detector and the
          positions of the various detectors. The electromagnetic
          calorimeter is not shown in this picture. From
          \cite{ALICEWeb}.
          }
          \protect\label{fig:ALICE}
        \end{figure}

The ALICE detector is currently under construction within a cavern 40 m
underground at Intersection Point 2 of the LHC at CERN
\cite{ALICEppr}. It has been designed to detect hadrons, 
leptons and photons over 
a broad momentum range (100 MeV/$c$$<$$p_{T}$$<$100 GeV/$c$) in an environment
with charged particle 
multiplicities of up to 8000 per unit of rapidity at central rapidities.
The layout of the ALICE apparatus with its various detectors, is shown in
Fig.~\ref{fig:ALICE} from \cite{ALICEWeb}.

Most of the ALICE detectors in Fig.~\ref{fig:ALICE}, with the exception of the muon
spectrometer, are positioned at midrapidity (-0.9$\leq\eta\leq$0.9) within
the large L3 magnet which is designed to provide
a magnetic field of 0.5 T to enable measurement of particle momenta
and identities by the various detectors. The magnet is 12 m long and
has a radius of 5 
m. Starting at the interaction vertex and proceeding outwards, 
the ALICE central detector system includes the:
\begin{itemize}
\item \textbf{Inner Tracking System (ITS)}
consisting of six layers of high-resolution silicon tracking detectors,
\item \textbf{Time-Projection Chamber (TPC)} which is the primary tracking detector
in ALICE,
\item \textbf{Transition Radiation Detector (TRD)} for the
  identification of electrons,
\item \textbf{Time Of Flight Detector (TOF)} for particle
  identification,
\item \textbf{High-Momentum Particle Identification Detector (HMPID)} which is an
  array of ring-imaging Cherenkov detectors for high-momentum particle identification,
\item \textbf{Photon Spectrometer (PHOS)} which consists of a small area
  lead-glass crystal electromagnetic calorimeter (-0.12$<\eta<$0.12,
  $\Delta\phi=100^{\rm{o}}$) (for detection of photons and neutral 
  mesons through their decay into photons) and a charged
  particle detector (CPV) which acts as a veto detector for charged
  particles which deposit energy in the calorimeter \cite{ALICEPhos},
\item \textbf{Proposed Electromagnetic Calorimeter (EMCal)} (not shown
  here) which is a large
  area (-0.7$<\eta<$0.7, $\Delta\phi=120^{\rm{o}}$) lead-scintillator
  sampling calorimeter.    
\end{itemize}

The Muon Spectrometer covers forward rapidity. It consists of
a dipole magnet,
tracking stations, a muon filter, trigger stations and an absorber close to the
vertex which acts as a shield for the spectrometer. Other detectors
which are not shown in Fig.~\ref{fig:ALICE} include a
Photon Multiplicity Detector (PMD) which counts photons and a Forward
Multiplicity Detector (FMD). Two Zero Degree Calorimeters (ZDC),
positioned at $0^{\rm{o}}$ about 90 m from the interaction vertex, will
be used to measure event centrality.

\begin{figure}[!hbt]
          \center
          \includegraphics[scale=0.35]{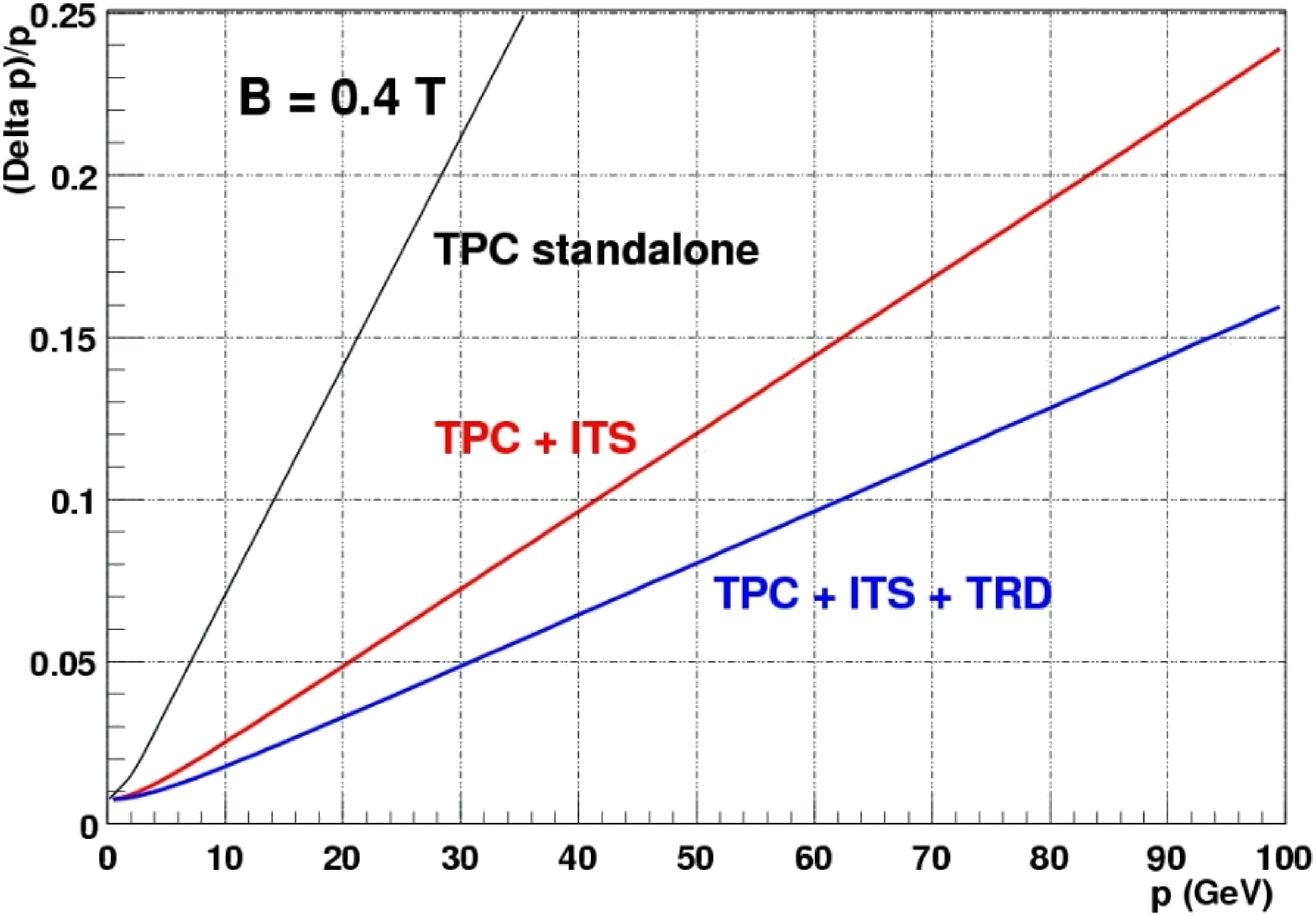}
          \protect\caption{Momentum resolution of the the ALICE
          tracking detectors. The black line shows the resolution
          using the TPC only, the red line shows resolution using TPC
          + ITS detectors and
          the blue line shows the improvement when using TPC, ITS and
          TRD detectors.
          From \cite{Glassel}.
          }
          \protect\label{fig:TrackRes}
        \end{figure}

The main ALICE tracking detector, the TPC, covers an area between -0.9$<\eta<$0.9 and
has full azimuthal coverage. The tracking efficiency of the TPC has
been tested using simulations of Pb+Pb collisions with charged
particle multiplicities of $dN_{ch}/d\eta=$ 8~000 resulting in
efficiencies of greater than $90\%$ \cite{ALICEppr}. The momentum
resolution of the TPC for tracks with 100 MeV/$c$$<p_{T}<$1 GeV/$c$ is
$\sim$1-2$\%$ \cite{ALICEppr}. For tracks with $p_{T}>1$GeV/$c$, the TPC can be used in
combination with the other tracking detectors for optimal resolution
as shown in Fig.~\ref{fig:TrackRes} from \cite{Glassel}. For example
for 30 GeV tracks, the addition of further tracking information from
the ITS and TRD, in addition to the TPC, improves the momentum
resolution from $\sim$21$\%$ (TPC only) to $\sim$5$\%$.

\begin{figure}[!hbt]
          \center
          \includegraphics[scale=0.4]{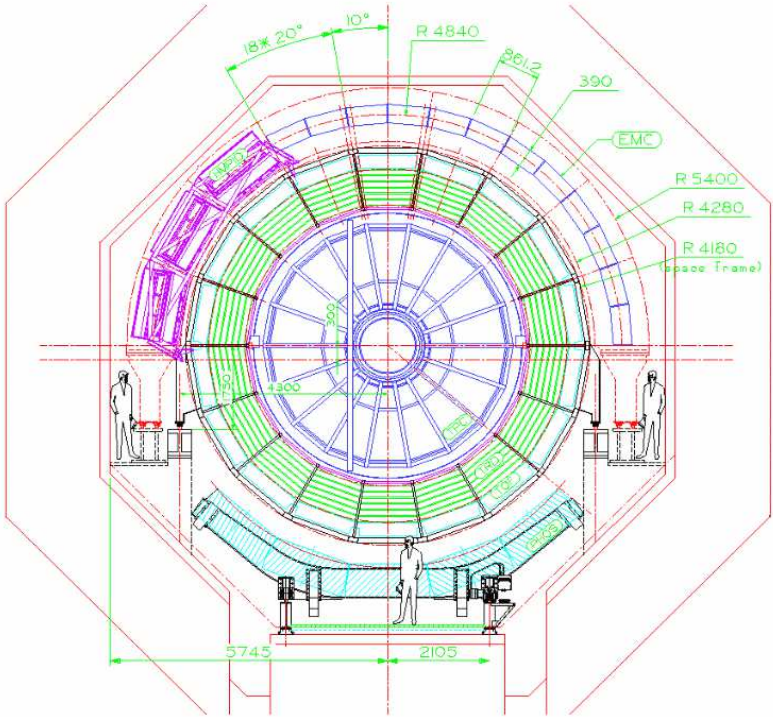}
          \protect\caption{Technical drawing of cross-section through
          ALICE. The electromagnetic calorimeter is shown in the
          outermost layer lying just below the octagonal magnet and
          spanning from $0^{\rm{o}}-120^{\rm{o}}$ (where the
          diagram is looking towards positive $z$ here). From  
          \cite{ALICEWeb}.
          }
          \protect\label{fig:EMCal}
        \end{figure}

The electromagnetic calorimeter (EMCal) will
measure the neutral component of jet energy as discussed in
Section~\ref{sec:ALICEJets}. It will also be used to trigger on
jets. The EMCal is a Pb-scintillator sampling
calorimeter which covers an area of -0.7$<\eta<$0.7 and
$60^{\rm{o}}<\phi<180^{\rm{o}}$, see Fig.~\ref{fig:EMCal}
\cite{ALICEWeb}. It consists of 12 
super-modules which altogether make up a total of
$114(\eta)$ x $168(\phi)=19152$
towers with projective geometry.
From simulations, the EMCal energy resolution is $\sim$15$\%/\sqrt{E}$
\cite{EMCalProposal}.

\chapter{Reconstructing Jets at LHC Energies}

\section{Current Jet Finding Methods}
Jets have been studied for many years in the field of high energy
physics at hadron-hadron colliders and $e^+e^-$
colliders and various algorithms have been developed to identify them
and measure their properties. With the high
energies used to collide heavy ions at the Relativistic Heavy Ion
Collider (RHIC), it has recently become possible to study jets
experimentally in
heavy-ion collisions as well.

\subsection{Methods Used in High-Energy Physics}\label{sec:PPAlgos}
The algorithms used to find and reconstruct jets were first developed
in the field of high-energy physics. Jet production cross-sections can
be predicted using pQCD calculations. Experimental measurements of
the jet cross sections can test the predictions. 
Jet cross-sections have been measured at both $e^+e^-$
and hadron-hadron colliders. However, due to the different event
structures in the two cases, different jet definitions and 
algorithms to reconstruct them, have been developed
\cite{SuccessiveJetCombo}. The aim of both however, is to provide a
mapping between the observed high-energy hadrons and the initial
state energetic partons which were involved in the hard scattering.

In $e^+$+$e^-$ collisions, at the lowest order the initial state is purely electromagnetic
and therefore all the final state hadrons
are due to the annihilation process giving rise to a virtual photon which splits
into a $q\overline{q}$ pair. In hadron-hadron collisions
there are a large number of initial state partons but generally only 
one parton from each of the incident hadrons is involved in a hard
scattering. Therefore, out of all the final state hadrons produced in the
collision, a fraction can be associated with the hard scattering process
and the rest are considered part of the `underlying event'. The final state
hadrons forming 
the `underlying event' contribution are due to soft interactions and
rescattering between
the remaining partons in the incident hadrons. Another difference in the
event structures is that in hadron-hadron collisions, the partons
involved in the hard scattering process produce radiation in the form
of initial state bremsstrahlung and this radiation combined with the
underlying event, produces the `beam jets' observed at hadron
colliders. 


\subsubsection{The Cone Jet Algorithm}
A cone jet definition, originally defined in
\cite{JetsFromQCD} has been used to reconstruct jets in hadron-hadron
collisions \cite{RunII, UA11}. A cone jet is defined as a group of particles whose
3-momenta all lie within a cone of a certain angular size. `Beam
jets' are suppressed by this definition because only a fraction of the
low-$p_{T}$ `beam jet' hadrons will fall inside the cone of a
high-$p_{T}$ jet. Analytically, a cone algorithm groups together all
particles, $i$, whose trajectories lie within a cone, $C$, of a certain radius,
$R$, in ($\eta $,$\phi$)-space where

\begin{equation}
\sqrt{(\eta ^i - \eta ^C)^2 + (\phi ^i - \phi^C)^2} \leq R .
\end{equation}

Experimentally, since the data is usually in the form of output from
electromagnetic or hadronic calorimeters, or a combination of both, the
cone is projected to two 
dimensions in the form of a circle of radius $R$, in ($\eta$,
$\phi$)-space where the calorimeter tower positions can be used to
represent the particle trajectories.
Cone algorithm iterations begin with a trial cone centre around a seed
tower, usually with energy above some threshold (so-called `seed' energy).
The energy of particles inside the cone is used to calculate the
$E_{T}$-weighted centroid of the cone which then becomes the new cone
centre. The same process is repeated until a `stable' cone centre
(when the change in position of the centroid between successive
iterations is small or below a chosen parameter) is found. 
The particles falling inside the stable cone are then
classified as part of the jet. A recombination scheme, for example the original
Snowmass scheme \cite{Snowmass1990}, is defined
to reconstruct the transverse energy and direction of the cone jet as
follows:

\begin{eqnarray}
E^{J}_{T} & = & \sum_{i\subset J=C}E^{i}_{T} = E^{C}_{T} \label{eq:jetEt}\\
\eta ^{J} & = & \frac{1}{E^{J}_{T}} \sum_{i\subset J=C} E^{i}_{T}\eta
^{i} \label{eq:jetEta}\\
\phi ^{J} & = & \frac{1}{E^{J}_{T}} \sum_{i\subset J=C} E^{i}_{T}\phi
^{i} \label{eq:jetPhi}
\end{eqnarray}

A number of corrections are applied to the reconstructed 
jet energy to take into account the non-uniform response of different
types of calorimeters and edges and gaps in the
calorimeters. Attention is also given to the
non-linear response of the calorimeter to low momentum
particles. Further corrections involve
a subtraction of the estimated energy from the `underlying
event' and an addition to take into account the jet energy that lies outside the
chosen cone radius \cite{CDFBocci, CDFDittmann}.  

\begin{figure}
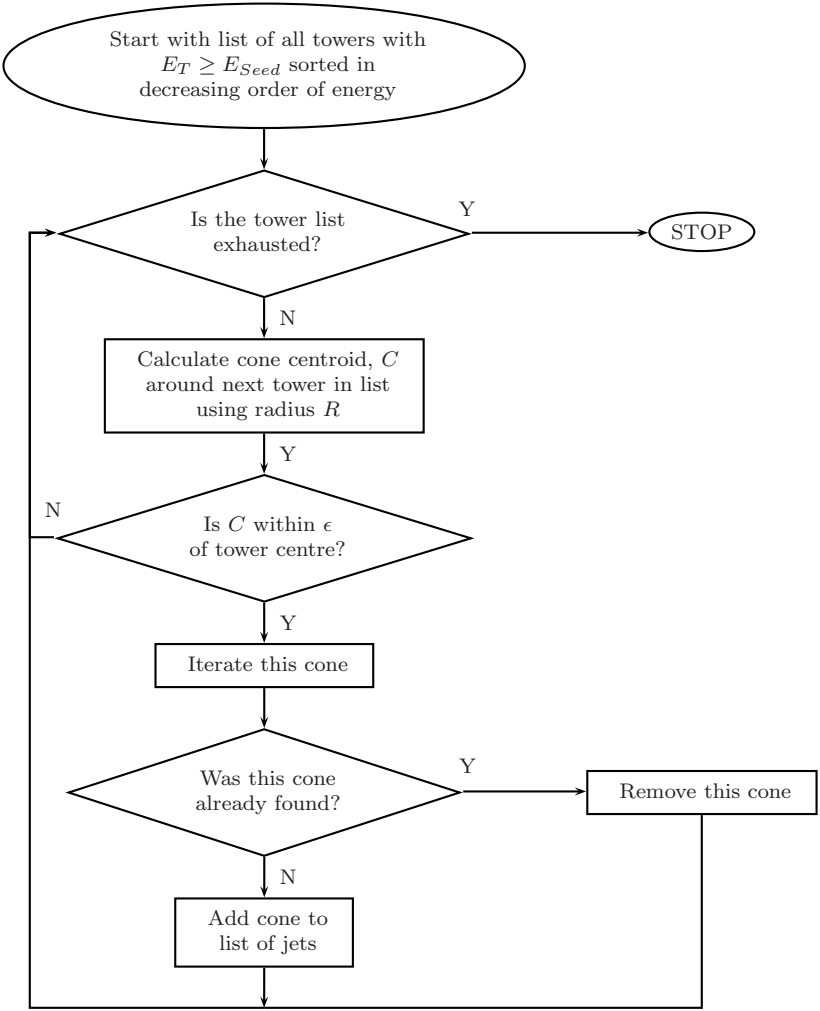
\center\scriptsize
    \begin{psmatrix}[rowsep=0.5cm,colsep=0.8cm]
      \psovalbox[]{
        \begin{tabular}{c}
        Start with list of all towers with\\
        $E_{T} \geq E_{Seed}$ sorted in \\
        decreasing order of energy
      \end{tabular}}\\
      \psdiabox[]{
        \begin{tabular}{c}
          Is the tower list\\
          exhausted?
        \end{tabular}} &
         \psovalbox{STOP} \\
      \psframebox{
        \begin{tabular}{c}
        Calculate cone centroid, $C$\\
        around next tower in list\\
        using radius $R$
        \end{tabular}}\\
      \psdiabox[]{
        \begin{tabular}{c}
          Is $C$ within $\epsilon$ \\
          of tower centre?
        \end{tabular}} \\
      \psframebox{
        \begin{tabular}{c}
          Iterate this cone
        \end{tabular}}\\
      \psdiabox[]{
        \begin{tabular}{c}
          Was this cone\\
          already found?
        \end{tabular}} &
      \psframebox{
        \begin{tabular}{c}
          Remove this cone
        \end{tabular}}\\
      \psframebox{
        \begin{tabular}{c}
          Add cone to\\
          list of jets
        \end{tabular}} 
      \ncline{->}{1,1}{2,1}
      \ncline{->}{2,1}{2,2}
      \rput(1.5, 9.7){Y}
      \ncline{->}{2,1}{3,1}
      \Aput*{N}
      \ncline{->}{3,1}{4,1}
      \Aput*{Y}
      \ncline{->}{4,1}{5,1}
      \Aput*{Y}
      \ncangle[angleA=180,angleB=180]{->}{4,1}{2,1}
      \rput(-4,5.7){N}  \\    
      \ncline{->}{5,1}{6,1}
      \ncline{->}{6,1}{6,2}
      \rput(2.7, 3.3){Y}           
      \ncline{->}{7,1}{8,1}
      \ncangle[angleA=180,angleB=180]{->}{8,1}{2,1}
      \ncangle[angleA=-90]{6,2}{8,1}
      \ncline{->}{6,1}{7,1}
      \Aput*{N}
    \end{psmatrix}
    \caption{Cone jet finding algorithm used at the Tevatron at
      Fermilab, from
      \cite{RunII}.
      }
    \label{fig:ConeAlgoCDF}
    \end{figure}

Fig.~\ref{fig:ConeAlgoCDF} shows the flow chart representation of the
cone algorithm used at the CDF experiment
\cite{RunII}. This cone algorithm does not take into account that jets
may overlap and thus one calorimeter tower may be included in more
than one final jet. In order to take care of this problem, further
`splitting' and `merging' algorithms have been developed
\cite{RunII}. The choice of cone radius differs from experiment to
experiment, for example, $R=1.0$ was originally used at UA1
\cite{UA11} but smaller cone sizes of $R=0.4, 0.7$ have been used at CDF and D\O,
\cite{RunII,CDFJets,CDFLatino}. A discussion of the choice and
optimisation of algorithm parameters is in section~\ref{sec:AlgoOpt}.


\subsubsection{The $k_{T}$ Algorithm}
For the $e^+$+$e^-$ case, the commonly used
algorithm is called the $k_{T}$ algorithm. This method
successively groups sets of particles with `nearby' momenta into
larger sets of particles which are then classified as 
jets \cite{SuccessiveJetCombo}. The jets defined this way typically do
not have regular shapes compared to the cone jets defined in the
cone jet algorithm but with this definition there is no problem of
overlapping jets because each particle (or calorimeter tower) is
uniquely associated with one jet. 
The $k_{T}$ algorithm
has also been adapted for hadron-hadron collisions
\cite{RunII, SuccessiveJetCombo}. In the $k_{T}$ algorithm used at the 
Tevatron, \cite{RunII}, `preclusters' of grouped particles or calorimeter
towers are formed (similar to finding towers with energy above some
`seed' energy in the cone algorithm). Iterating over all preclusters
until there are none left resulted in each precluster having a
four-momentum vector assigned to it where

\begin{equation}
  (E, \textbf{p}) = E(1, \cos\phi \sin\theta , \sin\phi \sin\theta , \cos\theta)
\end{equation}

and $E$ is the energy of the precluster, $\theta$ is the polar angle
with respect to the beam axis, and $\phi$ is the azimuthal angle.
Thus the square of the precluster's momentum and its rapidity can
be calculated:

\begin{eqnarray}
  p_{T}^2 & = & p_{x}^2 + p_{y}^2 \\
  y & = & \frac{1}{2}\ln \frac{E + p_{z}}{E - p_{z}}.
\end{eqnarray}

Next, for each precluster:

\begin{equation}
  d_{i} = p_{T,i}^2
\end{equation}

\noindent and for each pair of preclusters where $i \neq j$

\begin{equation}
  d_{ij} = \min (p_{T,i}^2, p_{T,j}^2)\frac{(y_{i} - y_{j})^2 +
  (\phi_{i} - \phi_{j})^2}{D^2}
\end{equation}

\noindent is defined. $D \approx 1$ and is one of the algorithm parameters. The
minimum of all the $d_{i}$ and $d_{i,j}$ is then found and
labeled $d_{min}$. If $d_{min}$ was originally a $d_{ij}$ then
preclusters $i$ and $j$ are merged to form a new precluster with

\begin{eqnarray}
  E_{ij} & = & E_{i} + E_{j} \\
  p_{ij} & = & p_{i} + p_{j} .
\end{eqnarray}  

\begin{figure}
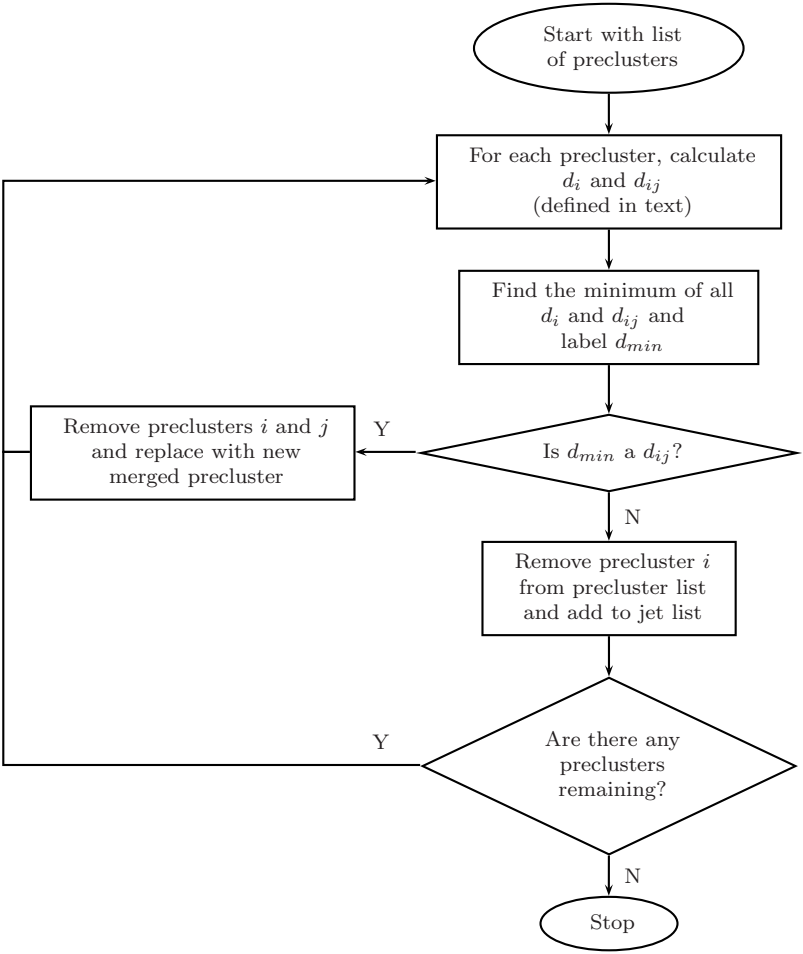
\center\scriptsize
    \begin{psmatrix}[rowsep=0.5cm,colsep=0.8cm]
      &
      \psovalbox[]{
        \begin{tabular}{c}
        Start with list \\
        of preclusters
        \end{tabular}}\\
      &
      \psframebox{
       \begin{tabular}{c}
         For each precluster, calculate \\
         $d_{i}$ and $d_{ij}$\\
         (defined in text)
       \end{tabular}}\\  
      &
      \psframebox{
        \begin{tabular}{c}
          Find the minimum of all\\
          $d_{i}$ and $d_{ij}$ and \\
          label $d_{min}$
        \end{tabular}}\\
      \psframebox{
        \begin{tabular}{c}
          Remove preclusters $i$ and $j$ \\
          and replace with new \\
          merged precluster 
        \end{tabular}}  &
      \psdiabox[]{
        \begin{tabular}{c}
          Is $d_{min}$ a $d_{ij}$?
        \end{tabular}} \\
      &
      \psframebox{
        \begin{tabular}{c}
          Remove precluster $i$ \\
          from precluster list \\
          and add to jet list
        \end{tabular}}\\
      &
      \psdiabox[]{
        \begin{tabular}{c}
       Are there any \\
       preclusters \\
       remaining?
        \end{tabular}}\\
      &
      \psovalbox[]{
        \begin{tabular}{c}
          Stop
        \end{tabular}}\\ 
      \ncline{->}{1,2}{2,2}
      \ncline{->}{2,2}{3,2}
      \ncline{->}{3,2}{4,2}
      \ncline{->}{4,2}{5,2}
      \Aput*{N}
      \ncangle[angleA=180]{->}{4,2}{4,1}
      \Aput*{Y}      
      \ncangles[angleA=180,angleB=180]{->}{4,1}{2,2}
      \ncline{->}{5,2}{6,2}
      \ncline{->}{6,2}{7,2}
      \Aput*{N}      
      \ncangles[angleA=180,angleB=180]{4,1}{6,2}
      \rput(2.5, 3.4){Y}
    \end{psmatrix}
    \caption{$k_{T}$ jet finding algorithm used at the Tevatron at
      Fermilab from 
      \cite{RunII}.
    }
    \label{fig:kTAlgo}
    \end{figure}

Otherwise, if $d_{min}$ is a $d_{i}$, then the precluster is
classified as `not mergeable' and is added to the list of jets. A
flowchart representation of the algorithm is shown in
Fig.~\ref{fig:kTAlgo} taken from \cite{RunII}.

\subsubsection{Energy Summation Methods}
The jets resulting from the fragmentation of hard scattered partons
are composed of charged and neutral hadrons and their hadronic,
leptonic and photonic decay
products i.e.

\begin{equation}
  E_{Jet} = \sum{E_{Charged}} + \sum{E_{\gamma,Leptons}} + \sum{E_{Neutral}}.
\end{equation}

\begin{figure}[!hbt]
          \center
          \includegraphics{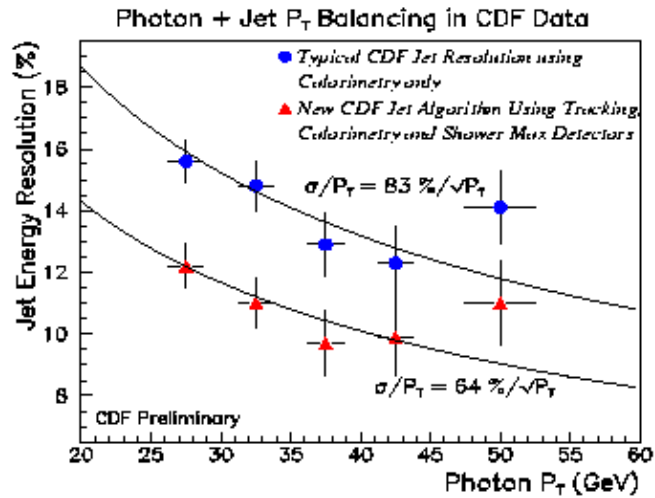}
          \caption{Jet energy resolution as a function of the photon
          energy in the sample of $\gamma$-jet events
          measured
          \cite{CDFLami}.
        }
          \label{fig:CDFResolution}
        \end{figure}

For optimal jet energy resolution, the various components of the
jets need to be taken into account in the jet finding
algorithms.
Therefore, in order to measure full jet energies experimentally,
detectors are needed which can measure the energies of the different
types of particles.
Usually, calorimetry (both electromagnetic and
hadronic) information has been used as input to jet finding algorithms
\cite{RunII,UA11}. 
However, some particle physics experiments, including ALEPH at LEP \cite{Aleph} and
CDF at the Tevatron \cite{CDFBocci}, have also included tracking
information, resulting in much improved jet energy resolution.
In the case of CDF, each track is
associated with a calorimeter tower and thus the combination of
calorimetric and tracking information is
used in jet finding. A process has been devised to take care of double
counting the energy from particles that leave tracks and also deposit
energy in the calorimeters. The improvements in jet energy
resolution at CDF due to the inclusion of tracking information in addition to shower maximum
and calorimetry information
is shown in Fig.~\ref{fig:CDFResolution} from \cite{CDFLami}.

It is also possible to measure certain components of jets, for
example charged particle jets have been measured at CDF
\cite{CDFJets}. Charged particle jets are defined in \cite{CDFJets} as
clusters of charged particles within circular regions of
($\eta$,$\phi$)-space of radius
$R=\sqrt{(\Delta\eta)^2+(\Delta\phi)^2}=0.7$.         
In contrast to using calorimeter
data, in this case a cone jet algorithm was used on tracking data from the
Central Tracking Chamber (CTC).

\subsection{Methods Used in Heavy-Ion Physics}\label{sec:HIMethods}
In the field of heavy-ion physics, jets are a relatively new
observable. At the lower energies of previous heavy-ion experiments,
the cross-sections for hard processes and jet production were low and
it was also difficult to identify low-energy jets above the
background in the events \cite{PRL908,PRL903}.
However, at RHIC, with the increase in collision energy, the jet
cross-section has increased relative to previous heavy-ion experiments
and jet-like signals have been observed at the STAR and 
PHENIX experiments by using a statistical two-particle angular correlation
method. Angular correlation methods were also used in the past for
p+$\overline{\rm{p}}$ collisions, prior to the
development of jet finding algorithms, to confirm the existence of
parton hard scattering and fragmentation into hadrons
\cite{UA13}. When a hard parton fragments, the 
resulting high-$p_{T}$ hadrons are correlated at small $\Delta \eta,
\Delta \phi$ \cite{UA13}.

At the STAR experiment at RHIC, for example, jet-like correlations have been measured
by constructing the
azimuthal correlations of high-$p_{T}$
charged particles detected in the STAR TPC \cite{PRL903}.   
Events containing a trigger particle with transverse momentum 
in a chosen range $trig_{min} < p_{T}(trig) < trig_{max}$ and within
a chosen region of pseudorapidity $|\eta| < x$ are selected. For these events, the
relative azimuthal distribution of other charged tracks with
$track_{min}<p_{T}<trig_{min}$ and in the same $|\eta|$-range,
normalised by the number of high-$p_{T}$ trigger particles, is
constructed. The correlation function is given by

\begin{equation}
  \frac{1}{N_{trigger}}\frac{dN}{d(\Delta \phi)}\equiv
  \frac{1}{N_{trigger}}\frac{1}{\epsilon} \int{d\Delta\eta
  N(\Delta\phi, \Delta\eta)}
\end{equation}

\noindent where $N_{trigger}$ is the number of tracks fulfilling the trigger
requirement, $\epsilon$ is the single track efficiency, and
$N(\Delta\eta, \Delta\phi)$ is the number of observed pairs of hadrons
as a function of relative pseudorapidity, $\Delta\eta$, and relative
azimuth, $\Delta\phi$.

Thus, so far at heavy-ion collision experiments, jets are found on a statistical
basis and the jet properties such as jet energy and direction have not
been able to be measured on an event-by-event basis in heavy-ion collisions.

\section{Specifications for a Jet Finding Algorithm for Use at ALICE}\label{sec:Specs}
At ALICE, the purpose (as discussed in Section~\ref{sec:ALICEJets}) of
finding and measuring jets is to be able to measure the jet fragmentation
functions and thus be able to infer properties of the medium produced
in the heavy-ion collision. The fragmentation function $D(z)$ was
defined in equation~(\ref{eqn:ffunc}) where $z=p_{L}/E_{Tjet}$. 
Replacing $z$ with the quantity
$p_{P}/E_{Tjet}$, where $p_{P}$ is the momentum component of a jet
particle in the direction perpendicular to the jet axis, is also of
interest when looking for modifications of jet fragmentation functions
due to jet-quenching.
Therefore, to
measure accurately the jet fragmentation functions, the \emph{jet direction}, which
approximates the original parton direction, and the \emph{jet
transverse energy}, which approximates the scattered parton transverse
energy, need to be measured. In order to be able to make a meaningful
measurement of the fragmentation functions, a large, unbiased sample
of jets is needed. 

The two-particle correlation methods
used currently at heavy-ion experiments are not appropriate in this
case since they are used to find the existence of jets but do not measure jet
energies or directions accurately. Also, by choosing only events that
satisfy a high-$p_{T}$ charged particle trigger, events where jets are
led by neutral particles can be excluded.
In addition, the trigger requirement biases the jet selection to jets
which fragment primarily to a single high-$p_{T}$ leading particle,
biasing the measured fragmentation function.

\section{Proposed Jet Finding Algorithm for Use at ALICE}

At the ALICE experiment, the concept of using a combination of data
from multiple detectors to find 
jets will be applied, similarly to what has been done in particle physics
experiments\cite{CDFBocci,Aleph}. The 
ALICE TPC will provide highly 
efficient particle tracking capabilities (Section~\ref{sec:ALICEexp}).
The EMCal will act as a fast jet trigger as well as providing
calorimetric data.
Since the cross-section for jet production at the LHC is
expected to be much higher than at RHIC (see
Section~\ref{sec:QGPSearch}) and the ratio of jet signal to background
to be much better, the use of algorithms similar to those used in
particle physics becomes more viable. However, some modifications need
to be implemented to take care of the large background signal and
the underlying event-by-event fluctuations.

The algorithm proposed here is based on a version of
the UA1 cone jet algorithm \cite{STARNOTE196}. Further modifications have 
been made to take into account the background characteristics expected
at the LHC. In its final version, it is very similar to the cone
algorithm presented in Section~\ref{sec:PPAlgos}. The cone algorithm
was chosen as a starting point for 
jet finding in heavy-ion collisions because it has been successfully used
in hadron-hadron collisions. With the very high
multiplicity backgrounds and the signal to background ratios expected in
heavy-ion collisions at the LHC, adapting a $k_{T}$ algorithm would be
more difficult.


\subsubsection{The ALICE Cone Jet Algorithm}
\begin{figure}
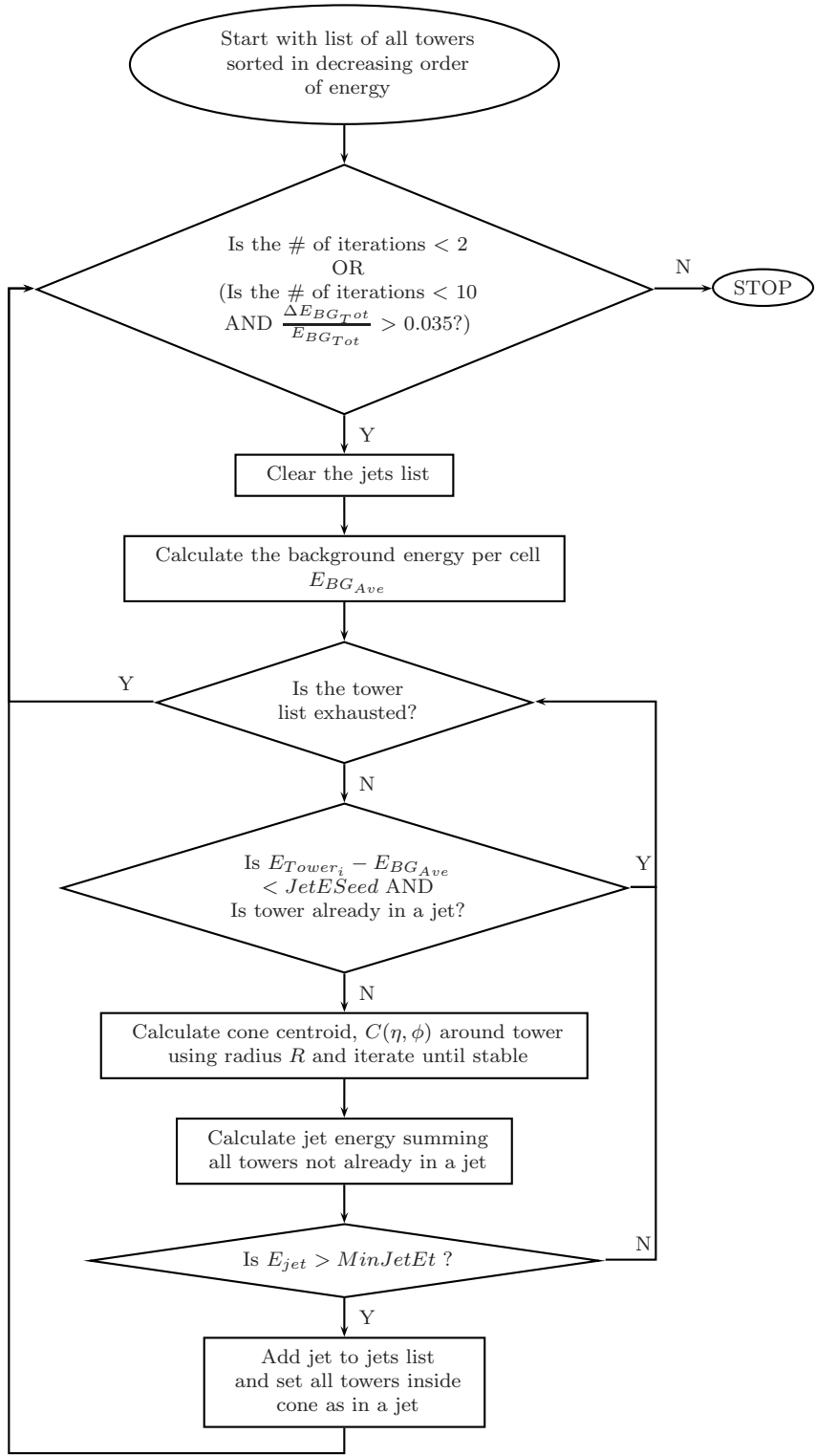
\center\scriptsize
    \begin{psmatrix}[rowsep=0.5cm,colsep=0.8cm]
      \psovalbox[]{
        \begin{tabular}{c}  
        Start with list of all towers\\
        sorted in decreasing order \\
        of energy
      \end{tabular}}\\
      \psdiabox[]{
        \begin{tabular}{c}
          Is the $\#$ of iterations $<$ 2 \\
          OR \\
          (Is the $\#$ of iterations $<$ 10 \\
          AND $\frac{\Delta E_{BG_Tot}}{E_{BG_{Tot}}}$ $>$ 0.035?)
        \end{tabular}} &
         \psovalbox{STOP} \\
         \psframebox{
           \begin{tabular}{c}
             Clear the jets list \\
           \end{tabular}}\\
         \psframebox{
           \begin{tabular}{c}
             Calculate the background energy per cell\\
             $E_{BG_{Ave}}$
           \end{tabular}}\\
      \psdiabox[]{
        \begin{tabular}{c}
          Is the tower \\
          list exhausted?
        \end{tabular}} \\
      \psdiabox[]{
        \begin{tabular}{c}
          Is $E_{Tower_{i}}-E_{BG_{Ave}}$ \\
          $< JetESeed$  AND \\
          Is tower already in a jet?
        \end{tabular}} \\               
      \psframebox{
        \begin{tabular}{c}
        Calculate cone centroid, $C(\eta ,\phi)$ around tower\\
        using radius $R$ and iterate until stable \\
        \end{tabular}}\\
      \psframebox{
        \begin{tabular}{c}
          Calculate jet energy summing \\
          all towers not already in a jet\\
        \end{tabular}}\\
      \psdiabox[]{
        \begin{tabular}{c}
          Is $E_{jet} > MinJetEt$ ?
        \end{tabular}} \\
      \psframebox{
        \begin{tabular}{c}
          Add jet to jets list \\
          and set all towers inside\\
          cone as in a jet
        \end{tabular}}
      \ncline{->}{1,1}{2,1}
      \ncline{->}{2,1}{2,2}
      \Aput*{N}
      \ncline{->}{2,1}{3,1}
      \Aput*{Y}
      \ncline{->}{3,1}{4,1}
      \ncline{->}{4,1}{5,1}
      \ncline{->}{5,1}{6,1}      
      \Aput*{N}
      \ncangle[angleA=180,angleB=180]{->}{5,1}{2,1}
      \rput(-5,9.8){Y}
      \ncline{->}{6,1}{7,1}
       \Aput*{N}     
      \ncangle[angleA=0,angleB=0]{<-}{5,1}{6,1}
       \rput(2.2,7.3){Y}     
      \ncline{->}{7,1}{8,1}
      \ncline{->}{8,1}{9,1}
      \ncline{->}{9,1}{10,1}
      \Aput*{Y}      
      \ncangle[angleA=0,angleB=0]{-}{9,1}{6,1}
       \rput(2.2,2){N}      
      \ncline{->}{10,1}{11,1}
      \ncangles[angleA=-90,angleB=180]{->}{10,1}{2,1}\\ \\    
    \end{psmatrix}
    \caption{Modified cone algorithm for the heavy-ion case at ALICE
      at the LHC.}
    \label{fig:ALICEConeAlgo}
    \end{figure}

The ALICE cone jet algorithm follows the same main steps as the
particle physics cone jet algorithm already discussed and a detailed
flow chart representation of the algorithm is shown in 
Fig.~\ref{fig:ALICEConeAlgo}.

The main
difference between the two algorithms is the addition of an iteration
(Box 2 in 
Fig.~\ref{fig:ALICEConeAlgo}) to ensure more accurate
calculation of the background or `underlying event' on an
event-by-event basis. The data input to the algorithm are a combination
of calorimeter data from the EMCal and tracking data with transverse
energies $E_{T}$ ($\approx p_{T}$) projected onto a grid in ($\eta , \phi$)-space with
cells of the same granularity as the EMCal towers.

Towers are then sorted in decreasing order of $E_{T}$ and the average
background energy per tower, $E_{BG_{Ave}}$ is calculated. Iterating
over all towers, if the tower energy after background subtraction is
greater than a set seed energy, $JetESeed$, and if the
tower is not already part of a jet, then iterations begin to calculate
the centroid of the jet cone. Centroid iterations continue until the
distance between the current centroid and the centroid from the
previous iteration is less than $0.05R$ and the distance between the
initiator tower and the calculated centroid is smaller than
$0.15R$. (These parameters are tunable and here the choice of parameters
in \cite{STARNOTE196} is followed.) Once the centroid
is found, the jet energy $E_{T}^{J}$, is 
calculated by summing the energy of each tower, after background
subtraction, that is within the cone radius, $R = \sqrt{(\eta^i -
  \eta^C)^2+(\phi^i - \phi^C)^2}$. If the calculated
jet energy is greater than a minimum value, $MinJetEt$, then the jet
is classified as valid and added to the list of jets. The scheme used
to calculate the jet cone centroid position in ($\eta ,\phi$)-space
and the jet transverse energy is the same as the original Snowmass
scheme \cite{Snowmass1990} used in particle physics as given by
equations~\ref{eq:jetEt}-\ref{eq:jetPhi}. 
There are
three parameters, namely the cone radius $R$, the jet seed energy,
$JetESeed$, and the minimum allowed jet energy, $MinJetEt$, that need
to be optimised to produce the optimal jet energy resolution and
jet finding efficiency. The algorithm can also be made into a seedless
algorithm by setting the $JetESeed = 0$. The optimisation analysis is presented in
Chapter~\ref{sec:Analysis} and results using the optimised parameters
are discussed in Chapter~\ref{sec:Results}.

\section{Simulation of jets in the ALICE framework}\label{section:Simulations}
Two Monte Carlo event generators were used to simulate the data for
high-$p_{T}$ jets in heavy-ion collisions at $\sqrt{s_{NN}} = 5.5$ TeV
at ALICE \cite{AliceWorkshop}. Within the ALICE software framework, namely AliRoot
\cite{AliRoot}, PYTHIA 6.2 \cite{Pythia} was used to simulate the
high-$p_{T}$ jets 
and HIJING 1.36 \cite{Hijing} was used to simulate the high
multiplicity background or `underlying event'. In order to simulate a
jet event in a heavy-ion collision, a PYTHIA p+p jet event was
superimposed on a HIJING Pb+Pb event. The passage of particles
through the materials of the ALICE detectors was simulated using GEANT3 \cite{GEANT}.
The EMCal response was
simulated using GEANT3 and a parameterised detector response was used
for charged tracks in the ALICE tracking system.

\subsubsection{Simulations Note}\label{Birk}
Immediately prior to the submission of this thesis, it was found that
Birks' formula\footnote{$\frac{dL}{dx}=\frac{A(dE/dx)}{1+kB(dE/dx)}$
  where $dL/dx$ is the light output per unit length, $dE/dx$ is the
  energy deposited per unit length, $A$ is the
  absolute scintillation efficiency and $kB$ is a parameter which
  relates the density of ionisation centres to $dE/dx$ \cite{Leo}.}
\cite{Leo}, describing the deviation from a 
linear response of the light emitted due to the
energy deposited by a particle in the scintillator, had not been applied in the
simulations \cite{AliceWorkshop}. The deviations for heavier, higher
ionising particles are larger than for lighter 
particles. The light response of the scintillator is
dependent on the type of particle depositing energy, the amount of
energy deposited and the ionisation of the particle.
Preliminary simulations \cite{Heather} have shown that this is a small
effect ($<10\%$) and is not expected to have a large impact on the analysis
and results presented in this thesis.

\subsection{PYTHIA Events}
PYTHIA is a Monte Carlo event generator which generates high energy
physics events \cite{Pythia}.
The model has been tuned to reproduce the same detail and results as
obtained from experimental measurements.
PYTHIA includes QCD
and QED as well as other theories beyond the standard model. In
this case PYTHIA was used to generate jets from hard scatterings in
p+p collisions at $\sqrt{s_{NN}} = 5.5$ TeV. Certain nominal jet
energies were chosen at which to simulate jets so that the results
produced by the algorithm could be reconciled with the input jet
energies. Jets were generated within narrow windows of $\pm 2$ GeV
around the nominal energies of 30 GeV, 50 GeV, 75 GeV, and 100 GeV. Since
the statistical error $\sigma \sim$1$/\sqrt{n}$ where 
$n$ is the number of events, 10~000
events were generated at each energy to keep the statistical error to
the $1\%$ level. A condition applied to the events ensured that at
least one jet was pointing towards the fiducial volume of the
EMCal. Initial and final state radiation were included because
the simulated interaction was at a high energy where the emission of
hard gluons plays a large role, relative to fragmentation, in determining the
structure of events \cite{PYTHIAManual}. Initial state radiation is
the radiation of gluons from a parton before it interacts in a hard
scattering and final state radiation describes the radiation of gluons
from the hard scattered parton before it fragments into a jet. More
specifically, the parameters for the generation of the 
PYTHIA events with jets of nominal energy $x$ GeV were as follows:

\begin{itemize}
  \item Initial state and final state gluon radiation ON
  \item $\sqrt{s_{NN}} = 5.5 \ \rm{TeV}$  
  \item $x - 2 \ \rm{GeV}$ $< E_{T_{Jet}} < x + 2 \ \rm{GeV}$
  \item $\eta_{Jet} < |0.3|$
  \item $75^{\rm{o}} < \phi_{Jet} < 165^{\rm{o}}$
  \item Structure function is GRVLO98
  \item $x \ \rm{GeV}$ $< p_{T_{Hard Scattering}} < x+0.1  \ \rm{GeV}$
  \item All decay types ON  
\end{itemize}

\subsection{HIJING Events}
HIJING (Heavy Ion Jet Interaction Generator) is a Monte Carlo event
generator for A+A collisions \cite{Hijing}. It uses a perturbative QCD approach,
based on PYTHIA, 
to model multiple jet production in the collisions. HIJING also
includes the production of multiple mini-jets which dominate events
with high multiplicities \cite{Hijing} and which 
can lead to correlations in transverse momentum observables and
fluctuation enhancements. HIJING therefore attempts to model the
observed event-by-event fluctuations in the `underlying event' in
heavy-ion collisions. 

\begin{figure}[!hbt]  
          \center
          \includegraphics[scale=0.6]{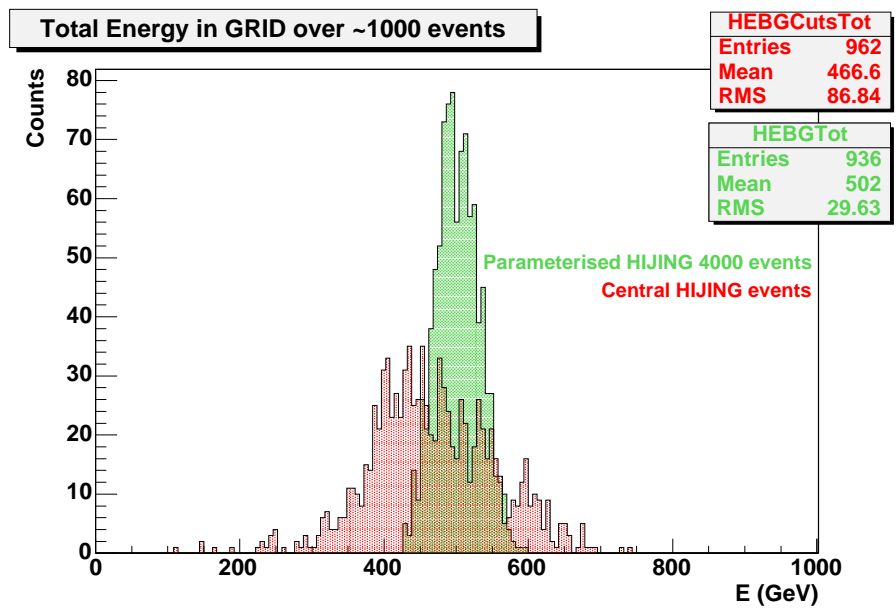}
          \caption{\label{fig:CentralParamHIJING}Total GRID energy for
          Parameterised HIJING 4000 events (shown in green) and for
          Central HIJING events (shown in red).}
        \end{figure}

Two types of HIJING events were simulated, namely
central HIJING events and parameterised HIJING events. Parameterised
HIJING events are a parameterised version of
real HIJING events containing only pions and kaons sampled from the
$p_{T}$ and $\eta$ distributions of particles in real HIJING. The multiplicity
of the events can be set to any $dN/dy$ and for the purpose of this
thesis, a multiplicity of $dN/dy =$ 4~000 was used and the event type
will be referred to as parameterised HIJING 4000 from now on.
As shown in Fig.~\ref{fig:Multiplicity}, the predicted charged
particle multiplicity for central 
Pb+Pb collisions at $\sqrt{s_{NN}}=5.5$ TeV is $\sim$2~500. Therefore,
in order to be conservative in the simulations, a multiplicity of 4~000
was simulated in the events. An
advantage of using parameterised HIJING events is that they do not
require detailed calculations to be performed since they sample
different distributions and the computing time required is relatively
short.
However, since these events are a parameterisation, they do not accurately
simulate the event-by-event background fluctuations as seen in full
HIJING simulations and real data. Fig.~\ref{fig:CentralParamHIJING}
shows the energy distributions in a grid in ($\eta, \phi$)-space with
cells of the same granularity as the EMCal from full central HIJING events
(shown in red) and parameterised HIJING 4000 events (shown in
green). The distributions consist of a combination of energy from
charged particles leaving tracks
in the TPC and which were pointing in the direction of the
EMCal, and from particles which deposited energy in the
EMCal. The parameterised HIJING 4000 events show a much narrower
distribution than the full central HIJING events with fluctuations of the
order of $6\%$ compared to the full central HIJING fluctuations which are
of the order of $19\%$.  

Almost all simulations of jet events with background shown here
consist of PYTHIA events mixed with full HIJING events. The cases
where parameterised HIJING 4000 events are used, are clearly noted.

The full HIJING events that were simulated will be denoted `central HIJING'
events from now on. Multiplicity in heavy-ion collisions is
proportional to the event centrality and thus the highest multiplicity
events were generated in order to test the algorithm in a very conservative 
situation. Since full HIJING event simulations take many
hours to run and have a size of the order of 400 megabytes per event, 1~000 events of
this type were simulated. When performing analysis, the events were
mixed randomly with the PYTHIA 
events. The central HIJING event generation parameters were as follows:

\begin{itemize}
  \item $\sqrt{s_{NN}} = 5.5 \ \rm{TeV}$
  \item Jet-quenching for new LHC parameters with log(e) dependence ON \\
  (\emph{Special parameters from ALICE workshop 2003})
  \item Gluon shadowing ON  
  \item Impact parameter $b \leq 5$ fm
  \item Initial state and final state radiation ON
  \item Decays of $\pi^0$, $K^{0}_{s}$, $D^\pm$, $\Lambda$,
  $\Sigma^{\pm}$ OFF 
\end{itemize}





\chapter{Analysis}\label{sec:Analysis}

The jet data analysed in this section were simulated using PYTHIA 6.2
and HIJING 1.36 Monte Carlo event generators.
The process used to reconstruct jets and a description of
the optimisation of the jet finding algorithm
will be described in the following sections.

\section{Input Data and Preparation for Jet Finding}

\subsection{Jet Finding Data Preparation}

\begin{figure}[!hbt]
          \center
          \includegraphics[scale=0.9]{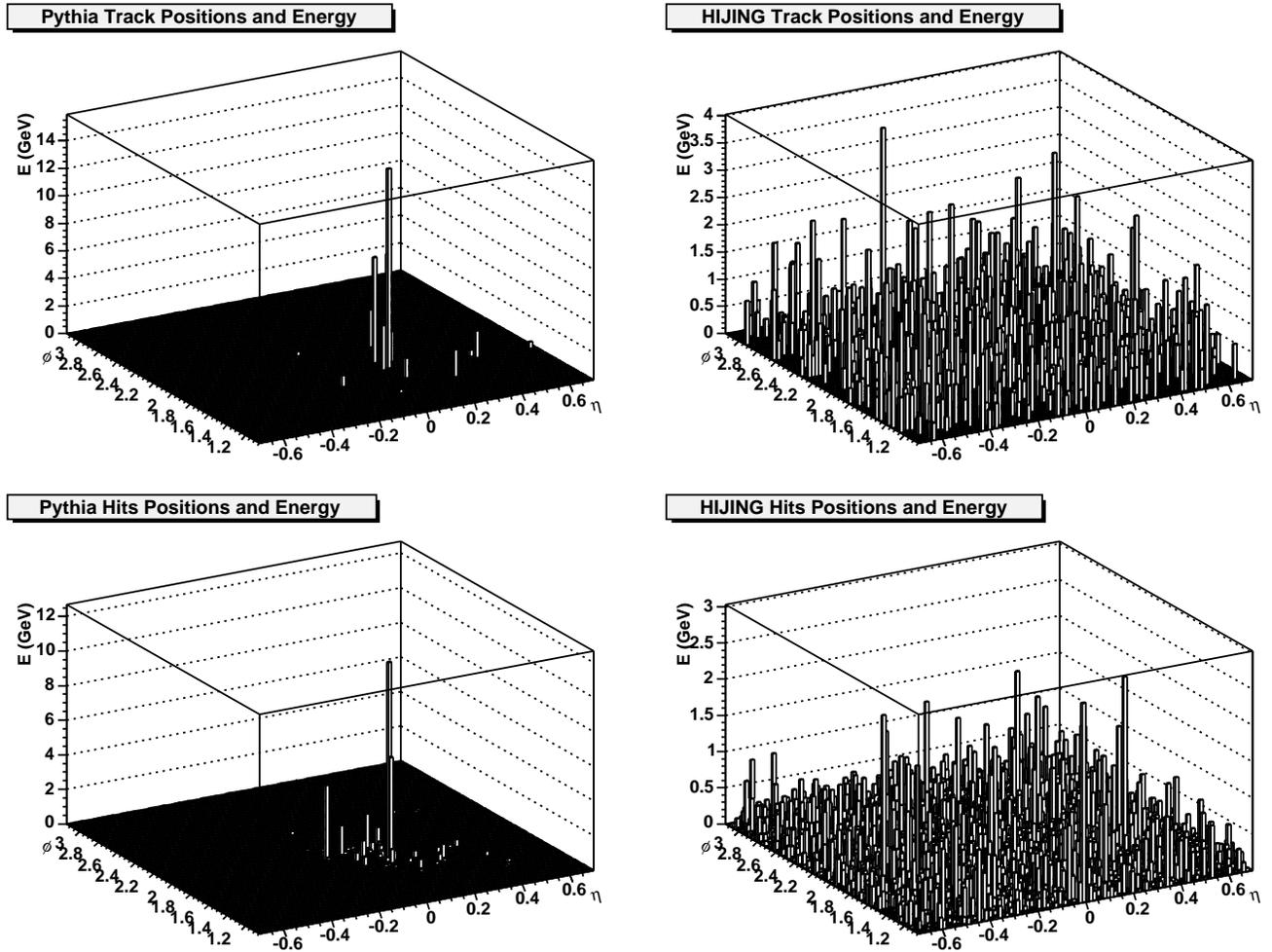}
          \caption{\label{fig:TracksAndHits} Left hand top(bottom) plot
          shows the energy from PYTHIA charged tracks(hits) in the grid for a 100 GeV PYTHIA
          event before data preparation is performed. Right hand
          top(bottom) plot shows the energy from Central HIJING
          charged tracks(hits) in the grid 
          before any data preparation. }
        \end{figure}

\begin{figure}[!hbt]
          \center
          \includegraphics[scale=0.9]{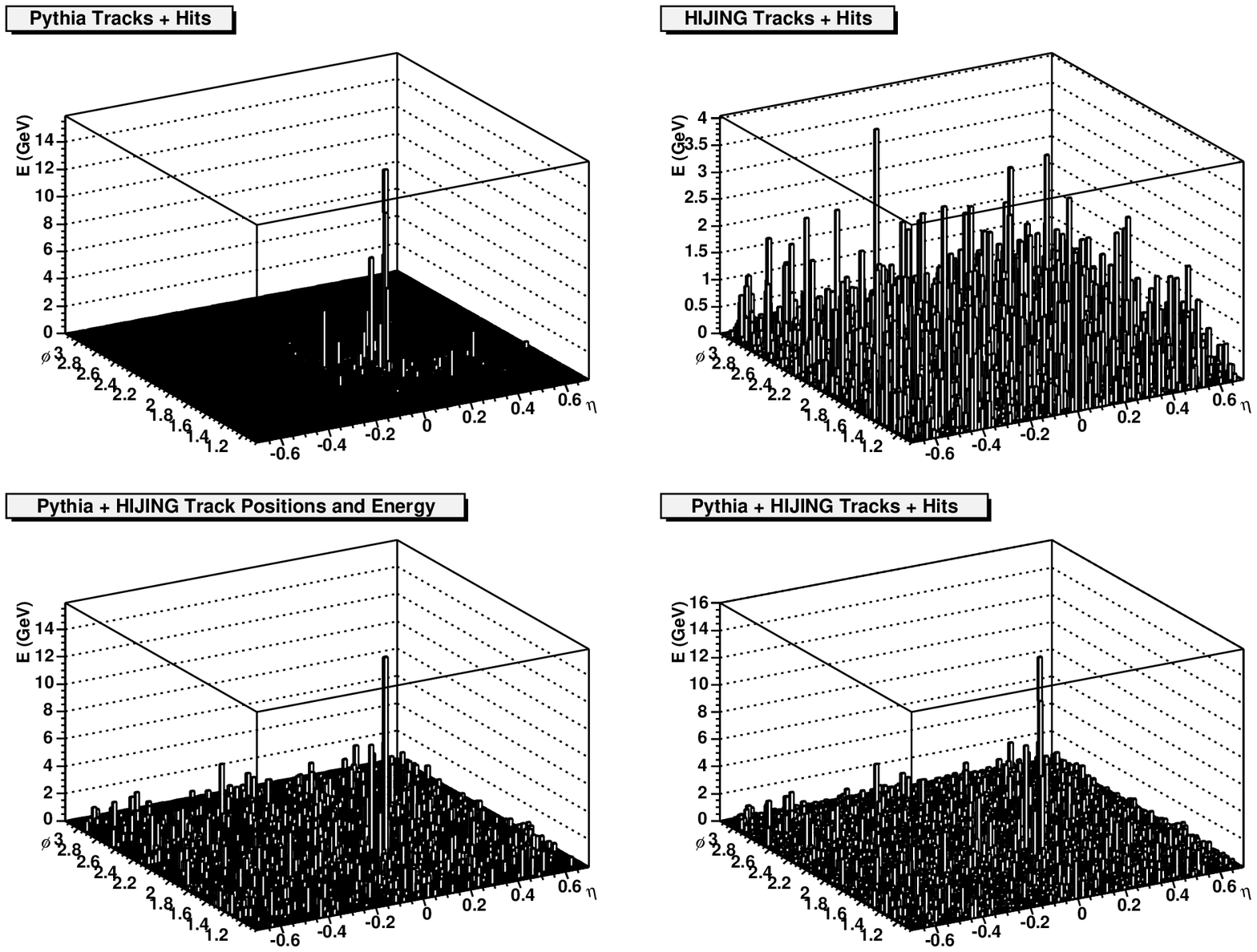}
          \caption{\label{fig:TotalTracksAndHits} Top left hand plot
          shows the energy from tracks plus hits for a 100 GeV Pythia 
            event, top right hand plot shows the same for a Central
            HIJING event. Bottom left shows 
            sum of track energies from a 100 GeV Pythia event and a Central
            HIJING event, while bottom right hand plot 
            shows the sum of all track and hit energies from the 100 GeV
            Pythia and HIJING events.} 
        \end{figure}

        Since information from both the tracking detectors and the EMCal
        needed to be combined for jet finding, an energy grid was created with
        the same fiducial range as the EMCal ($\vert\eta\vert <0.7$,
        $\pi/3\leq\phi\leq\pi$) and with the same number of cells as
        the EMCal towers ($13824 = 96(\eta) \times 144(\phi)$).
        After some data preparation (detail to follow), the grid was
        then filled with energy from
        calorimeter hits and from charged tracks
        from primary particles measured in the TPC that were
        originally pointing towards the EMCal fiducial area, see
        Fig.~\ref{fig:TracksAndHits} and
        Fig.~\ref{fig:TotalTracksAndHits}.

        \subsubsection{Time cut}
\begin{figure}[!hbt]
          \centering
          \includegraphics[scale=0.8]{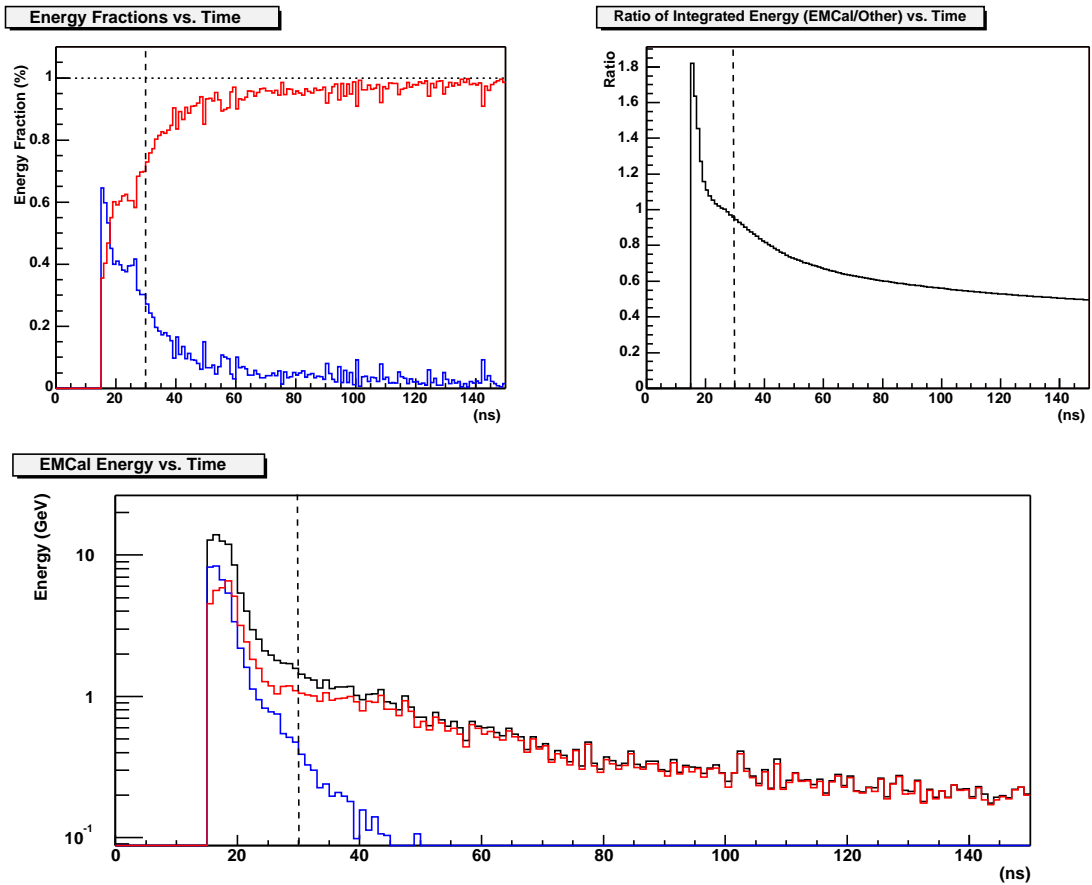} 
          \caption{ Top left plot shows the source of the energy in the
          EMCal as a function of time after a collision. The blue
            histogram represents the energy deposited due to particles
            initially pointing 
            into the EMCal fiducial volume. The red histogram
            shows the energy fraction due to particles that were orginally
            pointing outside the EMCal fiducial volume. The top right
            hand plot shows the ratio of the energy due to primary particles
            initially inside the EMCal acceptance to the energy due to
            primaries outside the EMCal acceptance. The bottom plot
            shows the energy deposited in the EMCal from the two
            sources as a function of time. The vertical dashed line in
            all plots indicates the time cut of 30 ns. From
            \cite{Mark}.
          }
          \label{fig:Marktimecut}
        \end{figure}

        The first step in preparing the data for jet finding was to
        apply a time cut on the hits measured in the EMCal,
        i.e. all hits
        measured after 30 ns were not added to the grid. The time cut
        suppressed the amount of background energy measured due
        to backscattering in the detector. There will also be a
        hardware time cut implemented to do this on real data. Fig.~\ref{fig:Marktimecut}
        shows the energy deposited in the EMCal as a function of time
        after a collision. In all panels, the dashed vertical line
        indicates the time cut at 30 ns.  The top left hand plot shows the source of
        the energy in the 
        EMCal as a function of time where the blue histogram
        represents the fraction of energy deposited in 
        the EMCal by primary particles that were originally
        pointing into the EMCal fiducial volume and the red histogram
        shows the fraction due to particles that were orginally
        pointing outside the EMCal fiducial volume. It can be seen that
        most of the energy from primary particles initially pointing in the
        correct direction, is deposited early while the major
        energy contribution after $\sim$30 ns  is due to primary particles
        that were not pointing in the direction of the EMCal
        initially. The top right hand plot in
        Fig.~\ref{fig:Marktimecut} shows the ratio of the energy due
        to primary particles
        initially inside the EMCal acceptance to the energy due to
        primary particles outside the EMCal acceptance (i.e. the ratio of the blue
        histogram to the red histogram in the left hand figure). The
        energy due to particles orginally outside the EMCal acceptance
        overtakes the energy due to particles inside the acceptance
        range after $\sim$30 ns as can be seen by the ratio dropping
        below 1 around $30$ ns. The bottom plot in
        Fig.~\ref{fig:Marktimecut} shows the absolute energy deposit
        from the different primary particles 
        in units of GeV as a function of time. The effect of the time
        cut on the energy deposited by primary particles pointing into the
        EMCal acceptance was to exclude $\sim$10.4$\%$ of the energy.

        The time cut was found
        to reduce the energy measured in the grid by a factor of
        $\sim$1/3 and to increase the ratio of energy due to primary
        particles inside the EMCal acceptance to energy due to primary
        particles outside the EMCal acceptance to $\sim$1, see
        Fig.~\ref{fig:Marktimecut}\footnote{Preliminary simulations
        \cite{Heather} which include Birks' formula (see
        section~\ref{Birk}) show that the energy deposition before 30
        ns remains approximately the same whereas the deposition after
        the time cut is reduced since Birks' formula affects mainly
        heavier particles. For this thesis all jet analysis
        is based on energy deposited \emph{before} 30 ns and therefore
        the presented jet reconstruction results are not expected to be affected.}.

        \subsubsection{Hadron Correction}
        The next correction to be made was to eliminate the double counting of
        energy from 
        charged hadrons measured in both the TPC and the EMCal. All
        charged hadrons leave tracks in the TPC and will deposit energy
        in the EMCal if they are within its fiducial volume. Therefore,
        when adding track energy and hit 
        energy, some double counting of energy occurs.
        To correct for this, an average amount of energy,
        $\langle E_{HC}(\eta,p_{T})\rangle$, was subtracted
        from all the grid cells whose coordinates in ($\eta , 
        \phi$)-space matched the direction in which a primary track was
        initially pointing. In order to calculate 
        $\langle E_{HC}(\eta,p_{T})\rangle$ for each primary track, 10~000
        events each were simulated for seven different particle momenta and six different
        $\eta$-directions, where a
        single charged pion was detected in both the TPC and the
        EMCal \cite{Mark}.

       \begin{figure}[!hbt]
          \centering
          \includegraphics[scale=0.6]{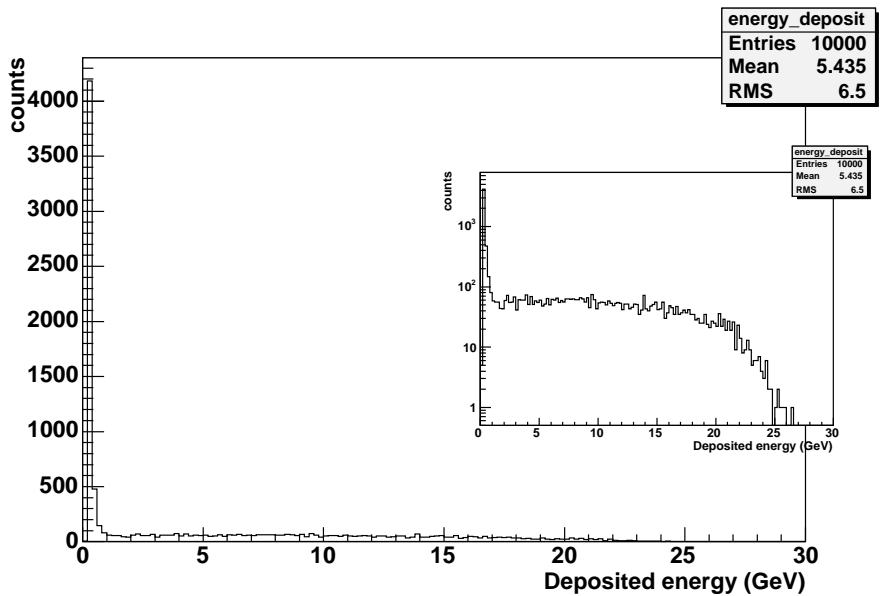}
          \caption{Main plot shows the distribution of deposited
            energy in the EMCal from 10~000 
          $\pi^{+}$ each with an energy of 25 GeV and $\eta$-direction
          = 0.0. In this case
          $\langle E_{dep}\rangle=5.4$ GeV. The inset shows the same
          plot with a log scale. From
          \cite{Mark}. 
          }
          \label{fig:MipPeak}
        \end{figure}

        Fig.~\ref{fig:MipPeak} shows that the mean energy deposited by
        a 25 Gev $\pi^{+}$ in the calorimeter is 5.4 GeV. 
        For each case of particle momentum and direction, the
        average energy deposited $\langle E_{dep}\rangle$ was
        calculated. A parameterisation of 
        the energy deposition as a function of particle $p_{T}$ and
        $\eta$-direction was obtained from fitting the $\langle
        E_{dep}\rangle$ as a function of $\eta$ and $p_{T}$ \cite{Mark}. For each
        charged track in the EMCal fiducial volume, $\langle
        E_{HC}(\eta,p_{T})\rangle$ was calculated using the
        parameterisation and then subtracted from the relevant grid
        cell\footnote{With the inclusion of Birks' formula, the energy 
        deposition of hadrons in the calorimeter is expected to
        be reduced by a small amount ($\sim$few $\%$ effect
        \cite{Heather}). However, since Birks' formula was not taken
        into account when calculating the total energy deposition of
        hadrons in the calorimeter or in the hadron correction which
        subtracts this energy, the
        effect should be effectively cancelled out in the jet analysis
        which follows.}.

        \subsubsection{Track $p_{T}$-cut}
\begin{figure}[!hbt]
          \center
          \includegraphics[scale=0.6]{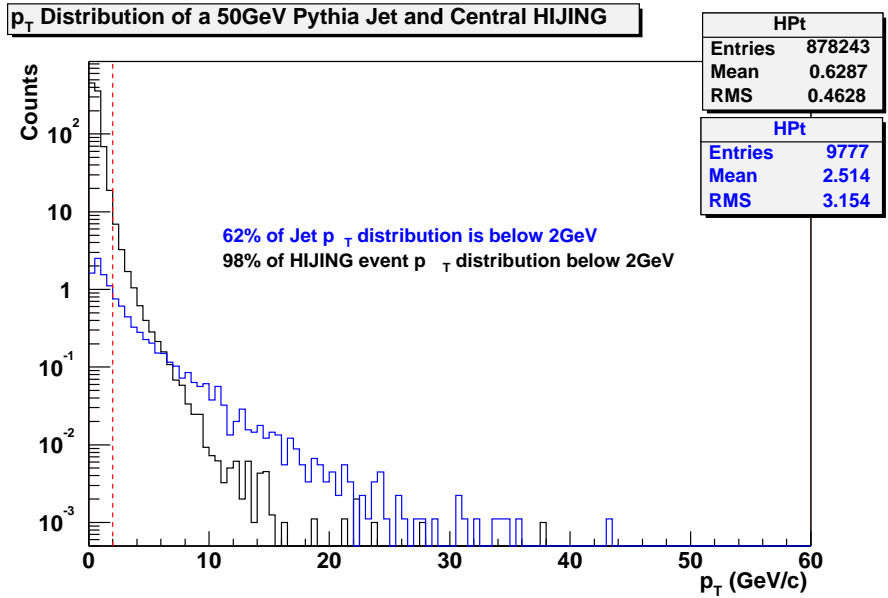}
          \caption{\label{fig:PtDist} $p_{T}$ Distributions averaged
          over $\sim1000$ events for 50 GeV  
            PYTHIA events (in blue) and Central HIJING events (in
            black) respectively.
            The dotted line indicates the $p_{T}$-cut at
            2.0 GeV/$c$.}
        \end{figure}

        A further method used to reduce the background energy
        in the grid from the underlying event, was to perform a
        $p_{T}$-cut on all tracks in the grid.
        The average $p_{T}$ distributions for charged tracks for a jet event and
        background
        event (averaged over many events) can be seen in Fig.~\ref{fig:PtDist}.
        The average jet track $p_{T} = 2.5$ GeV/$c$ whereas the average
        background track $p_{T} = 0.63$ GeV/$c$. Thus, a $p_{T}$-cut of
        $2.0$ GeV/$c$
        excluded $98\%$ of background tracks and $62\%$ of jet tracks,
        see Fig.~\ref{fig:PtDist}.


        \subsection{Study of the Characteristics of the `Underlying Event'}\label{sec:BGFluct}

        \subsubsection{Event-by-Event Background Energy Fluctuations}
        The main cause of complexity in jet finding in heavy-ion
        collisions, is the size of the
        event-by-event fluctuations of
        energy deposited in the detectors by particles from the
        `underlying event'. Therefore a more detailed study was performed to
        quantify the amount, and spread, of the background energy
        event-by-event.
        The impact that a $p_{T}$-cut on the track
        momenta and a $30$ ns time cut on hits in the EMCal would have
        on the background energy distribution in
        the grid was also studied.

        \begin{figure}[!hbt]
          \center
          \includegraphics[scale=0.6]{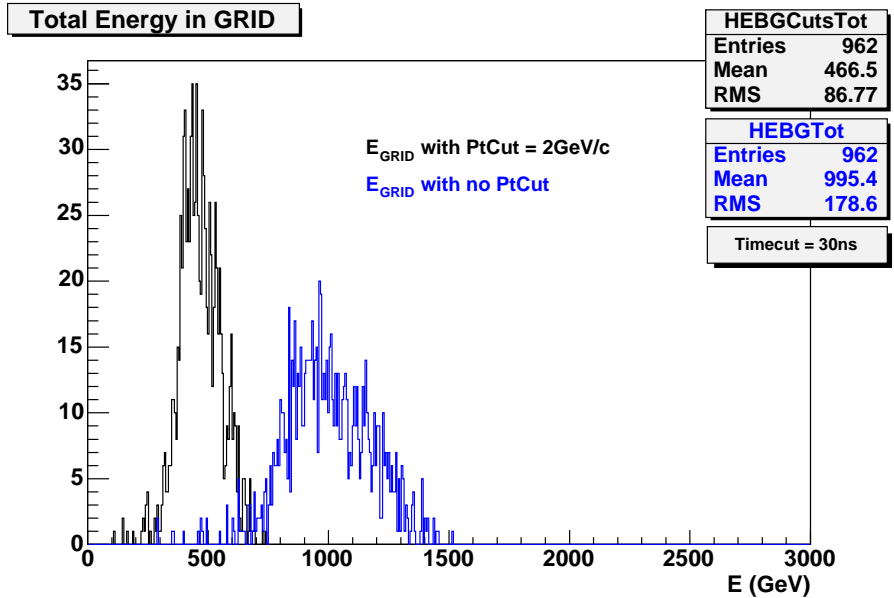}
          \caption{\label{fig:BGFluct} Energy in the grid from 1~000
            Central HIJING
            background events. Blue indicates the distribution without a 
            $p_{T}$-cut and black indicates the resulting distribution with
            $p_{T}$-cut = 2 GeV.}
        \end{figure}
        
        Fig.~\ref{fig:BGFluct}
        shows the distribution of total energy deposited in the
        grid for 1000 Central HIJING events with and without a
        $p_{T}$-cut. For both cases, a time cut of $30$ ns was
        applied. The mean energy deposited in the grid with a
        $p_{T}$-cut imposed, was 467 GeV with large event-by-event
        fluctuations of the order of $20\%$. The fluctuations without a
        $p_{T}$-cut were also of the order of $20\%$ but the mean
        deposited energy was 995 GeV. Thus the addition of the 2 GeV $ p_{T}$-cut,
        while excluding $98\%$ of background tracks, reduced the background
        grid energy by a factor of $2$.

\begin{figure}[!hbt]
          \center
          \includegraphics[scale=0.6]{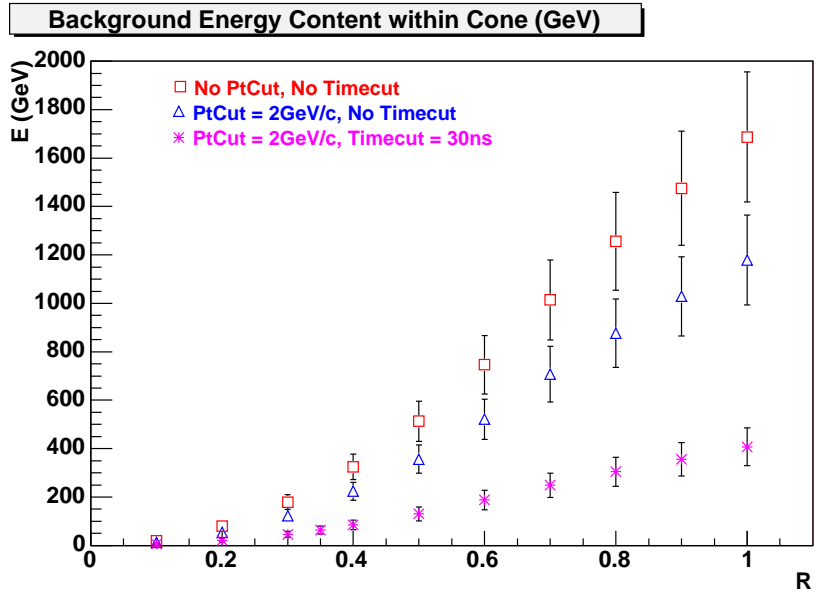}
          \caption{\label{fig:BGEnergyInCone} Background energy
          contained within a cone as a function of cone radius,
          $R$. The symbols represent the 
          mean of the distributions for set $R$ and the error bars
          represent the RMS for each distribution. Squares represent
          no $p_{T}$-cut or time cut, triangles represent $p_{T}$-cut
          only and stars represent $p_{T}$-cut and a time cut on the data. }
        \end{figure}

        The effect of a $p_{T}$-cut and a time cut on the background
        energy as a function of cone radius $R$ was also
        studied. For 1~000 Central HIJING events, the amount of
        energy was summed within different sized cones which were randomly positioned
        on the grid. The results can be seen in
        Fig~\ref{fig:BGEnergyInCone}. The addition of the $p_{T}$-cut
        and time cut was found to reduce the
        absolute size of the fluctuations and the mean energy in the cone
        by $\sim$70$\%$ with the time cut having a larger effect ($\sim$40$\%$) than
        the $p_{T}$-cut ($\sim$30$\%$).

\begin{figure}[!hbt]
          \center
          \includegraphics[scale=0.6]{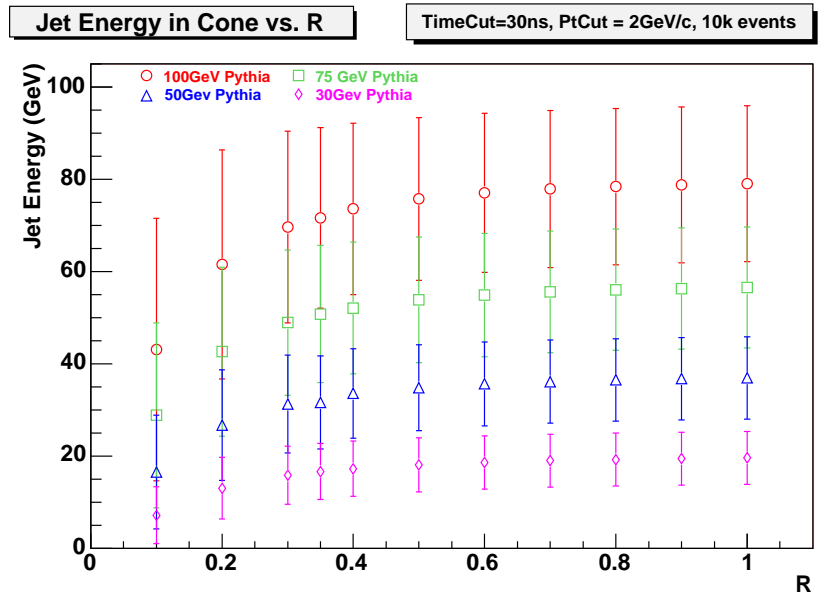}
          \protect\caption{ Sum of $E_{T}$ for all tracks plus hits for pure
          PYTHIA jets of energies 100 GeV, 75 GeV, 50 GeV, 30 GeV respectively, within
          varying cone radii $R$. }
          \protect\label{fig:JetECone}
        \end{figure}

        The next step in the
        background study was to determine how the background
        energy contained within various cone radii, $R$, compared to the
        amount of jet energy contained within the same $R$.
        For 10~000 PYTHIA events of different energies
        (30 GeV, 50 GeV, 75 GeV, 100 GeV), the energy from the
        tracks and hits was summed within cones of varying $R$, see
        Fig.~\ref{fig:JetECone}. The symbols in the
        plot represent the means and the error bars represent the RMS of
        the distributions. The amount of energy inside the jet cone
        can be seen to saturate for $R \geq 0.6$.

\begin{figure}[!hbt]
          \center
          \includegraphics[scale=0.6]{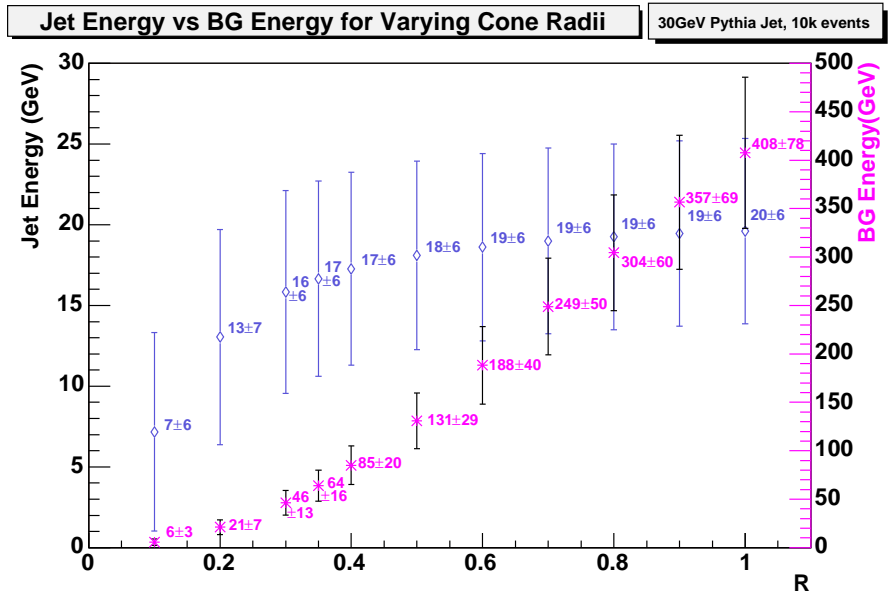}
          \caption{\label{fig:JetAndBGCompare30} Energy
          within varying cone radii, $R$, for 30 GeV jets compared to
          background energy. Symbols represent the mean of the distribution and
          error bars represent the RMS. The axis on the left is the
          jet energy scale and on the right is the background energy scale.}
        \end{figure}

\begin{figure}[!hbt]
          \center
          \includegraphics[scale=0.6]{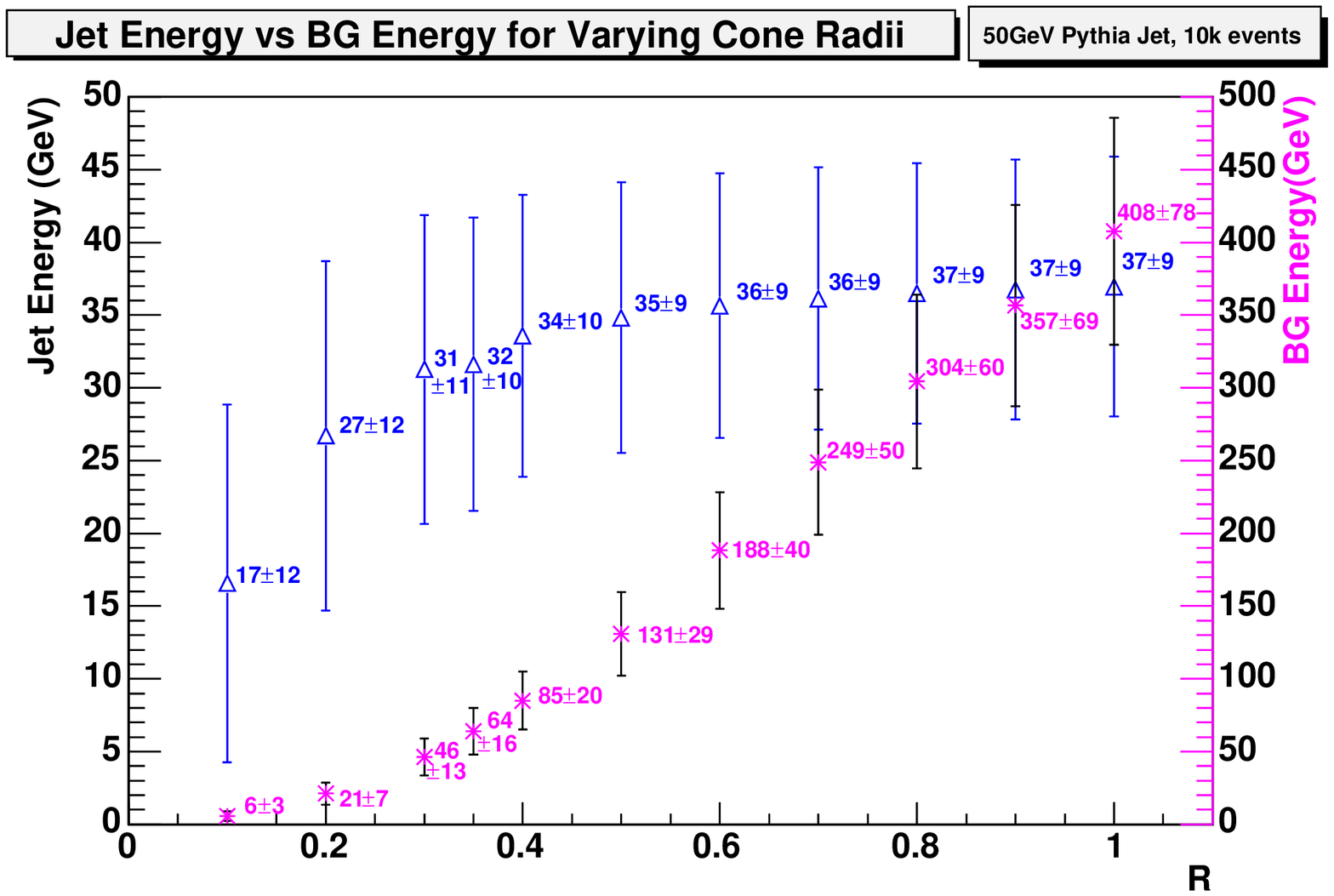}
          \caption{\label{fig:JetAndBGCompare50} As for
          Fig.~\ref{fig:JetAndBGCompare30} but jet energy = 50 GeV.}
        \end{figure}

\begin{figure}[!hbt]
          \center
          \includegraphics[scale=0.6]{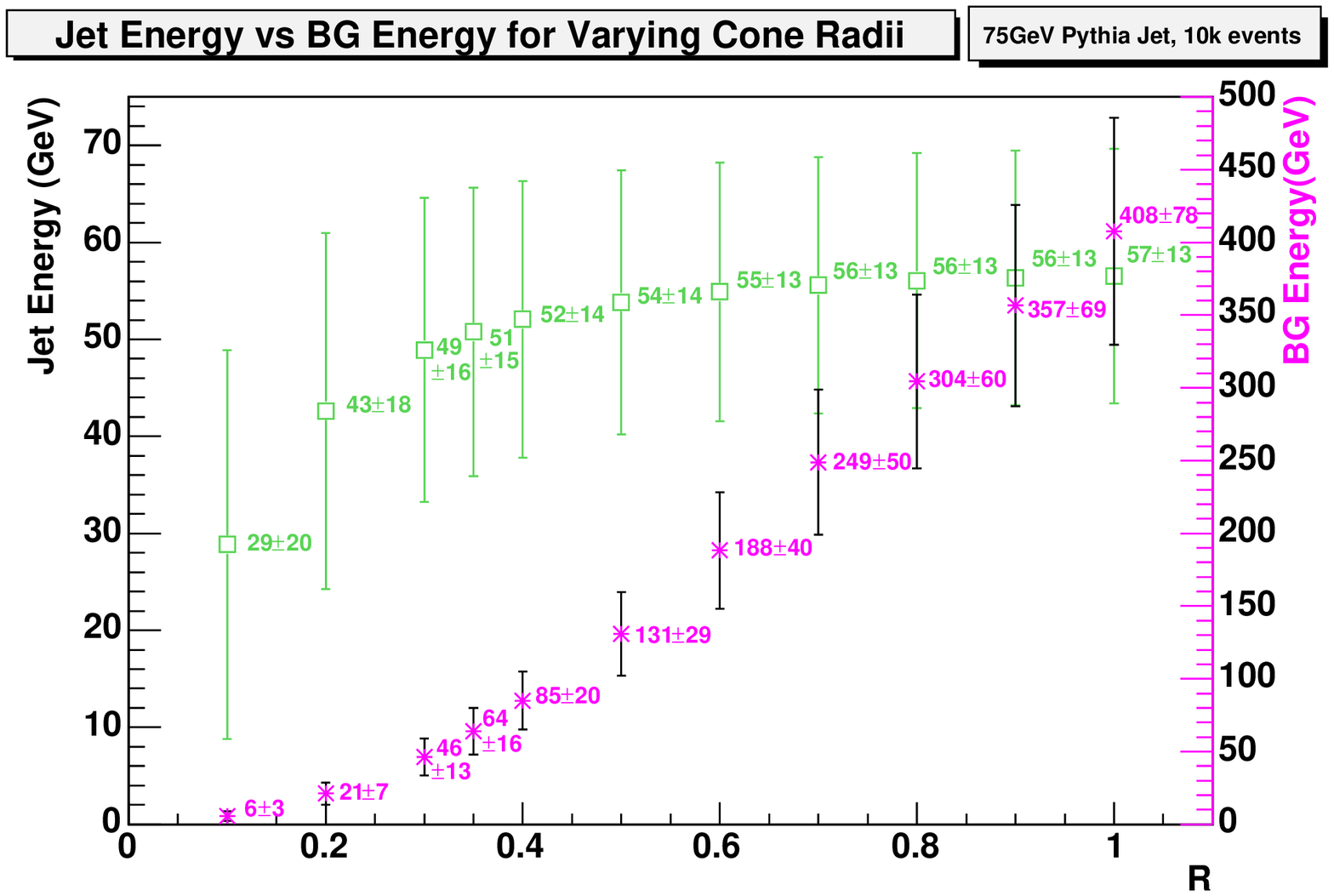}
          \caption{\label{fig:JetAndBGCompare75} As for
          Fig.~\ref{fig:JetAndBGCompare30} but jet energy = 75 GeV.}
        \end{figure}

\begin{figure}[!hbt]
          \center
          \includegraphics[scale=0.6]{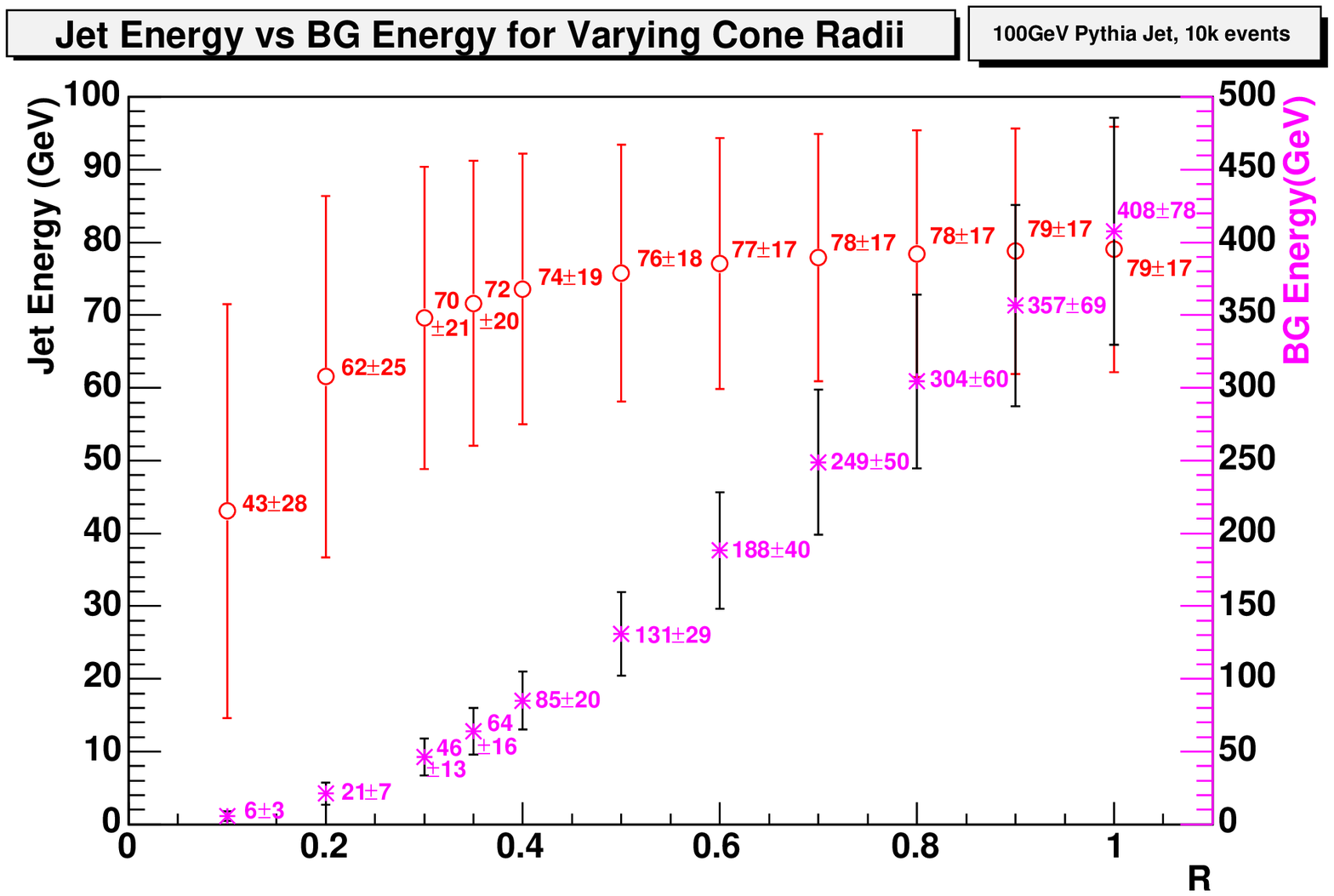}
          \caption{\label{fig:JetAndBGCompare100}As for
          Fig.~\ref{fig:JetAndBGCompare30} but jet energy = 100 GeV.}
        \end{figure}

        Figs.~\ref{fig:JetAndBGCompare30}-\ref{fig:JetAndBGCompare100}
        compare the amount of background energy and jet
        energy contained in the cone for varying $R$ (Note the
        different y-axis scales for the jet energy (left) and the background
        energy (right)). The symbols in the 
        plots indicate the means and the error bars indicate the RMS of
        the distributions. For all jet energies presented, the mean
        background energy is greater than the jet energy in the cone
        for $R\geq0.4$.

\begin{figure}[!hbt]
          \center
          \includegraphics[scale=0.6]{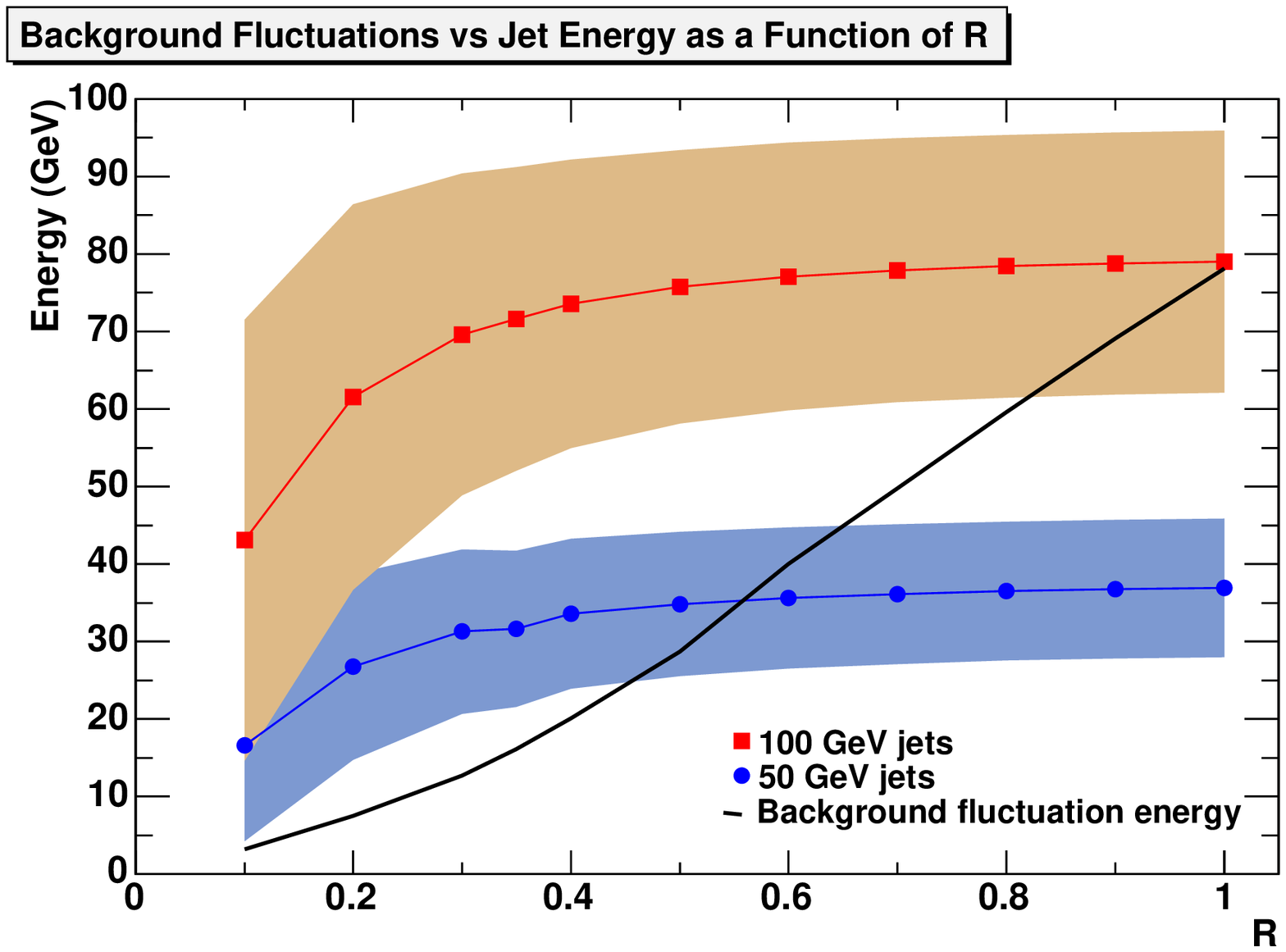}
          \caption{Fluctuations in background energy within cone
          (solid line) compared to jet energy within cone for 50 GeV
          jets (circles) and 100 GeV jets (squares). The shaded area
          represents the RMS of the jet energy distributions.\label{fig:BGFluctVsJetE}}
        \end{figure}

        However, it is the size of the event-by-event
        fluctuations that provide the challenge to reconstructing the
        jet energies. In Fig.~\ref{fig:BGFluctVsJetE} the background
        fluctuation energies are plotted versus the jet energy in the
        cone for 50 GeV and 100 GeV jets. For 50 GeV jets for $R>0.5$, the event-by-event
        background energy fluctuations are larger than the jet energy
        in the cone making jet energy reconstruction impossible using $R>0.5$.
        This indicates that in order to find
        jets of lower energies ($E_{TJet}\sim $50 GeV), a relatively small
        cone size is required.

        \subsubsection{Background Energy Fluctuations with Respect to Position on Grid}
\begin{figure}[!hbt]
          \center
          \includegraphics[scale=0.6]{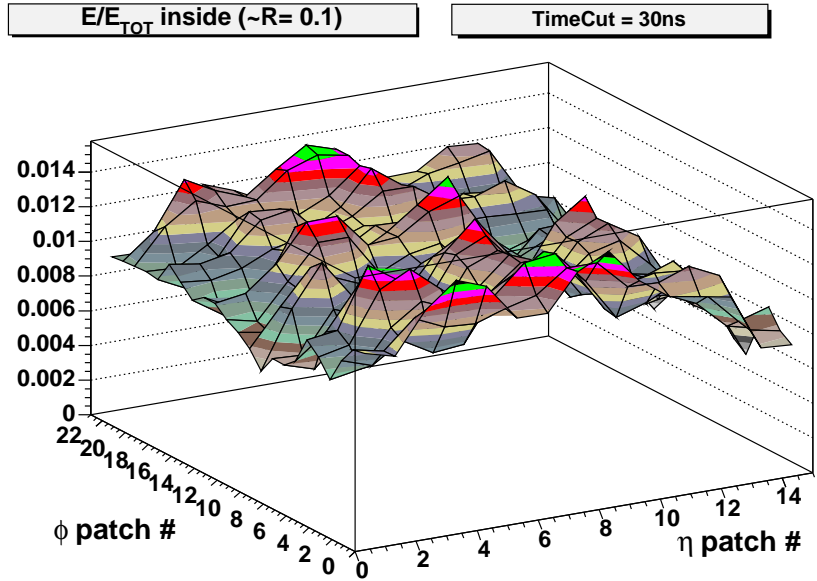}
          \protect\caption{Plots show energy contained in the patch,
          with equivalent area to $R = 0.1$, divided by the total
          energy in the grid as a function of patch position.}
          \protect\label{fig:etaphi}
        \end{figure}

        A further background energy study was performed to
        check that there was no significant bias in the grid energy 
        deposit in terms of position on the grid. The detector setup
        in ALICE is not symmetrical and material in front of the EMCal could cause
        different particle absorption or scattering in certain
        directions. A `sliding patch' method was used to calculate the
        size of the fluctuations by summing all the
        energy in the patch and plotting it divided by total grid
        energy as a function of patch
        position. The `patch' was then slid to a new position such
        that the new patch position overlapped the old by half of the
        patch area. This process was repeated over the area of the grid.
        Fig.~\ref{fig:etaphi} shows that for a patch 
        size equivalent to a cone radius of $R=0.1$, the fluctuations
        from place to place on the grid 
        were of the order of $1\%$ and were due to statistical fluctuations.

        \section{Algorithm Optimisation}\label{sec:AlgoOpt}

        \subsection{Parameter Optimisation}
        In order to optimise the reconstructed jet resolution and the
        jet finding efficiency, and to  
        minimise the number of `fake' jets reconstructed, the main
        algorithm parameters needed to be fine-tuned. These parameters
        are cone radius ($R$), minimum
        accepted jet seed tower energy ($JetESeed$) and the minimum accepted jet energy
        ($MinJetEt$). The parameters were optimised for the case of 50 GeV jets so that
        the majority of 
        jets above this threshold would also be found and
        reconstructed. In order to reconstruct jets with $E_{T}<50$ GeV, alternative methods
        can be implemented, such as particle correlation methods
        which are currently used at experiments at RHIC \cite{PRL903}
        as discussed in Chapter~\ref{sec:HIMethods}.


        \subsubsection{Cone Radius ($R$)}
        From the study of background
        fluctuations in heavy-ion collisions, in
        Section~\ref{sec:BGFluct}, it was seen that for large cone radii, the
        absolute value of the background fluctuations is of the order of the jet
        energy contained in the cone. This implies that the use of smaller cone radii
        is preferable. On the other hand, very small cone radii ($R
        \leq 0.2$) contain around half of the total jet energy and
        this could lead to inaccuracies in reconstructing full jet
        energies. Thus various quantities, such as the amount of jet
        energy contained within different radii, and the resulting
        resolution of the reconstructed jet energy for different radii
        needed to be studied in order to find the optimal $R$ to be
        used in the cone algorithm for heavy-ion collisions.
        
\begin{figure}[!hbt]
          \center
          \includegraphics[scale=0.6]{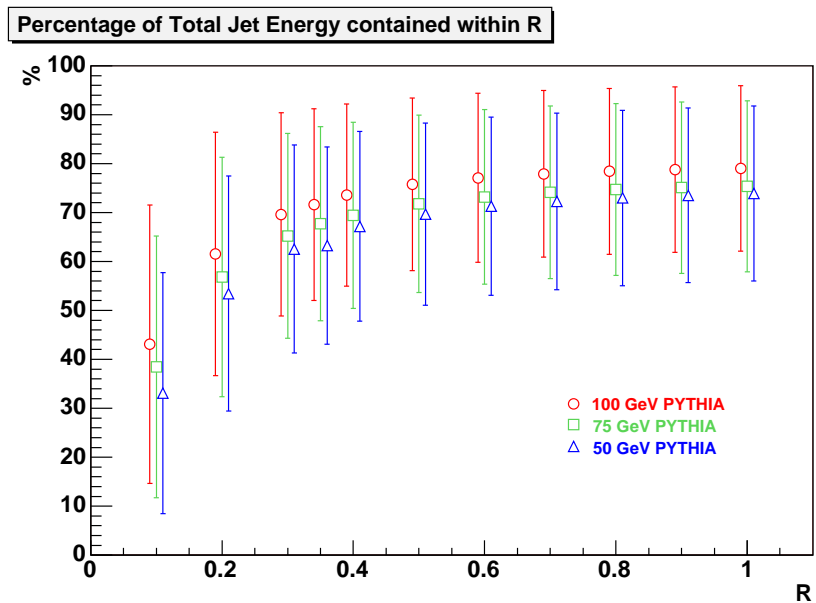} 
          \protect\caption{Percentage of jet energy contained within
          different cone radii for 100 GeV (red circles), 75 GeV
          (green squares) and 50 GeV (blue triangles)
          jets. The symbols represent the mean and the
          error bars represent the RMS of the distributions. Symbols
          for 50 GeV and 100 GeV jets have been offset for clarity.}
          \protect\label{fig:JetConePercentage}
        \end{figure}

        Fig.~\ref{fig:JetConePercentage} shows the percentage of jet energy contained
        within the cone as a function of cone radius $R$. The amount
        of jet energy contained inside the cone is found to saturate
        for $R \geq 0.6$. Note that even for $R=1.0$ the mean values are not
        expected to reach $100\%$ since energy due to neutrons is not measured
        and the $p_{T}$-cut also excludes some of the jet
        energy. It can also be seen that the $p_{T}$-cut has a larger
        effect on the lower energy jets since a larger fraction of
        particle energies from these jets are below the cut than is
        the case for the higher energy jets.

\begin{figure}[!hbt]
          \center
          \includegraphics[scale=0.6]{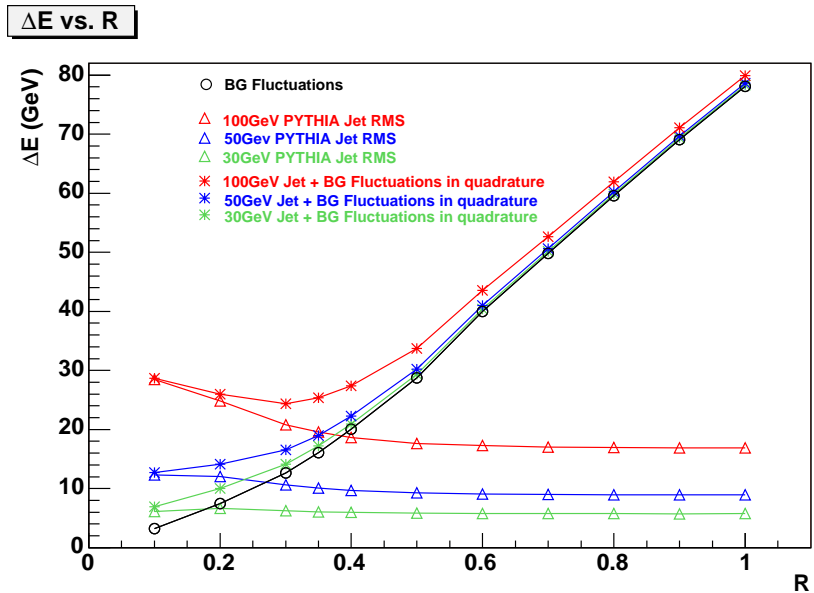}
          \protect\caption{$\Delta E$ as a function of cone radius
          $R$. The black circles represent the background
          fluctuations, the red/blue/green triangles represent the RMS
          of the 100GeV/50GeV/30GeV PYTHIA jet distributions and the
          coloured stars represent the sum in quadrature of the
          background fluctuations and the relevant jet energy widths. }
          \protect\label{fig:DeltaE}
        \end{figure}

        Another point to note, as can be seen in
        Fig.~\ref{fig:DeltaE}, is that the absolute value of the width
        of the jet energy
        distribution decreases as a function of $R$ (represented by the
        coloured triangles) which is the
        opposite behaviour to the background fluctuations (represented
        by the black circles). Therefore,
        in order to optimise the resolution of the reconstructed jets,
        the trade-off between the $R$-dependence of the background
        fluctuations and the jet energy widths needed to be
        analysed. The coloured stars in Fig.~\ref{fig:DeltaE}
        represent the addition in quadrature of the RMS of the jet
        energy distributions and the background fluctuations.

\begin{figure}[!hbt]
          \center
          \includegraphics[scale=0.6]{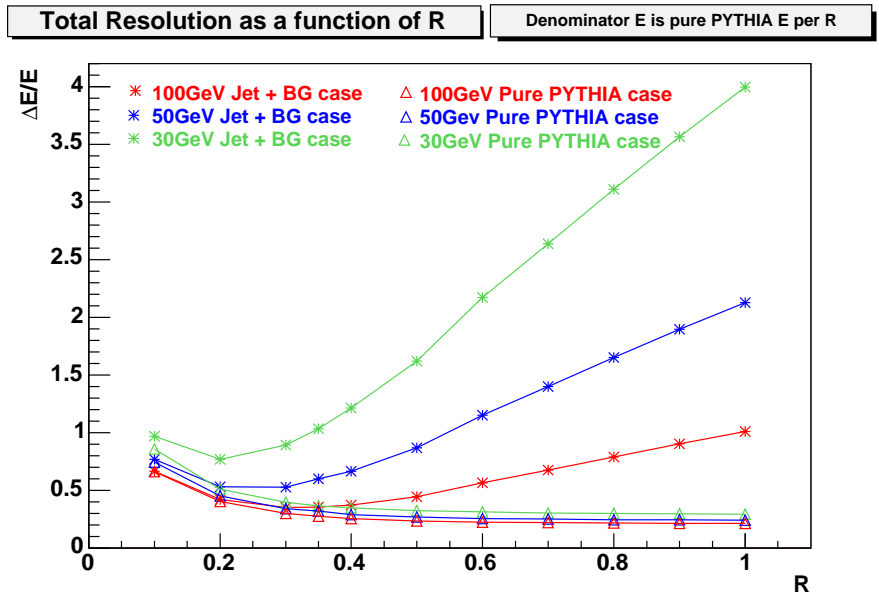}
          \protect\caption{$\Delta E/E$ as a function of $R$ where the
          $\Delta E$ are shown in Fig.~\ref{fig:DeltaE} and where $E$
          is the amount of pure PYTHIA energy contained in the
          respective $R$. The red/blue/green stars represent the 100
          GeV/50 GeV/30 GeV PTYHIA plus background cases and the
          red/blue/green
          triangles represent the 100
          GeV/50 GeV/30 GeV pure PYTHIA cases.}
          \protect\label{fig:WorstCaseDeltaEoverE}
        \end{figure}

        Fig.~\ref{fig:WorstCaseDeltaEoverE} shows the resulting
        resolution, $\Delta E/E$, when the $\Delta E$ in
        Fig.~\ref{fig:DeltaE} are divided by the pure PYTHIA
        jet energy contained in $R$. The stars represent what was classified as the
        `worst case' scenario because the total value of the background
        fluctuations was added in quadrature to the jet energy widths
        when performing this calculation. When reconstructing
        the jet energy, the background energy was typically calculated
        and subtracted from the total energy in the jet cone and
        therefore using the total value of the background fluctuations is an
        overestimate of the contribution to the resolution from the
        background. However, it was important to find the upper 
        limit on the jet energy resolution that was possible when
        using tracking plus calorimetry
        data. Fig.~\ref{fig:WorstCaseDeltaEoverE} also shows the 
        `best case' scenario for resolution (shown by the coloured
        triangles) which represents the $\Delta E/E$ of the pure PYTHIA
        distributions as a function of $R$. In this case, the energy
        from all detectable particles
        inside the cone of radius $R$ was summed around the known jet
        centre to find $\Delta E$ and $E$ (not jet finding
        results). This is therefore the limiting case
        on the best possible resolution using the cone method for the
        various detectors and the goal for which to aim when
        reconstructing jet energies.

        As can also be seen from
        Fig.~\ref{fig:WorstCaseDeltaEoverE}, the resolution for 30
        GeV jets is very poor for almost all values of $R$ ($(\Delta
        E /E)_{30\rm{GeV}}>1$
        for $R > 0.3$) and thus optimisation was performed for the
        case of 50 GeV jets. For all values of $R$, the respective
        resolution for the case of 100 GeV jets is always better
        than that of 50 GeV jets. For 50 GeV jets the lowest value of $\Delta E/E =
        0.528$ when $R = 0.3$. At $R=0.3$ the values of $\Delta E/E$
        for the other jets are also close to their minimum values and
        therefore a cone radius of $R = 0.3$ was chosen for jet reconstruction.

        \subsubsection{Jet Seed Energy ($JetESeed$) and Minimum
          Acceptable Jet Energy ($MinJetEt$)}
        In order to maximise the jet finding efficiency (fraction of
        input jets that are found and reconstructed) of the
        algorithm and minimise the number of `fake' jets (jets found
        by the algorithm but that are not part of the input
        distribution) reconstructed, two further parameters needed to be
        tuned. The jet seed tower energy, $JetESeed$, and the minimum
        accepted jet energy, $MinJetEt$, were optimised by finding the
        best combination of the two parameters that give the optimal
        efficiency and minimal `fake' jet rate. The $JetESeed$ and $MinJetEt$
        act as two filters and are dependent
        on each other. If the $JetESeed$ is set low, in order not
        to exclude real jet seeds, this could allow a large
        number of `fake' towers to be classified as jet seeds consequently
        requiring the $MinJetEt$ to be set with a high threshold to
        eliminate `fake' jets while in turn preserving the real jets.
        It is also not optimal to set the $JetESeed$ too high because
        it was found to introduce a bias on the type of fragmentation process
        involved in forming the jet. i.e. this preselects processes which give
        rise to high-$p_{T}$
        leading particles, biasing the resulting fragmentation function. 

        Note that
        `fake' jets as classified above, are not always due to
        anomalous energy deposition (from fluctuations in the
        background which are then treated as jets by the algorithm),
        but can also be due to the presence of jets
        in the Central HIJING event. Although these are real jets, for
        the purpose of this analysis, all jets that 
        were found that were not part of the input PYTHIA
        distribution were classified as `fakes'. For more detail on
        `fake' jet classification in this thesis, refer to
        Chapter~\ref{sec:Efficiency}.
        Further study will need to be performed in order to separate
        and reconstruct multiple jets of different energies in the
        same event, but this is beyond the scope of this thesis.

\begin{figure}[!hbt]
          \center
          \includegraphics[scale=0.6]{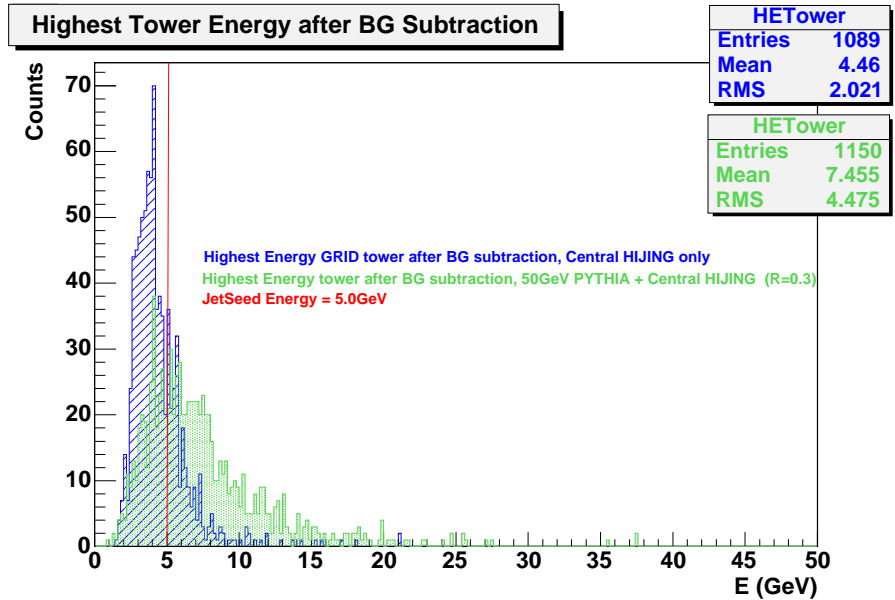}
          \protect\caption{Distribution of the energy of the highest
          energy tower per event after background subtraction, for
          Central HIJING events (in blue) and 50 GeV PYTHIA events
          combined with Central HIJING events (in green). The red line
          illustrates how a filter on seed tower energy would
          affect the inclusion/exclusion of background/jet events.}
          \protect\label{fig:TowerE}
        \end{figure}

\begin{figure}[!hbt]
          \center
          \includegraphics[scale=0.6]{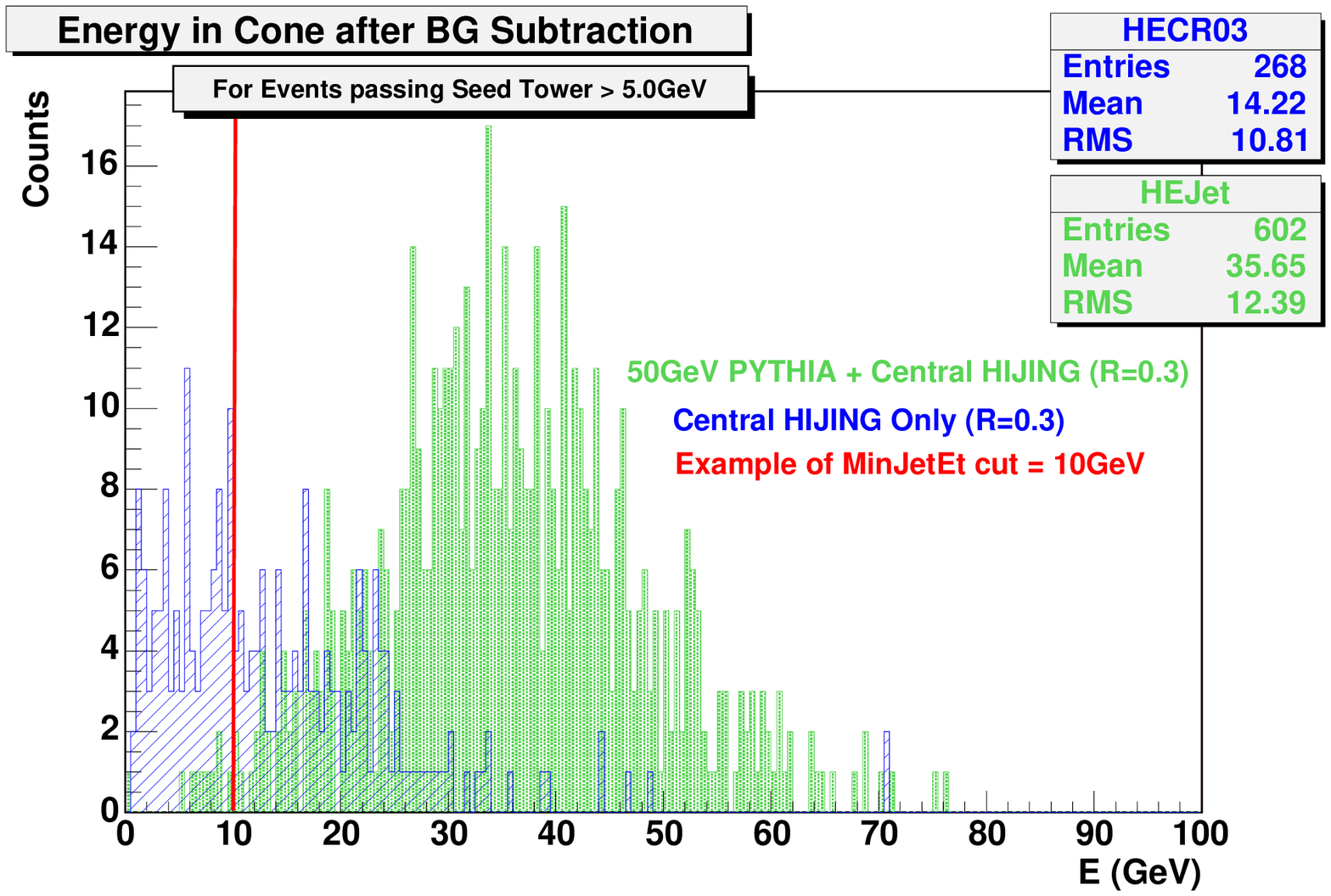}
          \protect\caption{Distribution of the energy inside a cone
          with $R = 0.3$ after background subtraction for events that
          pass a filter of $JetESeed = 5.0$ GeV. The case for Central HIJING
          events is shown in blue and the 50GeV PYTHIA combined with
          Central HIJING events is shown in green. The red line
          illustrates how a filter on minimum allowed jet $E_{T}$
          would affect the inclusion/exclusion of background/jet events.}
          \protect\label{fig:MinJetEt}
        \end{figure}

        Figs.~\ref{fig:TowerE} and \ref{fig:MinJetEt} are graphical
        representations of the method used to find the optimal values
        for $JetESeed$ and $MinJetEt$. Fig.~\ref{fig:TowerE} shows the
        distribution of highest energy grid towers per event after background
        subtraction was performed for Central HIJING events alone
        (shown in blue) and for 50 GeV PYTHIA events combined with
        Central HIJING events (shown in green). The two distributions
        were found to
        overlap up to $\sim 20$ GeV although there was a 3 GeV
        difference in the means of the distributions with the PYTHIA
        events having a greater highest tower energy value than 
        the pure HIJING events. The percentage of real jet seed exclusion
        and `fake' jet seed inclusion for the first filter ($JetESeed$)
        could then be calculated for 
        different values of $JetESeed$ by integrating the number of
        events that were above or below a chosen seed energy (an
        example of a 5 GeV seed
        energy is shown as the red line in Fig.~\ref{fig:TowerE}).

        Since the two parameters depend on each other, the
        optimal value for $MinJetEt$ cannot be chosen from the
        first piece of 
        analysis alone. Further analysis involved choosing a range of 
        values of $JetESeed$ and, for each one, plotting the value of
        the energy, after background subtraction,
        inside a cone of radius $R = 0.3$ for the case of Central
        HIJING events alone and for 50 GeV PYTHIA
        events combined with Central HIJING events. The situation for
        events passing a $JetESeed \geq 5$ GeV is shown in
        Fig.~\ref{fig:MinJetEt} where the pure Central HIJING events
        are shown in blue and the mixed events are shown in green. The
        real jet exclusion and `fake' jet inclusion rates could then
        be calculated further for the case of different choices of
        $MinJetEt$ (per set value of $JetESeed$) by integrating the
        events above the relevant 
        threshold (an example of $MinJetEt = 10$ GeV is shown by the
        red line in the figure). Finally, the resulting inclusion
        and exclusion rates were
        tabulated as a function of the values of $JetESeed$ and
        $MinJetEt$. The optimal combination of parameters was chosen
        so as
        to produce the lowest inclusion rate of Central HIJING events
        while the inclusion rate of Parameterised HIJING 4000 events
        was less than $5\%$ and the exclusion of 50 GeV jets was less
        than $30\%$. The combination of parameters satisfying all these
        criteria can be found from Tables~\ref{tab:ParamTableJets50} -
        \ref{tab:ParamTableHijingParam} and are $JetESeed = 4.6$ GeV
        and $MinJetEt = 14.0$ GeV.

        \begin{table}[!hbt]
          \center \small
          \begin{tabular}{|r|r||r|r|r||r||}  \hline
            \multicolumn{2}{|c||} {\textbf{Parameter Values}}      &
            \multicolumn{4}{|c||} {\textbf{50 GeV PYTHIA on Central
            HIJING}} 
            \\ \hline
           \begin{tabular}{c}
              $JetESeed$\\
              (GeV)\\
            \end{tabular}
            &
            \begin{tabular}{c}
              $MinJetEt$\\
              (GeV)\\
              \end{tabular}
            &
            \begin{tabular}{c}
              Fraction\\
              passing\\
              $JetESeed$\\
              ($\%$)\\
            \end{tabular}
              &
              \begin{tabular}{c}
                Fraction \\
                passing \\
                $MinJetEt$\\
                ($\%$)\\
              \end{tabular}  
             &
             \begin{tabular}{c}
                Fraction \\
                passing \\
                both \\
                filters ($\%$)\\                
              \end{tabular}  
             &
             \begin{tabular}{c}
                Fraction \\
                excluded \\
                ($\%$)\\
              \end{tabular}  
             \\ \hline
        
            4.5   & 10.0 &74.6  &98.2  &73.2  &26.8  \\
                  & 10.5 &      &97.8  &72.9  &27.1  \\
                  & 11.0 &      &97.6  &72.8  &27.2  \\
                  & 12.0 &      &97.3  &72.6  &27.4  \\
                  & 13.0 &      &96.4  &71.9  &28.1  \\            
                  & 14.0 &      &95.1  &70.9  &29.1  \\
            \hline
            4.6   & 10.0 &73.2  &98.5  &72.1  &27.9  \\
                  & 10.5 &      &98.0  &71.8  &28.2  \\
                  & 11.0 &      &97.7  &71.6  &28.4  \\
                  & 12.0 &      &97.0  &71.0  &29.0  \\
                  & 13.0 &      &96.5  &70.7  &29.3  \\            
                  & 14.0 &      &95.9  &70.2  &29.8  \\
            \hline
            4.7   & 10.0 &71.2  &98.5  &70.1  &29.9  \\
                  & 10.5 &      &98.5  &70.1  &29.9  \\
                  & 11.0 &      &97.7  &69.6  &30.4  \\
                  & 12.0 &      &97.2  &69.2  &30.8  \\
                  & 13.0 &      &96.9  &69.0  &31.0  \\            
                  & 14.0 &      &95.7  &68.1  &31.9  \\
            \hline
            4.8   & 10.0 &70.3  &98.1  &69.0  &31.0  \\
                  & 10.5 &      &98.1  &69.0  &31.0  \\
                  & 11.0 &      &97.8  &68.8  &31.2  \\
                  & 12.0 &      &97.3  &68.5  &31.5  \\
                  & 13.0 &      &96.9  &68.1  &31.9  \\            
                  & 14.0 &      &96.6  &67.9  &32.1  \\
            \hline
            4.9   & 10.0 &69.0  &98.6  &68.0  &32.0  \\
                  & 10.5 &      &98.2  &67.8  &32.2  \\
                  & 11.0 &      &97.4  &67.2  &32.8  \\
                  & 12.0 &      &96.6  &66.7  &33.3  \\
                  & 13.0 &      &96.3  &66.4  &33.6  \\            
                  & 14.0 &      &95.7  &66.0  &34.0  \\
            \hline
            5.0   & 10.0 &67.3  &98.5  &66.3  &33.7  \\
                  & 10.5 &      &98.4  &66.2  &33.8  \\
                  & 11.0 &      &98.2  &66.1  &33.9  \\
                  & 12.0 &      &97.1  &65.4  &34.6  \\
                  & 13.0 &      &96.8  &65.1  &34.9  \\            
                  & 14.0 &      &95.8  &64.5  &35.5  \\
            \hline
            5.1   & 10.0 &66.1  &98.1  &64.9  &35.1  \\
                  & 10.5 &      &98.0  &64.8  &35.2  \\
                  & 11.0 &      &97.5  &64.4  &35.6  \\
                  & 12.0 &      &96.6  &63.9  &36.1  \\
                  & 13.0 &      &96.3  &63.7  &36.3  \\            
                  & 14.0 &      &95.4  &63.1  &36.9  \\             
            \hline
          \end{tabular}
        \protect\caption{Exclusion rates for 50 GeV jets using
        various combinations of $JetESeed$ and $MinJetEt$.(Ratio
        background subtraction method was used.)}
        \protect\label{tab:ParamTableJets50} 
        \end{table}

        \begin{table}[!hbt]
          \center \small
          \begin{tabular}{|r|r||r|r|r||r||}  \hline
            \multicolumn{2}{|c||} {\textbf{Parameter Values}}      &
            \multicolumn{4}{|c||} {\textbf{100 GeV PYTHIA on Central
            HIJING}} 
            \\ \hline
           \begin{tabular}{c}
              $JetESeed$\\
              (GeV)\\
            \end{tabular}
            &
            \begin{tabular}{c}
              $MinJetEt$\\
              (GeV)\\
              \end{tabular}
            &
            \begin{tabular}{c}
              Fraction\\
              passing\\
              $JetESeed$\\
              ($\%$)\\
            \end{tabular}
              &
              \begin{tabular}{c}
                Fraction \\
                passing \\
                $MinJetEt$\\
                ($\%$)\\
              \end{tabular}  
             &
             \begin{tabular}{c}
                Fraction \\
                passing \\
                both \\
                filters ($\%$)\\                
              \end{tabular}  
             &
             \begin{tabular}{c}
                Fraction \\
                excluded \\
                ($\%$)\\
              \end{tabular}  
             \\ \hline
        
            4.5   & 10.0 &97.3  &99.8 &97.1 &2.9  \\
                  & 10.5 &      &99.8 &97.1 &2.9  \\
                  & 11.0 &      &99.8 &97.1 &2.9  \\
                  & 12.0 &      &99.8 &97.1 &2.9  \\
                  & 13.0 &      &99.7 &97.0 &3.0  \\            
                  & 14.0 &      &99.5 &96.8 &3.2  \\
            \hline
            4.6   & 10.0 &96.8  &99.9 &96.7 &3.3  \\
                  & 10.5 &      &99.9 &96.7 &3.3  \\
                  & 11.0 &      &99.9 &96.7 &3.3  \\
                  & 12.0 &      &99.9 &96.7 &3.3  \\
                  & 13.0 &      &99.8 &96.6 &3.4  \\            
                  & 14.0 &      &99.7 &96.5 &3.5  \\
            \hline
            4.7   & 10.0 &96.5 &100.0 &96.5 &3.5  \\
                  & 10.5 &     &99.9 &96.4 &3.6  \\
                  & 11.0 &     &99.8 &96.3 &3.7  \\
                  & 12.0 &     &99.7 &96.2 &3.8  \\
                  & 13.0 &     &99.6 &96.1 &3.9  \\            
                  & 14.0 &     &99.6 &96.1 &3.9  \\
            \hline
            4.8   & 10.0 &96.4 &99.9 &96.3 &3.7  \\
                  & 10.5 &     &99.9 &96.3 &3.7  \\
                  & 11.0 &     &99.9 &96.3 &3.7  \\
                  & 12.0 &     &99.8 &96.2 &3.8  \\
                  & 13.0 &     &99.8 &96.2 &3.8  \\            
                  & 14.0 &     &99.8 &96.2 &3.8  \\
            \hline
            4.9   & 10.0 &96.0 &99.8 &95.8 &4.2  \\
                  & 10.5 &     &99.8 &95.8 &4.2  \\
                  & 11.0 &     &99.7 &95.7 &4.3  \\
                  & 12.0 &     &99.7 &95.7 &4.3  \\
                  & 13.0 &     &99.7 &95.7 &4.3  \\            
                  & 14.0 &     &99.7 &95.7 &4.3  \\
            \hline
            5.0   & 10.0 &95.8 &99.8 &95.6 &4.4  \\
                  & 10.5 &     &99.8 &95.6 &4.4  \\
                  & 11.0 &     &99.8 &95.6 &4.4  \\
                  & 12.0 &     &99.8 &95.6 &4.4  \\
                  & 13.0 &     &99.8 &95.6 &4.4  \\            
                  & 14.0 &     &99.6 &95.4 &4.6  \\
            \hline
            5.1   & 10.0 &95.3 &100.0 &95.3 &4.7  \\
                  & 10.5 &     &100.0 &95.3 &4.7  \\
                  & 11.0 &     &100.0 &95.3 &4.7  \\
                  & 12.0 &     &100.0 &95.3 &4.7  \\
                  & 13.0 &     &99.9 &95.2 &4.8  \\            
                  & 14.0 &     &99.9 &95.2 &4.8  \\
            \hline            
          \end{tabular}
        \protect\caption{Exclusion rates for 100 GeV jets using
        various combinations of $JetESeed$ and $MinJetEt$.}
        \protect\label{tab:ParamTableJets100} 
        \end{table}

        \begin{table}[!hbt]
          \center \small
          \begin{tabular}{|r|r||r|r|r|}  \hline
            \multicolumn{2}{|c||} {\textbf{Parameter Values}}      &
            \multicolumn{3}{|c|} {\textbf{Pure Central HIJING}}
            \\ \hline
           \begin{tabular}{c}
              $JetESeed$\\
              (GeV)\\
            \end{tabular}
            &
            \begin{tabular}{c}
              $MinJetEt$\\
              (GeV)\\
              \end{tabular}
            &
            \begin{tabular}{c}
              Fraction\\
              passing\\
              $JetESeed$\\
              ($\%$)\\
            \end{tabular}
              &
              \begin{tabular}{c}
                Fraction \\
                passing \\
                $MinJetEt$\\
                ($\%$)\\
              \end{tabular}  
             &
             \begin{tabular}{c}
                Fraction \\
                passing \\
                both \\
                filters ($\%$)\\                
              \end{tabular}  
             \\ \hline
        
            4.5   & 10.0 &36.9 &49.3 &18.2   \\
                  & 10.5 &     &48.2 &17.8    \\
                  & 11.0 &     &46.8 &17.3    \\
                  & 12.0 &     &43.9 &16.2    \\
                  & 13.0 &     &40.6 &15.0    \\            
                  & 14.0 &     &38.0 &14.0    \\
            \hline
            4.6   & 10.0 &34.1 &49.7 &16.9    \\
                  & 10.5 &     &48.5 &16.5    \\
                  & 11.0 &     &47.0 &16.0    \\
                  & 12.0 &     &44.8 &15.3    \\
                  & 13.0 &     &41.2 &14.0    \\            
                  & 14.0 &     &38.7 &13.2    \\
            \hline
            4.7   & 10.0 &32.6 &49.7 &16.2    \\
                  & 10.5 &     &48.4 &15.8    \\
                  & 11.0 &     &46.8 &15.3    \\
                  & 12.0 &     &44.6 &14.6    \\
                  & 13.0 &     &40.8 &13.3    \\            
                  & 14.0 &     &38.5 &12.6    \\
            \hline
            4.8   & 10.0 &30.0 &50.2 &15.1    \\
                  & 10.5 &     &48.8 &14.7    \\
                  & 11.0 &     &47.1 &14.1    \\
                  & 12.0 &     &44.6 &13.4    \\
                  & 13.0 &     &41.5 &12.5    \\            
                  & 14.0 &     &39.4 &11.9    \\
            \hline
            4.9   & 10.0 &28.7 &49.8 &14.3    \\
                  & 10.5 &     &48.3 &13.8    \\
                  & 11.0 &     &47.1 &13.5    \\
                  & 12.0 &     &44.8 &12.8    \\
                  & 13.0 &     &41.7 &12.0    \\            
                  & 14.0 &     &39.8 &11.4    \\
            \hline
            5.0   & 10.0 &27.9 &49.6 &13.8    \\
                  & 10.5 &     &48.1 &13.4    \\
                  & 11.0 &     &46.3 &12.9    \\
                  & 12.0 &     &44.0 &12.3    \\
                  & 13.0 &     &40.7 &11.3    \\            
                  & 14.0 &     &38.4 &10.7    \\
            \hline
            5.1   & 10.0 &25.2 &49.2 &12.4    \\
                  & 10.5 &     &47.9 &12.1    \\
                  & 11.0 &     &46.3 &11.6    \\
                  & 12.0 &     &43.8 &11.0    \\
                  & 13.0 &     &40.1 &10.1    \\            
                  & 14.0 &     &37.6 &9.5     \\
            \hline            
          \end{tabular}
        \protect\caption{Inclusion rates for pure Central HIJING
        events using
        various combinations of $JetESeed$ and $MinJetEt$.}
        \protect\label{tab:ParamTableHijing} 
        \end{table}

        \begin{table}[!hbt]
          \center \small
          \begin{tabular}{|r|r||r|r|r|}  \hline
            \multicolumn{2}{|c||} {\textbf{Parameter Values}}      &
            \multicolumn{3}{|c|} {\textbf{Parameterised HIJING 4000}}
            \\ \hline
           \begin{tabular}{c}
              $JetESeed$\\
              (GeV)\\
            \end{tabular}
            &
            \begin{tabular}{c}
              $MinJetEt$\\
              (GeV)\\
              \end{tabular}
            &
            \begin{tabular}{c}
              Fraction\\
              passing\\
              $JetESeed$\\
              ($\%$)\\
            \end{tabular}
              &
              \begin{tabular}{c}
                Fraction \\
                passing \\
                $MinJetEt$\\
                ($\%$)\\
              \end{tabular}  
             &
             \begin{tabular}{c}
                Fraction \\
                passing \\
                both \\
                filters ($\%$)\\                
              \end{tabular}  
             \\ \hline
        
            4.5   & 10.0 &18.5 &31.2 &5.8    \\
                  & 10.5 &     &27.7 &5.1    \\
                  & 11.0 &     &26.6 &4.9    \\
                  & 12.0 &     &24.9 &4.6    \\
                  & 13.0 &     &19.7 &3.6    \\            
                  & 14.0 &     &15.6 &2.9    \\
            \hline
            4.6   & 10.0 &16.6 &30.3 &5.0    \\
                  & 10.5 &     &27.1 &4.5    \\
                  & 11.0 &     &25.8 &4.3    \\
                  & 12.0 &     &23.9 &4.0    \\
                  & 13.0 &     &19.4 &3.2    \\            
                  & 14.0 &     &15.5 &2.6    \\
            \hline
            4.7   & 10.0 &14.0 &32.1 &4.5    \\
                  & 10.5 &     &28.2 &4.0    \\
                  & 11.0 &     &27.5 &3.8    \\
                  & 12.0 &     &25.2 &3.5    \\
                  & 13.0 &     &19.8 &2.8    \\            
                  & 14.0 &     &16.0 &2.2    \\
            \hline
            4.8   & 10.0 &12.4 &32.8 &4.1    \\
                  & 10.5 &     &28.4 &3.5    \\
                  & 11.0 &     &27.6 &3.4    \\
                  & 12.0 &     &25.9 &3.2    \\
                  & 13.0 &     &19.8 &2.5    \\            
                  & 14.0 &     &15.5 &1.9    \\
            \hline
            4.9   & 10.0 &11.3 &33.0 &3.7    \\
                  & 10.5 &     &29.2 &3.3    \\
                  & 11.0 &     &28.3 &3.2    \\
                  & 12.0 &     &26.4 &3.0    \\
                  & 13.0 &     &19.8 &2.2    \\            
                  & 14.0 &     &15.1 &1.7    \\
            \hline
            5.0   & 10.0 &10.4 &34.0 &3.5    \\
                  & 10.5 &     &29.9 &3.1    \\
                  & 11.0 &     &28.9 &3.0    \\
                  & 12.0 &     &26.8 &2.8    \\
                  & 13.0 &     &20.6 &2.1    \\            
                  & 14.0 &     &15.5 &1.6    \\
            \hline
            5.1   & 10.0 &9.7  &35.2 &3.4    \\
                  & 10.5 &     &30.8 &3.0    \\
                  & 11.0 &     &29.7 &2.9    \\
                  & 12.0 &     &27.5 &2.7    \\
                  & 13.0 &     &22.0 &2.1    \\            
                  & 14.0 &     &16.5 &1.6    \\
            \hline            
          \end{tabular}
        \protect\caption{Inclusion rates for parameterised HIJING 4000
        events using
        various combinations of $JetESeed$ and $MinJetEt$.}
        \protect\label{tab:ParamTableHijingParam} 
        \end{table}

        The case for a seedless algorithm ($JetESeed = 0$ GeV) was
        studied in the same way and the resulting inclusion and
        exclusion rates can be seen in
        Table~\ref{tab:SeedlessTable}. In order to compare jet finding results for
        the two cases of a seedless algorithm and seeded algorithm, a value
        of $MinJetEt$ was chosen such that the inclusion rate of
        Central HIJING events was approximately the same ($\sim$13$\%$) as for the case
        of the seeded algorithm where $JetESeed = 4.6$ GeV and
        $MinJetEt = 14.0$ GeV. The 
        value for the seedless case which fulfilled this criterion was
        $MinJetEt = 20.0$ GeV.

        \begin{table}[!hbt]
          \center \small
          \begin{tabular}{|r||r|r|r|r|}  \hline
            
            \begin{tabular}{c}
              \textbf{Parameter}\\
              \textbf{Value}\\              
              \end{tabular}
            &
            \begin{tabular}{c}
              \textbf{50 GeV}\\
              \textbf{PYTHIA on}\\              
              \textbf{Central HIJING}\\              
              \end{tabular}
            &
            \begin{tabular}{c}
              \textbf{100 GeV}\\
              \textbf{PYTHIA on}\\              
               \textbf{Central HIJING}\\             
              \end{tabular}
            &
            \begin{tabular}{c}
              \textbf{Central}\\
              \textbf{HIJING}\\              
              \end{tabular}
            &
            \begin{tabular}{c}
              \textbf{Param.}\\
              \textbf{HIJING}\\
              \textbf{4000}\\              
              \end{tabular}
            \\\hline
            \begin{tabular}{c}
              $MinJetEt$\\
              (GeV)\\
              \end{tabular}
            &
            \begin{tabular}{c}
              Exclusion rate ($\%$)\\
            \end{tabular}
              &
             \begin{tabular}{c}
              Exclusion rate ($\%$)\\             
            \end{tabular}
             &
             \begin{tabular}{c}
                Inclusion \\
              rate ($\%$)\\            
              \end{tabular}  
             &
             \begin{tabular}{c}
                 Inclusion \\
              rate ($\%$)\\ 
              \end{tabular}  
             \\\hline
             10.0 &6.2 &0.9 &37.7 &21.8\\
             10.5 &6.7 &1.0 &36.0 &19.3 \\
             11.0  &7.1 &1.2 &34.1 &18.1 \\
             11.5 &7.7 &1.3 &32.2 &16.9 \\
             12.0 &8.1 &1.3 &30.0 &15.4 \\
             12.5 &8.7 &1.4 &28.9 &14.5 \\
             13.0 &9.1 &1.5 &27.1 &13.2 \\
             13.5 &9.2 &1.5 &25.6 &12.1 \\
             14.0 &9.9 &1.6 &23.9 &11.1 \\
             14.5 &10.6 &1.6 &23.3 &10.3 \\
             15.0 &11.6 &1.6 &22.2 &9.3 \\
             15.5 &12.3 &1.7 &20.8 &8.7 \\
             16.0 &12.9 &1.7 &19.6 &8.1\\
             16.5 &13.6 &1.7 &18.5 &7.5 \\
             17.0 &13.9 &1.7 &17.8 &6.7 \\
             17.5 &14.5 &1.8 &17.3 &5.9 \\
             18.0 &15.5 &2.0 &16.7 &4.9 \\
             18.5 &16.2 &2.0 &16.1 &4.5 \\
             19.0 &17.1 &2.0 &15.0 &4.0 \\
             19.5 &18.2 &2.1 &13.8 &3.7 \\
             20.0 &19.1 &2.2 &13.1 &3.0 \\
             20.5 &20.1 &2.3 &12.4 &2.9 \\
             21.0 &20.9 &2.3 &11.4 &2.5 \\
             21.5 &22.0 &2.4 &10.4 &2.1 \\
             22.0 &22.8 &2.7 &10.1 &1.8 \\
             22.5 &23.6 &2.7 &9.6 &1.7 \\
             23.0 &25.0 &3.0 &9.0 &1.3 \\
             23.5 &26.8 &3.2 &8.0 &1.3 \\
             24.0 &28.1 &3.4 &7.6 &1.3 \\
             24.5 &29.0 &3.5 &7.2 &1.1 \\
             25.0 &30.5 &3.6 &6.4 &0.9 \\           
            \hline
          \end{tabular}
        \protect\caption{Exclusion and inclusion rates for the case of
        a seedless algorithm ($JetESeed = 0$ GeV).}
        \protect\label{tab:SeedlessTable} 
        \end{table}

        \subsection{Methods Optimisation}

        \subsubsection{Calculation of the Background Energy in
          the Jet Cone}
        Accurate calculation of the background energy contribution inside the
        jet cone was necessary to enable accurate reconstruction of the
        energies of the jets found by the jet finding algorithm. A
        number of approaches, including statistical, event-by-event
        and a combination of both were attempted in order to calculate
        most accurately the background energy inside the jet cone. 

\begin{figure}[!hbt]
          \center
          \includegraphics[scale=0.7]{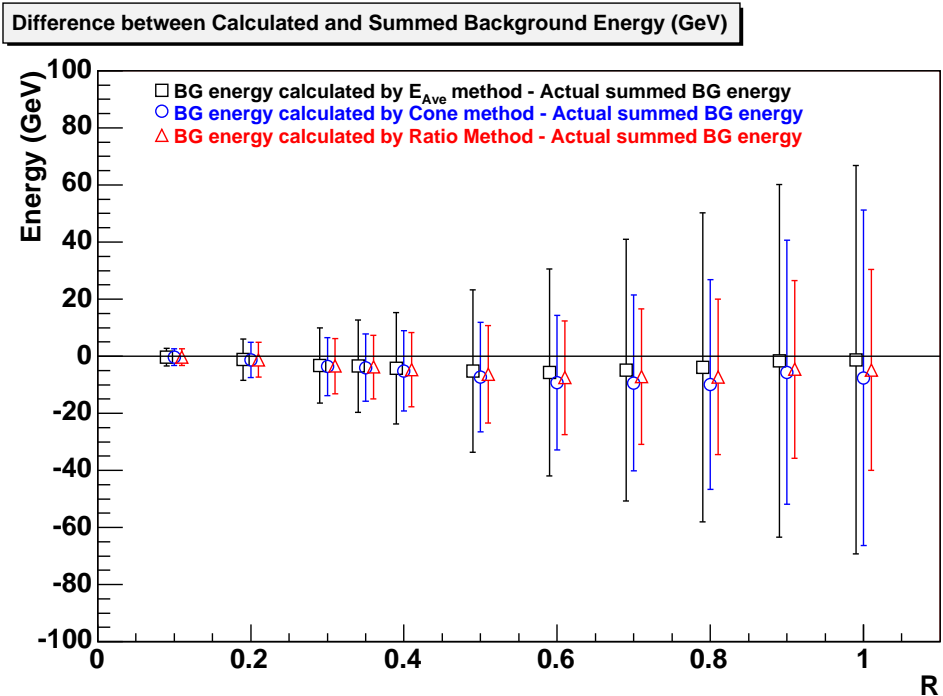}
          \protect\caption{Difference between background energy
          calculated using \emph{Statistical($E_{Ave}$)}
          /$Cone$/$Ratio$
          methods and the actual summed background energy inside
          varying cone radii shown by black/blue/red symbols. The
          error bars represent the RMS of the respective
          distributions. Symbols for \emph{Statistical}$(E_{Ave})$ and $Ratio$ methods have
          been offset for clarity.}
          \protect\label{fig:BGCompare}
        \end{figure}

        \begin{itemize}
          \item \textbf{Statistical Method} \\
        The \emph{Statistical Method}
        calculates the background energy contribution inside the jet
        cone on an average basis by 
        summing the energy in the grid
        from 1~000 Central HIJING events and finding the
        mean energy per event. This energy can then be scaled to
        the size of the cone radius used in jet finding and subtracted
        for each event on which
        jet finding is performed.
        However, when this calculated background energy is compared to
        the actual summed background energy event-by-event, it is
        found that the width of the distribution of the difference in
        the two quantities is large. This is due to the effect of event-by-event energy
        fluctuations. The average difference in the quantities is
        close to zero as shown in Fig.~\ref{fig:BGCompare} (black
        symbols) but the RMS is larger (represented by the error bars)
        than is the case for event-by-event calculation methods.
      \end{itemize}   
        The event-by-event methods were developed to take into account
        the fact that the event-by-event background fluctuations are
        large. The methods are based on the assumption that the distribution
        of the background energy in the grid is uniform, as was shown
        in Section~\ref{sec:BGFluct}.
        \begin{itemize}
          \item \textbf{Cone Method}\\
        The \emph{Cone Method} calculates the
        background contribution to the energy inside the jet cone as
        follows: once the jet cone is found by the jet finding
        algorithm, all the energy in the grid outside the jet cone is
        summed and then scaled for the size of the cone radius. This
        amount is then subtracted from the energy inside the jet cone
        to calculate the actual jet energy. This calculation is
        performed on an event-by-event basis (in fact, sometimes more than once
        per event depending on the number of iterations performed by
        the algorithm). The accuracy of the \emph{Cone Method} background
        calculation also depends on the size of the cone radius
        used since the bigger the cone radius, the less area there is outside the
        cone on which to base the background calculation, see Fig.~\ref{fig:BGCompare}. 
      \end{itemize}
        In addition to event-by-event background energy fluctuations,
        there are also statistical fluctuations in the background
        energy deposited in the grid from tower to tower. Further
        fluctuations on a tower level, caused by implementing the track
        $p_{T}$-cut, are also present.

\begin{figure}[!hbt]
          \center
          \includegraphics[scale=0.7]{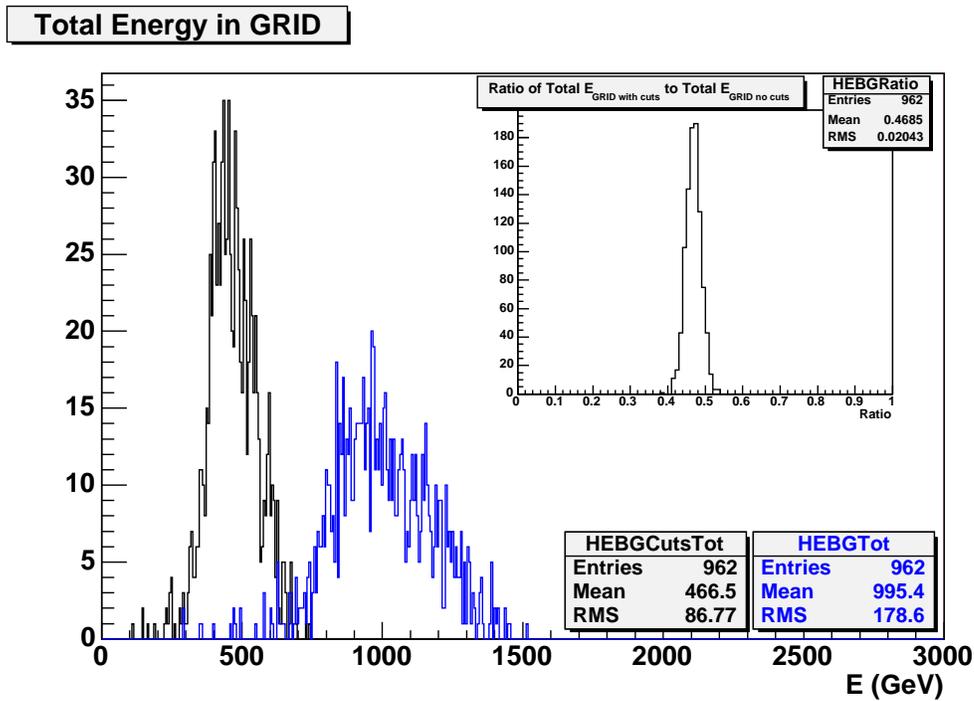}
          \protect\caption{Distribution of total energy contained in
          the grid from Central HIJING events. The blue histogram
          shows the energy in the grid from events without a
          $p_{T}$-cut and the black histogram shows the case with a
          $p_{T}$-cut $= 2$ GeV. The inset shows the ratio
          $E_{cuts}/E_{NoCuts} = 0.4685$ of the two histograms.}
          \protect\label{fig:RatioInset}
        \end{figure}

        \begin{itemize}
          \item \textbf{Ratio Method} \\
        A second
        event-by-event method, \emph{Ratio Method}, was developed to further reduce the
        effect of the fluctuations. The method is based on
        the \emph{Cone Method} calculation but with some differences:
        firstly, two grids are used to store the energy
        information. The first grid is the same as the grid used in
        the \emph{Cone Method} and is composed of the energy from the hits
        in the EMCal
        plus the energy from all the tracks which pass the $p_{T}$-cut. The second
        grid however, is composed of the energy from hits and all the
        track energy without a $p_{T}$-cut implemented. Jet finding is
        performed on the first grid but once the jet cone is found by
        the algorithm, the energy outside the cone in the second grid
        is summed and scaled for the cone radius,
        $E_{Cone_{NoCuts}}$. Next, the actual background energy
        contribution, $E_{Cone_{Cuts}}$, to be subtracted from the jet cone energy in the
        first grid is calculated by multiplying $E_{Cone_{NoCuts}}$ by
        the ratio $F = E_{BG_{Cuts}}/E_{BG_{NoCuts}}$. The ratio, $F$,
        was found by calculating the average, over 1000 Central HIJING events, of the
        energy in the grid with a $p_{T}$-cut divided by the energy in
        the grid without a $p_{T}$-cut as shown in
        Fig.~\ref{fig:RatioInset}.
        \end{itemize}

        A comparison of the 3 background energy calculation methods can be seen in
        Fig.~\ref{fig:BGCompare}. This shows the difference between the
        calculated values and the actual amount of background energy
        that was summed in the cone as a function of cone radius
        $R$. All three methods resulted in means that lay close to
        the true value of the background energy for all $R$ but the error (shown
        by error bars) was smallest for the \emph{Ratio Method}. Therefore the
        \emph{Ratio Method} was chosen as the preferred background
        calculation method to be used in the jet finding algorithm.

\chapter{Results with the Optimised ALICE Algorithm}\label{sec:Results}
This chapter is divided into four main sections,
\textbf{Reconstruction Accuracy}, \textbf{Jet Energy Correction},
\textbf{Reconstruction Efficiency}, and 
\textbf{Energy Resolution}. The section discussing energy 
resolution is placed last in order to be able to compare the
resolutions for different cases discussed in the previous
sections.

\section{Reconstruction Accuracy}\label{sec:Accuracy}
The jet reconstruction results using the ALICE jet finding algorithm in
both its seeded and seedless versions are presented.
The values of the
parameters used in both cases are shown in Table~\ref{tab:Param}.
The TPC efficiency quoted in the table is not a parameter but is a
conservative estimate of the TPC tracking efficiency which was set in the
analysis. For the case
of the seeded algorithm, for comparison, jet finding is performed on a
combination of tracking and calorimeter data and additionally, on
tracking data alone. For the seedless algorithm case, jet finding is
performed on a combination of tracking and calorimetry data.

\begin{table}[!hbt]
  \center
  \begin{tabular}{|l||c|c|c|c|c|}  \hline 
    \begin{tabular}{c}
      \textbf{Algorithm}\\
    \end{tabular}
    &
    \begin{tabular}{c}
      \textbf{$R$}\\
    \end{tabular}
    &
    \begin{tabular}{c}
      \textbf{$JetESeed$}\\
      \textbf{(GeV)}
    \end{tabular}
    &
    \begin{tabular}{c}
      \textbf{$MinJetEt$}\\
      \textbf{(GeV)}
    \end{tabular}
    &
    \begin{tabular}{c}
      \textbf{BG Method}\\
    \end{tabular}
    &
     \begin{tabular}{c}
      \textbf{TPC}\\
      \textbf{efficiency}
    \end{tabular}
   \\ \hline
   Seeded    &  0.3  &  4.6  &  14.0  &  Ratio  &  $90\%$ \\ \hline
   Seedless  &  0.3  &  0.0  &  20.0  &  Ratio  &  $90\%$ \\ \hline   
  \end{tabular}   
  \protect\caption{Optimised parameters for the seeded and seedless
  versions of the algorithm.}
  \protect\label{tab:Param} 
\end{table}

\subsection{Seeded Algorithm Results}
\subsubsection{Results for tracking plus calorimetry data}
\begin{figure}[!hbt]
          \center
          \includegraphics[scale=0.5]{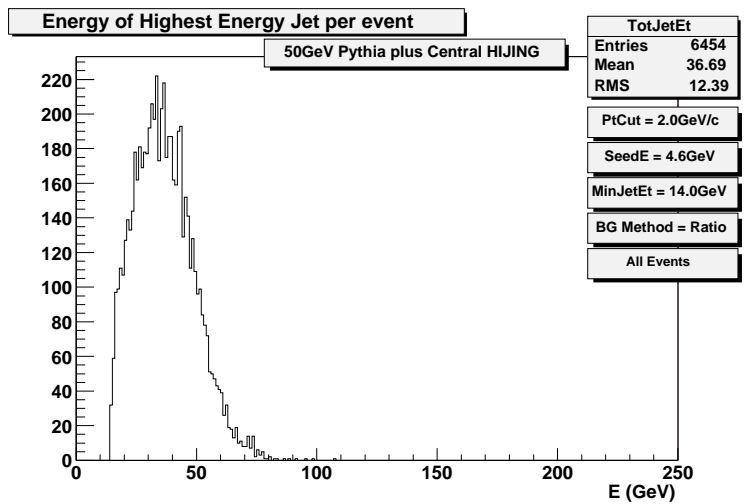}
          \protect\caption{Reconstructed jet energy distribution for
          input 50 GeV PYTHIA jets on Central HIJING background.}
          \protect\label{fig:AllJetEt50}
        \end{figure}

\begin{figure}[!hbt]
          \center
          \includegraphics[scale=0.5]{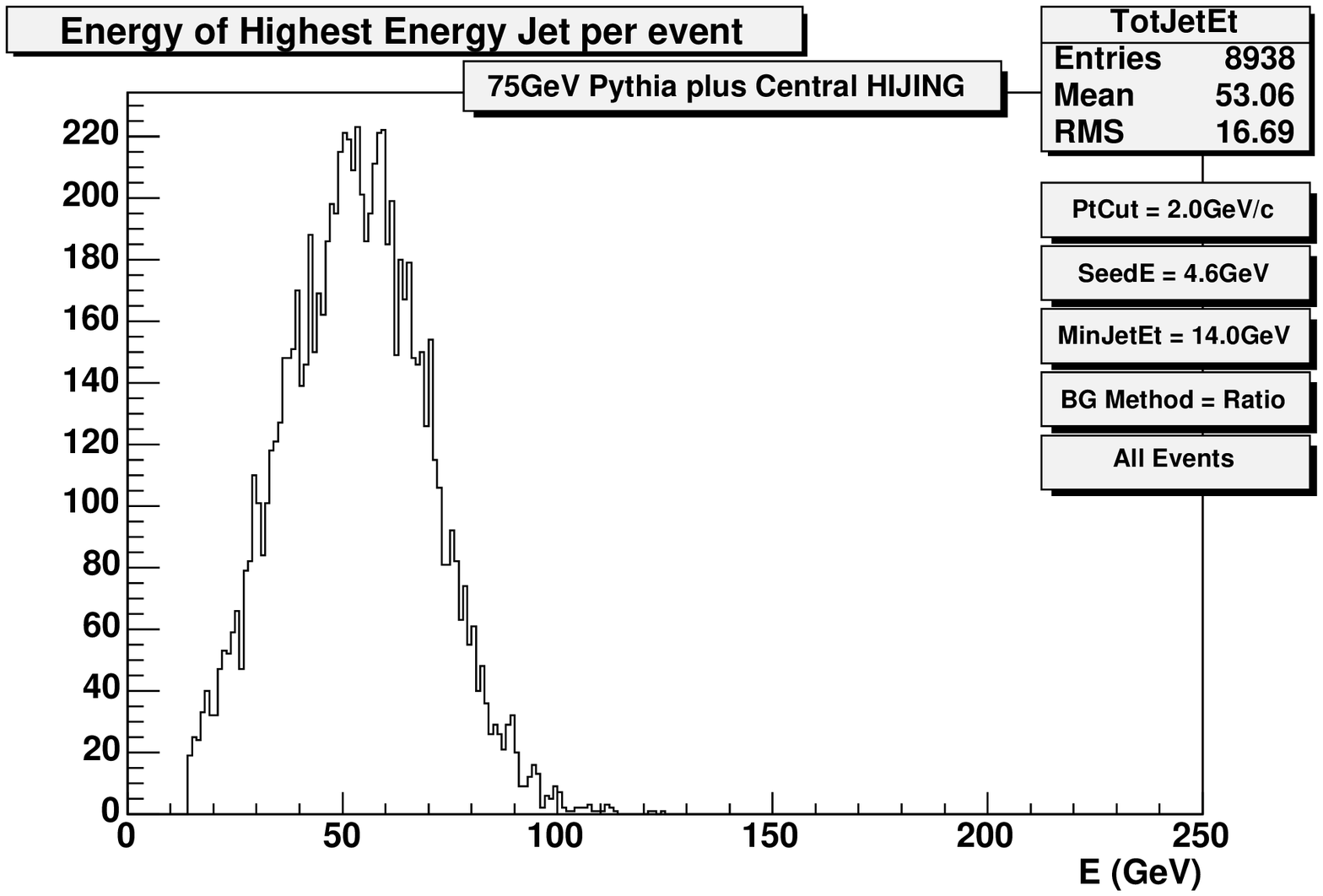}
          \protect\caption{Reconstructed jet energy distribution for
          input 75 GeV PYTHIA jets on Central HIJING background.}
          \protect\label{fig:AllJetEt75}
        \end{figure}

\begin{figure}[!hbt]
          \center
          \includegraphics[scale=0.5]{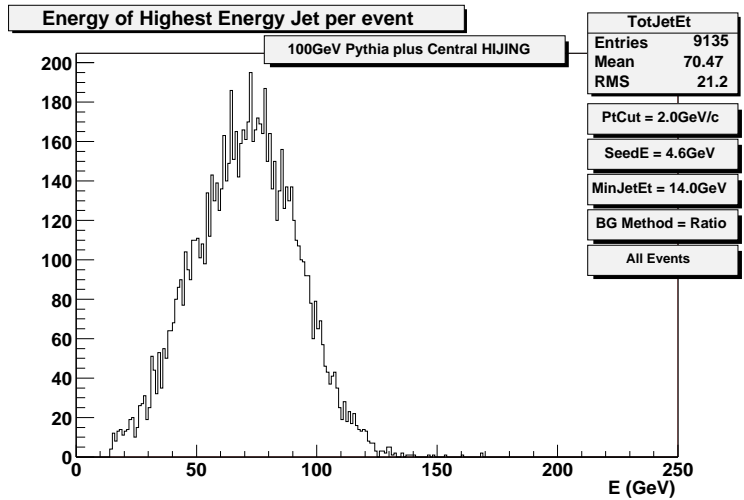}
          \protect\caption{Reconstructed jet energy distribution for
          input 100 GeV PYTHIA jets on Central HIJING background.}
          \protect\label{fig:AllJetEt100}
        \end{figure}

The reconstructed energy distributions for 50 GeV, 75 GeV and 100
GeV PYTHIA jets on Central HIJING background (i.e. \emph{combined} events) are shown in
Figs.~\ref{fig:AllJetEt50}-\ref{fig:AllJetEt100}. As expected, the
reconstructed jet energies, in all cases, are lower 
than the input jet
energies. This is due to the cone radius of $R=0.3$ putting a constraint on the
amount of jet energy contained within the cone, the energy from
neutral hadrons being excluded as it was not measured and the
$p_{T}$-cut on track energies. The sharp
cut-off in the jet energy distributions at 14 GeV is due to the
$MinJetEt$ parameter setting. 

For comparison, the algorithm was also applied to pure PYTHIA events
with no added background with identical parameters (except for background subtraction
which used the \emph{Cone Method}). Table~\ref{tab:SeededJetEnergy} shows,
for both the combined and pure PYTHIA events, the resulting mean jet
energies as a percentage of the input nominal jet energies. Thus the
algorithm produces results for the combined events which are within
$10\%$ of the results for pure PYTHIA events.

Corrections to the reconstructed jet energy can be made to account for
the losses due to the $p_{T}$-cut and small cone radius by using a
cross-section weighted multiplicative factor as discussed in
section~\ref{sec:EnergyCorrection}. When the experiment goes
live, $\gamma$-jet events will be used to calibrate
the reconstructed jet energies, see section~\ref{sec:EnergyCorrection}.
A detailed description of jet energy calibration will not be discussed
in detail in this thesis. 


\begin{table}[!hbt]
  \center
  \begin{tabular}{|l||c|c|c|}  \hline 
    \begin{tabular}{c}
      \textbf{Event Type}\\      
    \end{tabular}
    &
    \begin{tabular}{c}
     $\langle E_{Reco}\rangle / E_{Input} $ \textbf{for}\\
      50 GeV \textbf{input jet}\\
    \end{tabular}
    &
    \begin{tabular}{c}
      $\langle E_{Reco}\rangle / E_{Input} $ \textbf{for}\\
      75 GeV \textbf{input jet}\\
    \end{tabular}
    &
    \begin{tabular}{c}
      $\langle E_{Reco} \rangle/ E_{Input} $ \textbf{for}\\
      100 GeV \textbf{input jet}\\
    \end{tabular}    
   \\ \hline
   Combined        &  0.73  &  0.71  &  0.70  \\ \hline
   Pure PYTHIA  &  0.64  &  0.65  &  0.66  \\ \hline   
  \end{tabular}   
  \protect\caption{$\langle E_{Reco} \rangle/E_{Input}$ for combined events and pure
  PYTHIA events for the cases of 50 GeV, 75 GeV and 100 GeV input jets.}
  \protect\label{tab:SeededJetEnergy} 
\end{table}

\begin{figure}[!hbt]
          \center
          \includegraphics[scale=0.6]{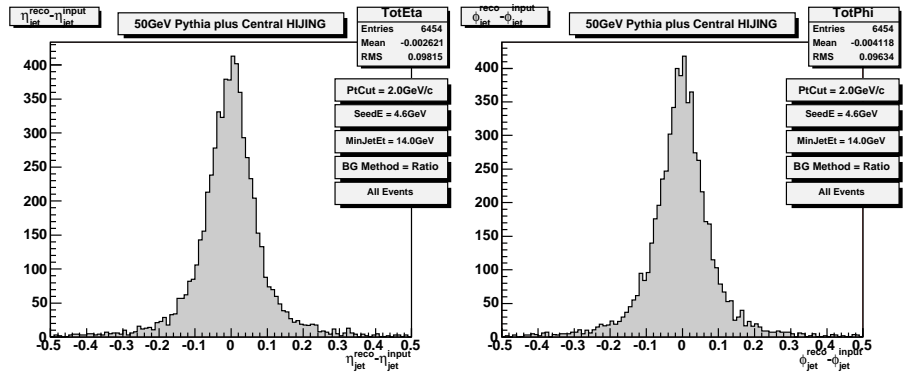}  
          \protect\caption{Difference between reconstructed jet
          directions and the input jet direction for 50 GeV PYTHIA
          events on Central HIJING background.}
          \protect\label{fig:AllJetEtaPhi50}
        \end{figure}

\begin{figure}[!hbt]
          \center
          \includegraphics[scale=0.6]{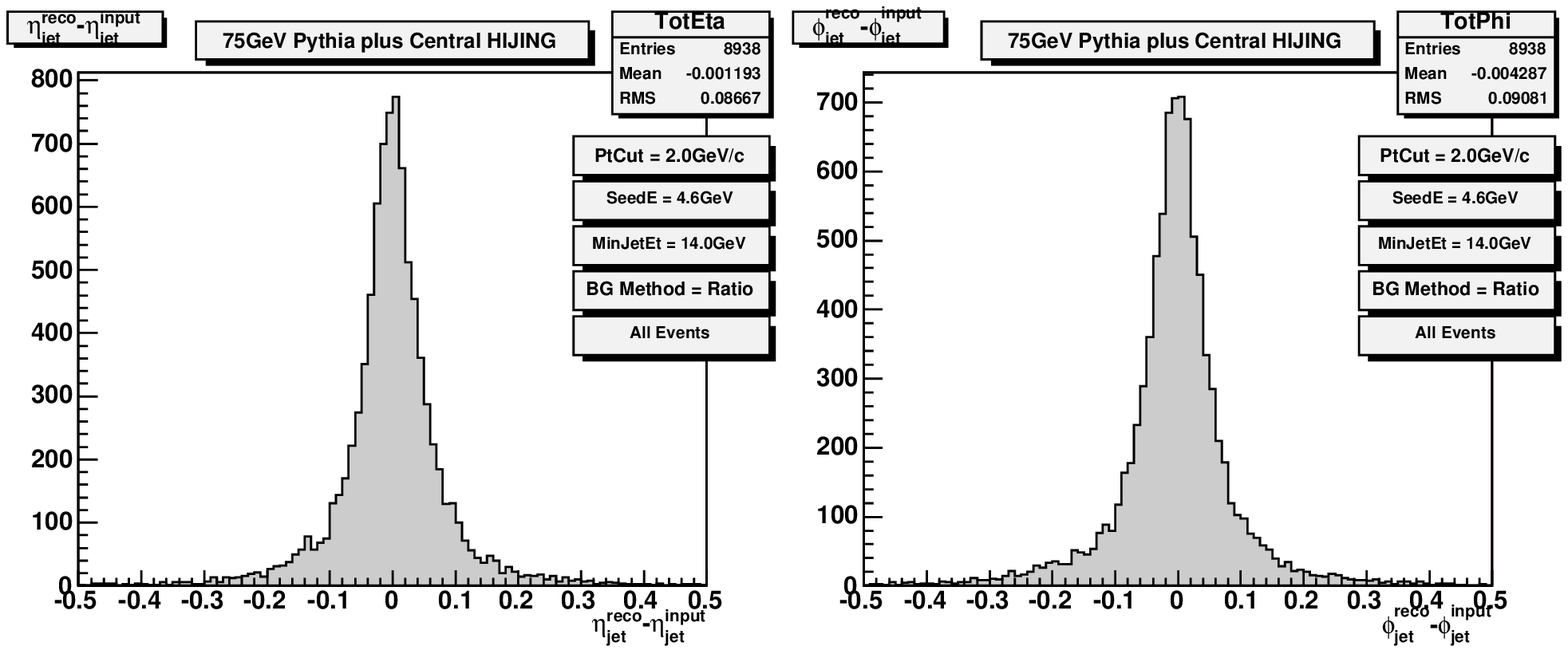}
          \protect\caption{Difference between reconstructed jet
          directions and the input jet direction for 75 GeV PYTHIA
          events on Central HIJING background. }
          \protect\label{fig:AllJetEtaPhi75}
        \end{figure}

\begin{figure}[!hbt]
          \center
          \includegraphics[scale=0.6]{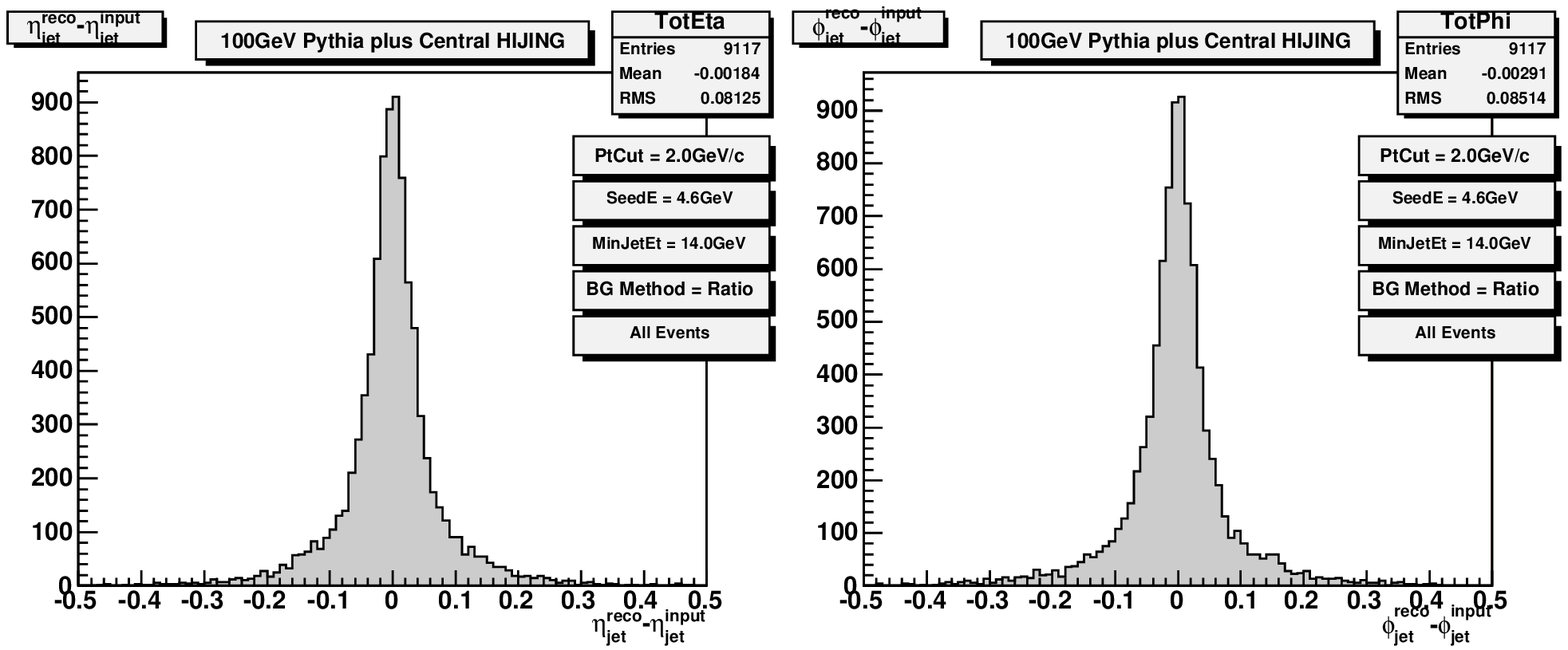}
          \protect\caption{Difference between reconstructed jet
          directions and the input jet direction for 100 GeV PYTHIA
          events on Central HIJING background. }
          \protect\label{fig:AllJetEtaPhi100}
        \end{figure}

\begin{figure}[!hbt]
          \center
          \includegraphics[scale=0.6]{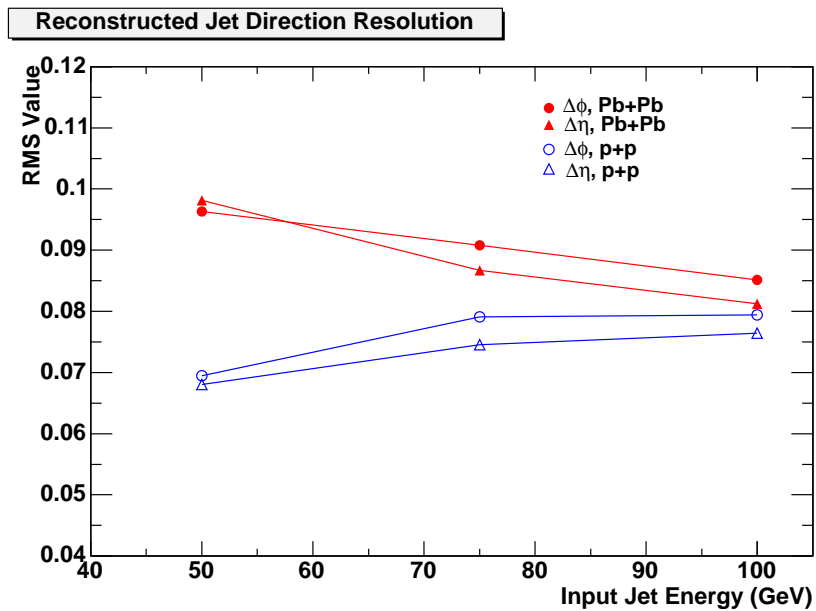}
          \protect\caption{RMS of reconstructed jet $\Delta\eta$ and
          $\Delta\phi$ distributions for the combined events (Pb+Pb) (closed
          symbols) case compared to 
          the pure PYTHIA (p+p) case (open symbols). }
          \protect\label{fig:JetDirResolution}
        \end{figure}

In order to measure the accuracy of the reconstructed jet directions
as calculated by the algorithm, the difference between the
reconstructed jet direction, in ($\eta,\phi$)-space, and the original
parton direction, from the PYTHIA event, was calculated for each combined event
on which jet finding was performed. 
Histograms showing $\Delta\eta =
\eta^{Reco}_{Jet}-\eta^{Input}_{Jet}$ and $\Delta\phi =
\phi^{Reco}_{Jet}-\phi^{Input}_{Jet}$ for the range of input jet
energies are shown in
Figs.~\ref{fig:AllJetEtaPhi50}-\ref{fig:AllJetEtaPhi100}. The maxima of the
histograms are peaked at zero, for all cases, showing that on average
the correct jet direction is found.
The accuracy of the direction reconstruction increases 
with increasing input jet energy as can be seen 
by comparing the RMS values which are plotted in
Fig.~\ref{fig:JetDirResolution}. Also shown in
Fig.~\ref{fig:JetDirResolution} are the RMS results for the algorithm
applied to pure PYTHIA events. The RMS values for the combined events
are larger than for the pure PYTHIA events, but approach the same
value as the input jet energy increases.

\subsubsection{Results for tracking data alone}
\begin{figure}[!hbt]
          \center
          \includegraphics[scale=0.5]{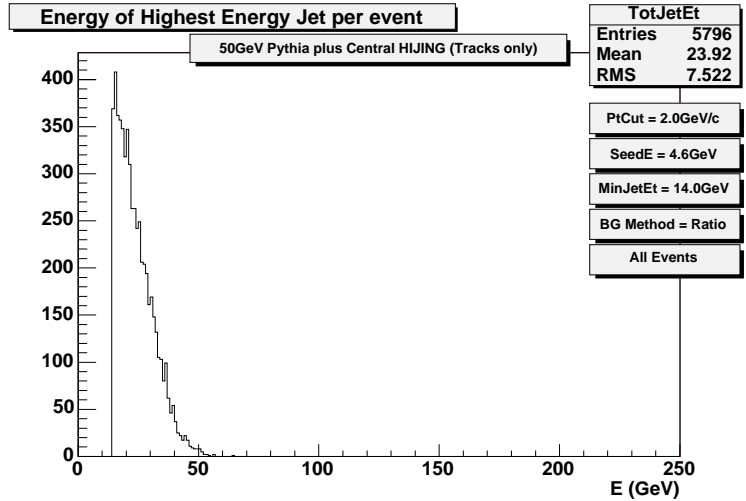}
          \protect\caption{Reconstructed jet energy distribution for
          input 50 GeV PYTHIA jets on Central HIJING background using
          tracking data alone.}
          \protect\label{fig:AllJetEt50Tracks}
        \end{figure}

\begin{figure}[!hbt]
          \center
          \includegraphics[scale=0.5]{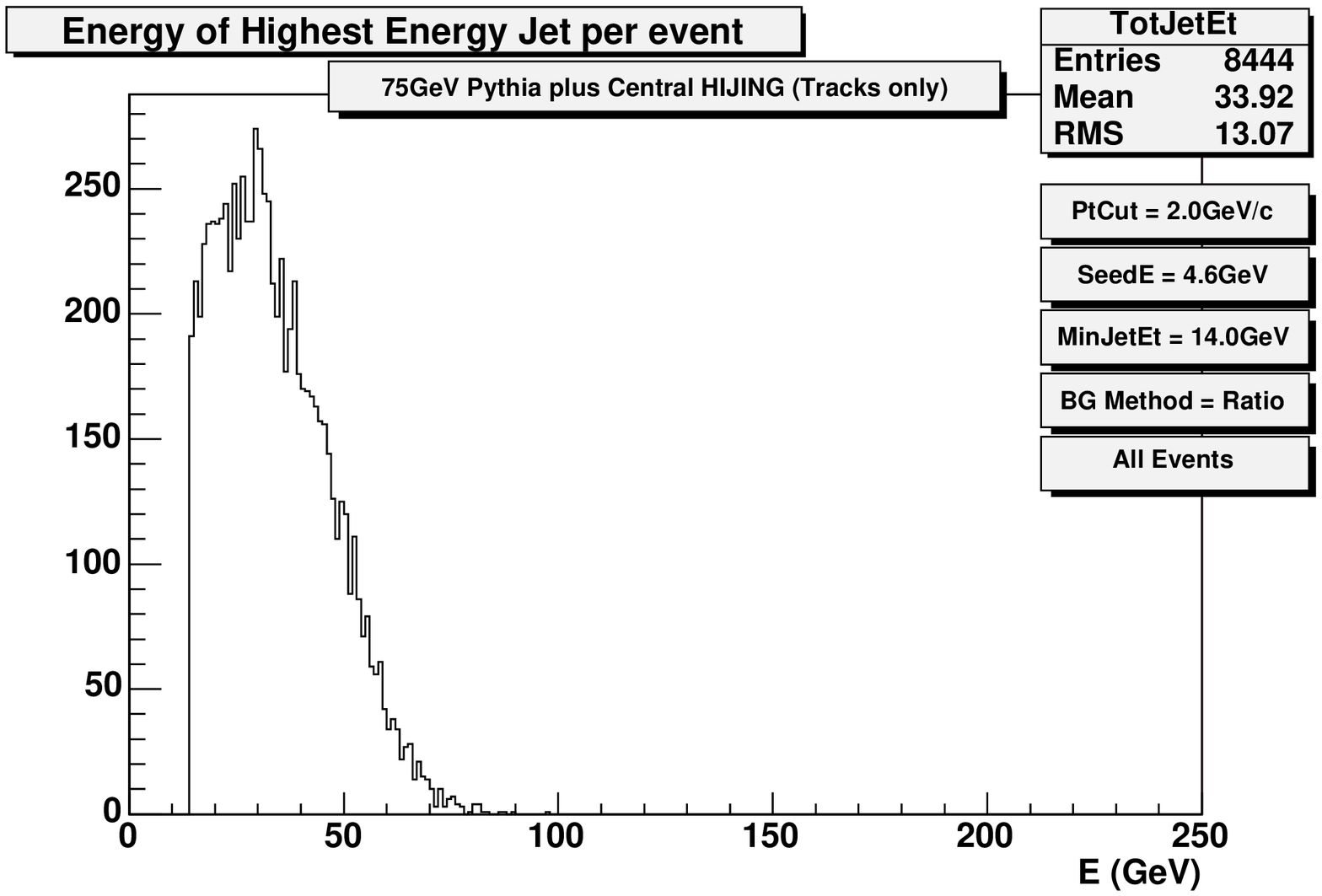}
          \protect\caption{Reconstructed jet energy distribution for
          input 75 GeV PYTHIA jets on Central HIJING background using
          tracking data alone.}
          \protect\label{fig:AllJetEt75Tracks}
        \end{figure}

\begin{figure}[!hbt]
          \center
          \includegraphics[scale=0.5]{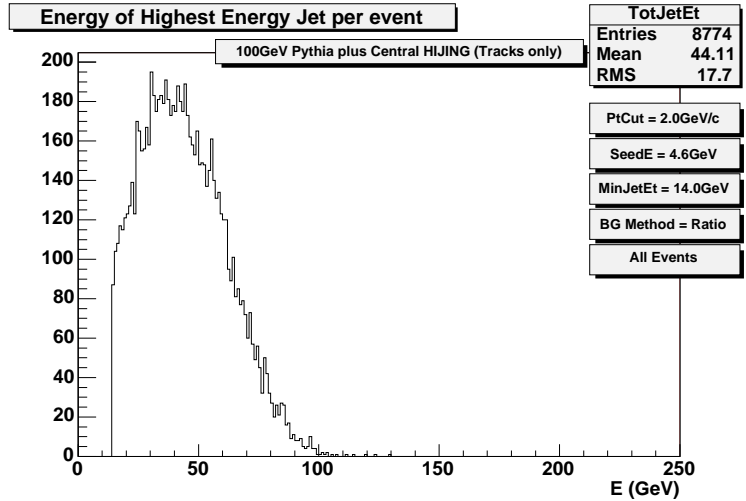}
          \protect\caption{Reconstructed jet energy distribution for
          input 100 GeV PYTHIA jets on Central HIJING background
          using tracking data alone.}
          \protect\label{fig:AllJetEt100Tracks}
        \end{figure}

In order to understand the accuracy and efficiency of the jet finding
algorithm in the case that only tracking information is available,
jet finding was performed on charged tracking data only, for PYTHIA
events combined with Central HIJING events. The
reconstructed energy distributions can be seen in
Figs.~\ref{fig:AllJetEt50Tracks}-\ref{fig:AllJetEt100Tracks}.  

Compared
to the reconstructed jet energy distributions from the tracking plus calorimetry
data, the distributions for tracking data alone have a larger shift towards lower
reconstructed jet energies. Table~\ref{tab:TracksVsHits} shows the
percentage of total nominal jet energy reconstructed using tracking
data alone compared to tracking plus calorimetry data. There is
$\sim$25$\%$ less jet energy reconstructed when using tracking data alone
compared to when using a combination of tracking plus calorimetry
data. However, it is still possible to reconstruct jets using the jet
finding algorithm with the seeded parameters on tracking data alone
although the resulting energy resolutions are worse than when a
combination of tracking and calorimetry data is used, see
Section~\ref{sec:Resolution}.

\begin{table}[!hbt]
  \center
  \begin{tabular}{|c||c|c|c|}  \hline 
    \begin{tabular}{c}
       \textbf{Data Type}\\      
    \end{tabular}
    &
    \begin{tabular}{c}
      $\langle E_{Reco}\rangle / E_{Input} $ \textbf{for}\\
      50 GeV \textbf{input jet}\\
    \end{tabular}
    &
    \begin{tabular}{c}
      $\langle E_{Reco}\rangle / E_{Input} $ \textbf{for}\\
      75 GeV \textbf{input jet}\\
    \end{tabular}
    &
    \begin{tabular}{c}
      $\langle E_{Reco} \rangle / E_{Input} $ \textbf{for}\\
      100 GeV \textbf{input jet}\\
    \end{tabular}    
   \\ \hline
   \begin{tabular}{l}
     Tracking plus\\
     calorimetry
    \end{tabular} 
     &  0.73  &  0.71  &  0.70  \\ \hline
   \begin{tabular}{l}
     Tracking alone\\
    \end{tabular}
     &  0.48  &  0.45  &  0.44  \\ \hline   
  \end{tabular}   
  \protect\caption{$\langle E_{Reco}\rangle / E_{Input}$ for the seeded
  algorithm case using tracking plus calorimetry data vs. tracking
  data alone. Both sets of events used were combined events.  }
  \protect\label{tab:TracksVsHits} 
\end{table} 

\begin{figure}[!hbt]
          \center
          \includegraphics[scale=0.6]{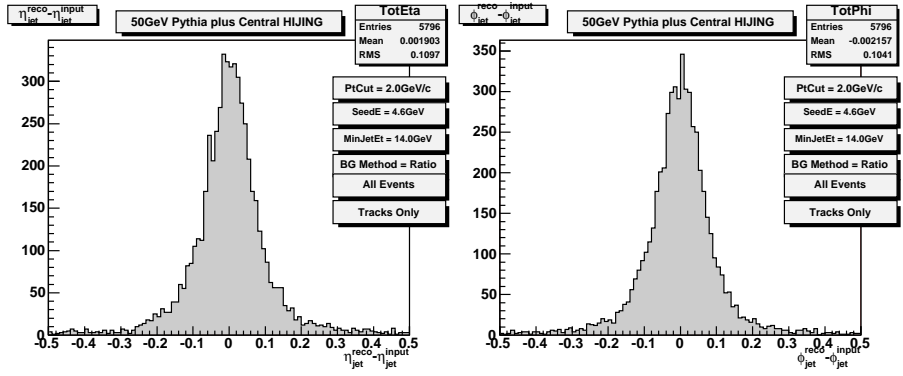}
          \protect\caption{Difference between reconstructed jet
          directions and the input jet direction for 50 GeV PYTHIA
          events on Central HIJING background using tracking data alone.}
          \protect\label{fig:AllJetEtaPhi50Tracks}
        \end{figure}

\begin{figure}[!hbt]
          \center
          \includegraphics[scale=0.6]{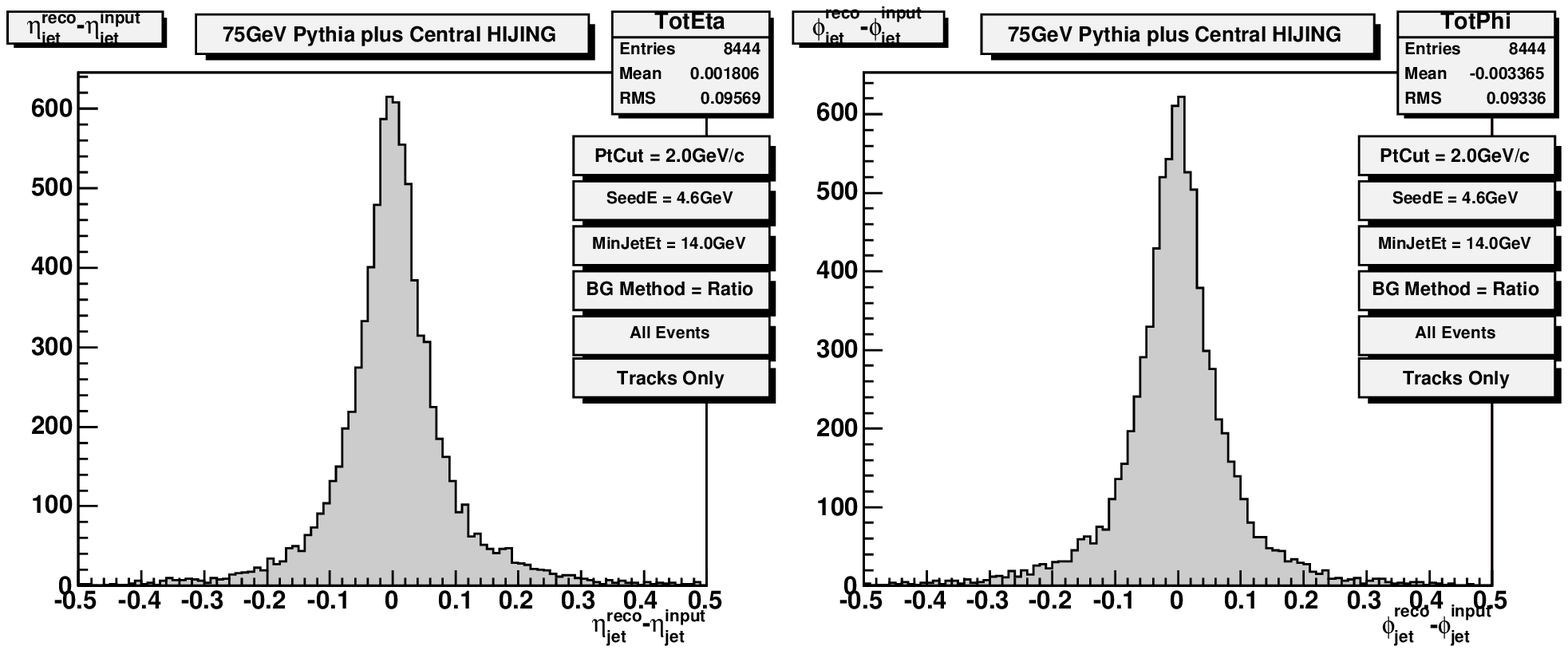}
          \protect\caption{Difference between reconstructed jet
          directions and the input jet direction for 75 GeV PYTHIA
          events on Central HIJING background using tracking data alone. }
          \protect\label{fig:AllJetEtaPhi75Tracks}
        \end{figure}

\begin{figure}[!hbt]
          \center
          \includegraphics[scale=0.6]{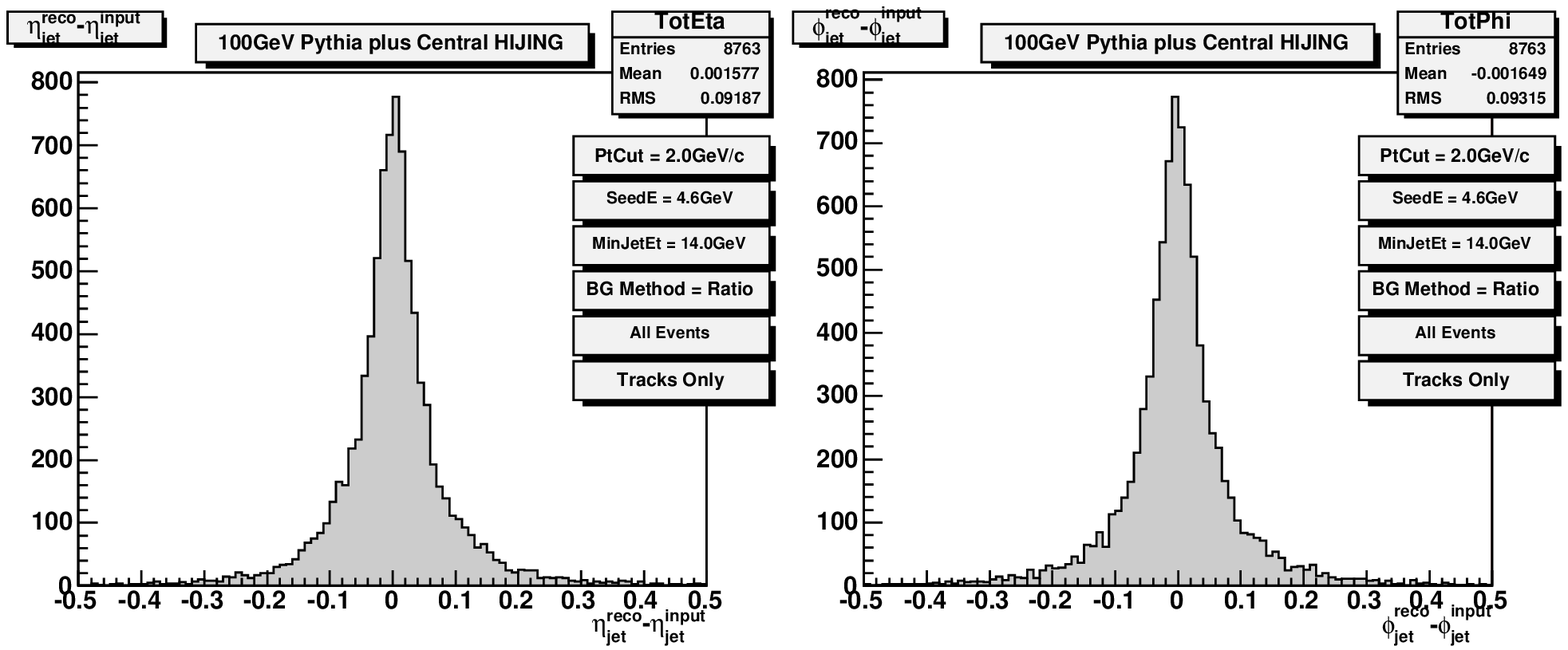}
          \protect\caption{Difference between reconstructed jet
          directions and the input jet direction for 100 GeV PYTHIA
          events on Central HIJING background using tracking data alone. }
          \protect\label{fig:AllJetEtaPhi100Tracks}
        \end{figure}

\begin{figure}[!hbt]
          \center
          \includegraphics[scale=0.6]{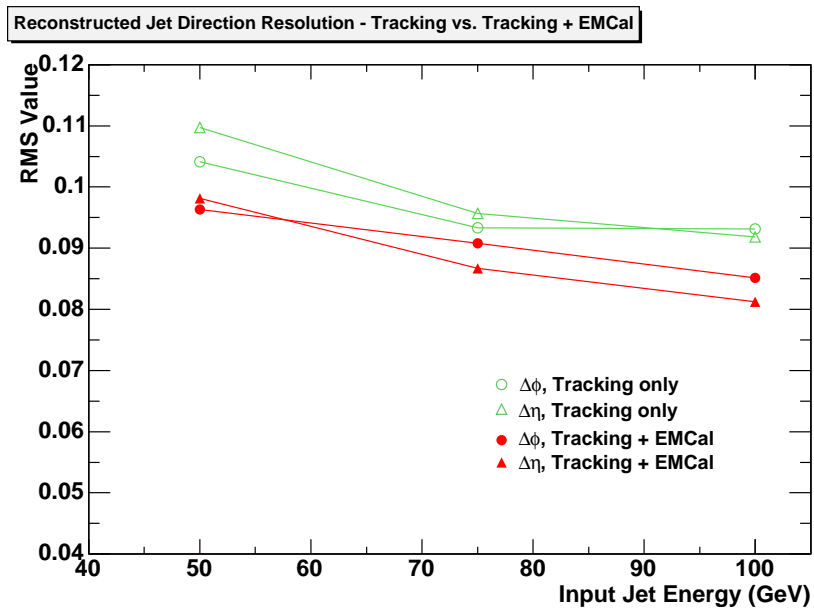}
          \protect\caption{RMS of the reconstructed jet $\Delta\eta$
          and $\Delta\phi$ distributions for combined events with
          tracking and calorimetry information (solid symbols)
          compared to events using only tracking information (open symbols). }
          \protect\label{fig:JetDirResTrackingOnly}
        \end{figure}

The accuracy with which the algorithm can reconstruct jet directions
using only tracking data can be seen in
Figs.~\ref{fig:AllJetEtaPhi50Tracks}-\ref{fig:AllJetEtaPhi100Tracks}.
The maxima of the histograms are peaked at zero as was the case for
the tracking plus calorimetry results but the widths of the
$\Delta\eta$ distributions are $\sim$12$\%$ greater than in the tracks
plus calorimetry case. The $\Delta\phi$ distributions are up to $9\%$
greater as shown in Table~\ref{tab:TracksVsHitsRMS}. Therefore,
direction reconstruction accuracy decreases when only tracking
information is used to reconstruct jets. A comparison
between the results using tracking plus calorimetry information and
tracking information only, is shown in Fig.~\ref{fig:JetDirResTrackingOnly}.

\begin{table}[!t]
  \center
  \begin{tabular}{|c||c|c|c|c|c|c|} \hline
    \textbf{Data Type}
    &
    \multicolumn{2}{|c|}{
     \begin{tabular}{c}
     \textbf{RMS Value} \\
     \textbf{for 50GeV}\\
     \textbf{distribution}
     \end{tabular}}
    &
    \multicolumn{2}{|c|}{
     \begin{tabular}{c}
     \textbf{RMS Value}\\
     \textbf{for 75GeV}\\
     \textbf{distribution}
     \end{tabular}}
    &
    \multicolumn{2}{|c|}{
     \begin{tabular}{c}
     \textbf{RMS Value}\\
     \textbf{for 100GeV}\\
     \textbf{distribution}
     \end{tabular}}    
    \\ \cline{2-7}
      & $\Delta\eta$ RMS $$ & $\Delta\phi$ RMS &
      $\Delta\eta$ RMS $$ & $\Delta\phi$ RMS & 
      $\Delta\eta$ RMS $$ & $\Delta\phi$ RMS \\ \hline
    \begin{tabular}{l}
     Tracking plus\\
     calorimetry 
    \end{tabular}
    & 0.098  & 0.096  & 0.087  & 0.091  & 0.081 & 0.085 \\ \hline
    \begin{tabular}{l}
     Tracking alone\\
    \end{tabular}  
    & 0.110 & 0.104  & 0.096  & 0.093  & 0.092  & 0.093 \\ \hline   
  \end{tabular}
   \protect\caption{Comparison of the RMS values for the $\Delta\eta$
  and $\Delta\phi$ distributions for jet finding results from tracking
  plus calorimetry data and tracking data alone.}
  \protect\label{tab:TracksVsHitsRMS} 
\end{table}

\newpage\subsection{Seedless Algorithm Results}

\subsubsection{Results for tracking plus calorimetry data}
\begin{figure}[!hbt]
          \center
          \includegraphics[scale=0.5]{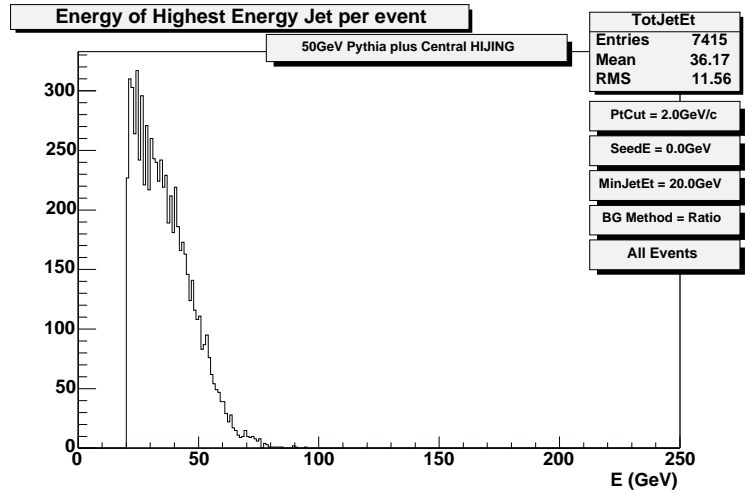}
          \protect\caption{Reconstructed jet energy distribution for
          input 50 GeV PYTHIA jets on Central HIJING background using the
          seedless algorithm.}
          \protect\label{fig:AllJetEt50NoSeed}
        \end{figure}

\begin{figure}[!hbt]
          \center
          \includegraphics[scale=0.5]{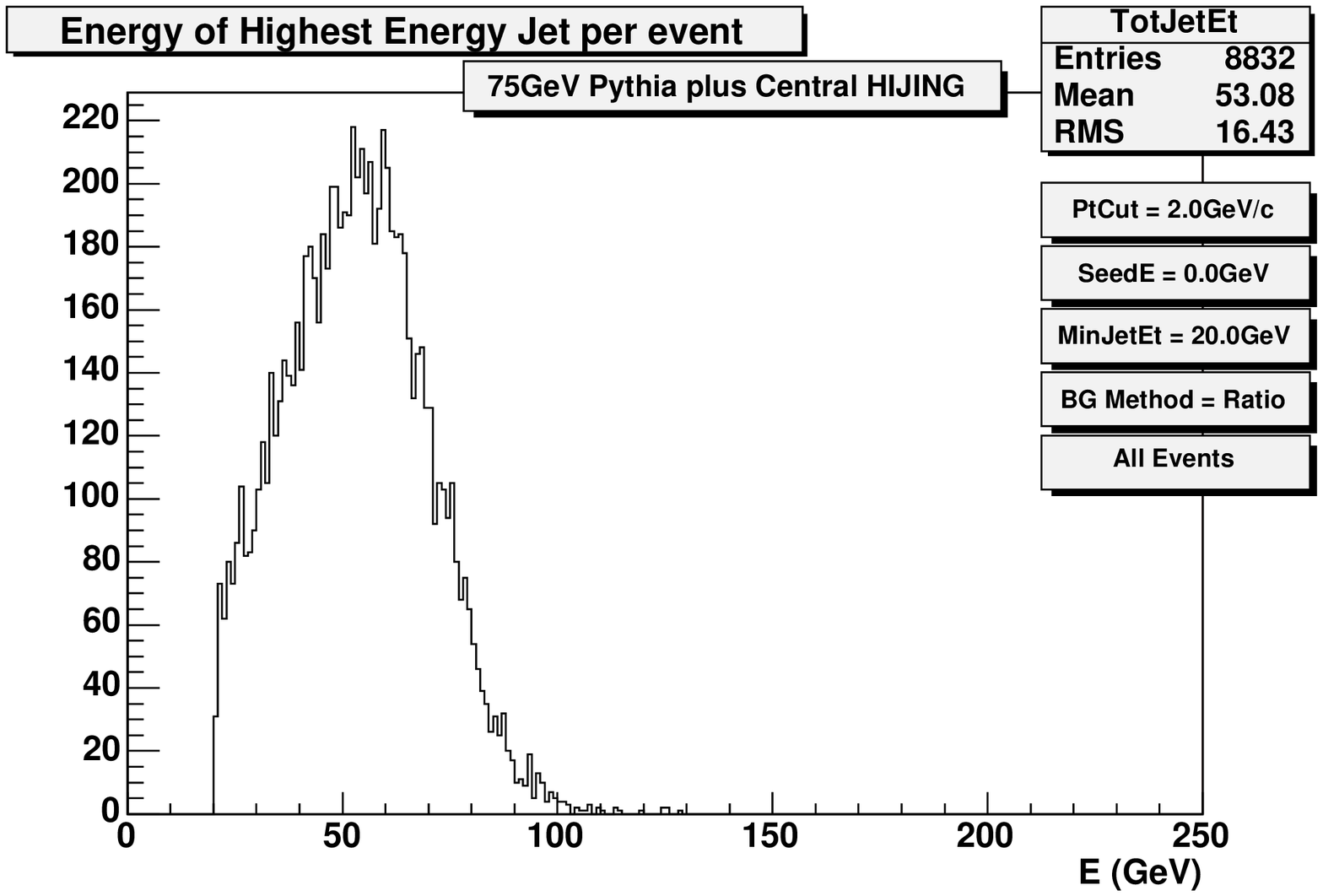}
         \protect\caption{Reconstructed jet energy distribution for
          input 75 GeV PYTHIA jets on Central HIJING background using the
          seedless algorithm.}
          \protect\label{fig:AllJetEt75NoSeed}
        \end{figure}

\begin{figure}[!hbt]
          \center
          \includegraphics[scale=0.5]{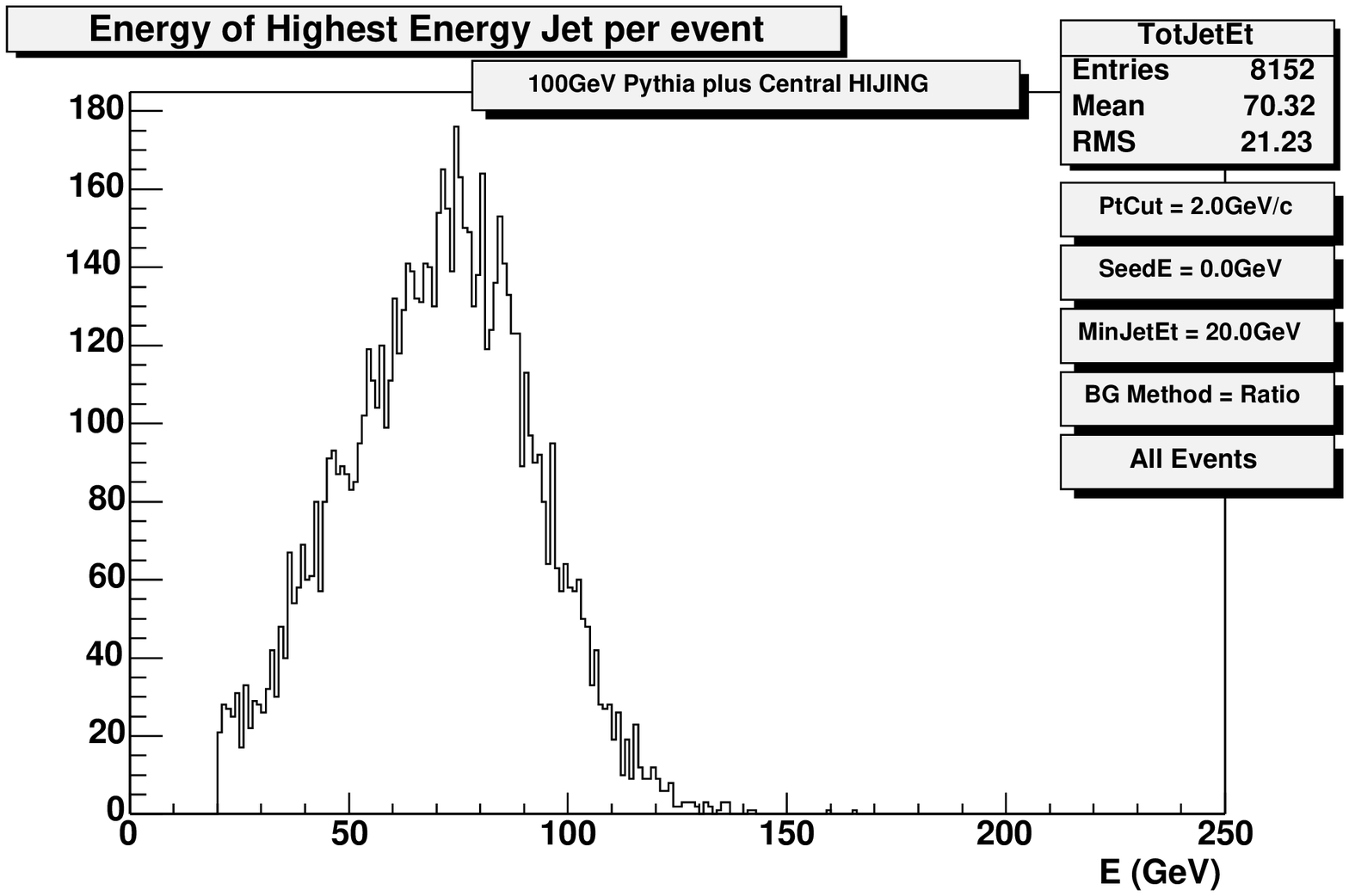}
          \protect\caption{Reconstructed jet energy distribution for
          input 100 GeV PYTHIA jets on Central HIJING background
          using the seedless algorithm.}
          \protect\label{fig:AllJetEt100NoSeed}
        \end{figure}

Figs.~\ref{fig:AllJetEt50NoSeed}-\ref{fig:AllJetEt100NoSeed} show the
reconstructed jet energy distributions for the case of zero seed
energy. The reconstructed mean energies are slightly lower than those for the
seeded algorithm case, see Table~\ref{tab:SeedVsSeedlessE}. However,
the shape of the seedless distributions is skewed to lower energies
compared to the seeded case. This is due to a higher number 
of `fake' jets being included by the algorithm as real jets in the
distribution, (discussed in Section~\ref{sec:Efficiency}), when the seed energy
is set to zero. The sharp cut-off in the jet energy distributions at
20 GeV is due to the \emph{MinJetEt} parameter setting.

\begin{table}[!hbt]
  \center
  \begin{tabular}{|l||c|c|c|}  \hline 
    \begin{tabular}{c}
       \textbf{Algorithm}\\
       \textbf{Type}       
    \end{tabular}
    &
    \begin{tabular}{c}
      $\langle E_{Reco}\rangle$ \textbf{for} 50GeV \\
      \textbf{input jet} \\
      (GeV)\\
    \end{tabular}
    &
    \begin{tabular}{c}
      $\langle E_{Reco}\rangle$ \textbf{for} 75GeV \\
      \textbf{input jet} \\
      (GeV) \\
    \end{tabular}
    &
    \begin{tabular}{c}
      $\langle E_{Reco}\rangle$ \textbf{for} 100GeV \\
      \textbf{input jet} \\
      (GeV)\\
    \end{tabular}    
   \\ \hline
   \begin{tabular}{l}
     Seeded\\
    \end{tabular} 
     & 36.69   & 53.06   &  70.47  \\ \hline
   \begin{tabular}{l}
     Seedless\\
    \end{tabular}
     & 36.17   & 53.08   &  70.32  \\ \hline   
  \end{tabular}   
  \protect\caption{Reconstructed mean jet energies for various input
  jet energies for the case of the seeded algorithm vs. the seedless algorithm. }
  \protect\label{tab:SeedVsSeedlessE} 
\end{table}

\begin{figure}[!hbt]
          \center
          \includegraphics[scale=0.6]{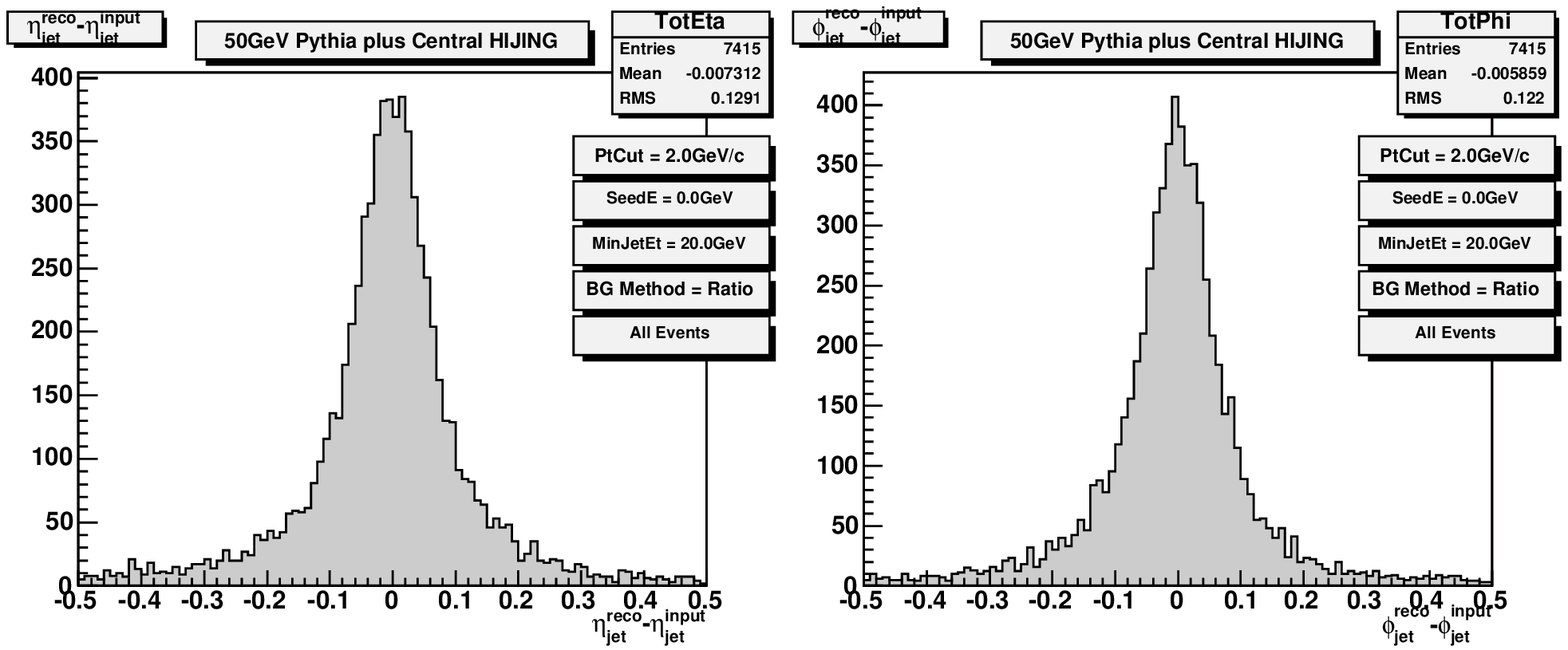}
          \protect\caption{Difference between reconstructed jet
          directions and the input jet direction for 50 GeV PYTHIA
          events on Central HIJING background using the seedless algorithm.}
          \protect\label{fig:AllJetEtaPhi50NoSeed}
        \end{figure}

\begin{figure}[!hbt]
          \center
          \includegraphics[scale=0.6]{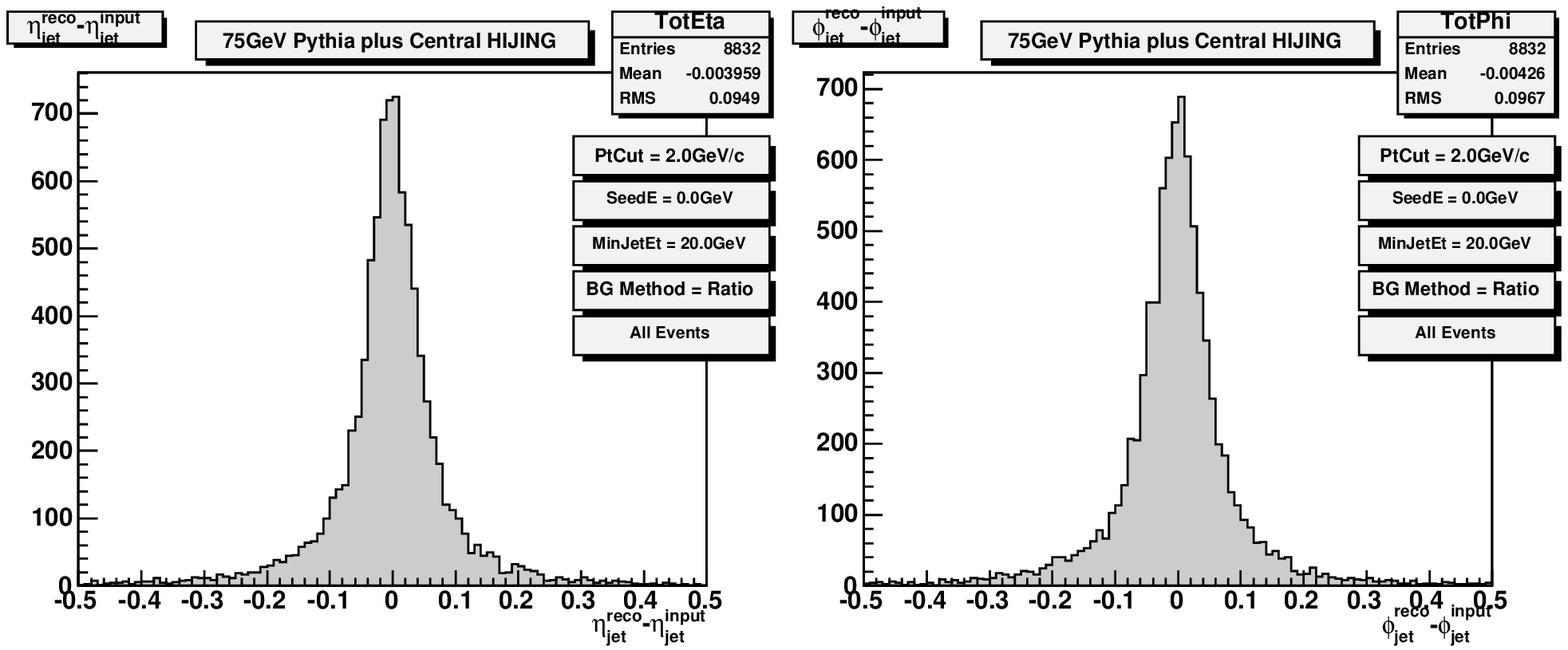}
          \protect\caption{Difference between reconstructed jet
          directions and the input jet direction for 75 GeV PYTHIA
          events on Central HIJING background using the seedless algorithm. }
          \protect\label{fig:AllJetEtaPhi75NoSeed}
        \end{figure}

\begin{figure}[!hbt]
          \center
          \includegraphics[scale=0.6]{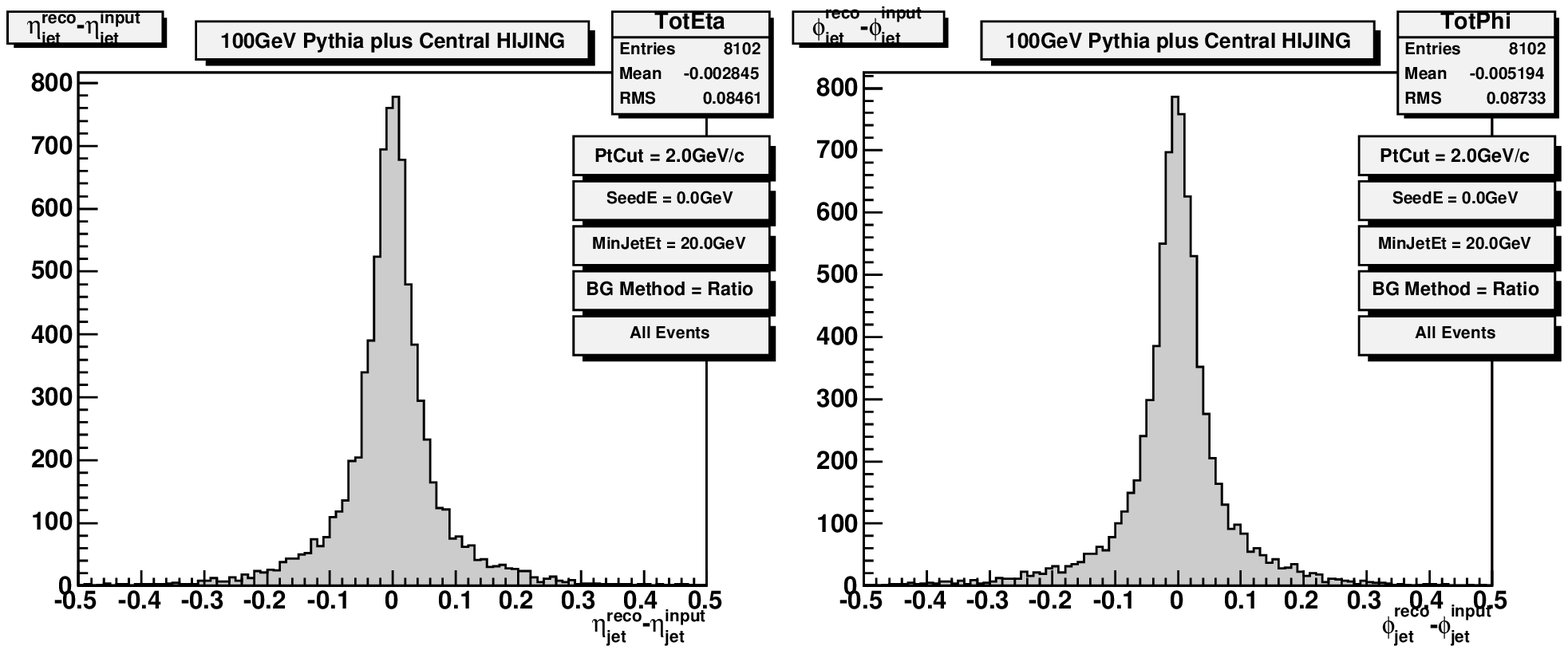}
          \protect\caption{Difference between reconstructed jet
          directions and the input jet direction for 100 GeV PYTHIA
          events on Central HIJING background using the seedless algorithm. }
          \protect\label{fig:AllJetEtaPhi100NoSeed}
        \end{figure}

The difference between the reconstructed jet directions and the input
directions are shown in
Figs.~\ref{fig:AllJetEtaPhi50NoSeed}-\ref{fig:AllJetEtaPhi100NoSeed}. The
widths of the $\Delta\eta$ and $\Delta\phi$ distributions are
$\sim$30$\%$ greater in the seedless case compared to the seeded case
for 
50 GeV input jets. However, the difference between the two cases
decreases with increasing 
input jet energy and for 100 GeV input jets, the spread is $\sim$3$\%$
greater for the seedless case than the seeded 
case. Table~\ref{tab:SeedVsSeedlessDir} shows the widths of the
reconstructed jet directions for the seedless case compared to the
seeded case. The top line of the table is taken from
Table~\ref{tab:TracksVsHitsRMS}.       
Therefore, from Table~\ref{tab:SeedVsSeedlessDir}, it can be seen that
the seeded algorithm reconstructs jet directions more
accurately than the seedless algorithm.

\begin{table}[!hbt]
  \center
  \begin{tabular}{|l||c|c|c|c|c|c|} \hline
    \begin{tabular}{c}
    \textbf{Algorithm}\\
    \textbf{Type}
    \end{tabular}
    &
    \multicolumn{2}{|c|}{
     \begin{tabular}{c}
     \textbf{RMS Value} \\
     \textbf{for 50GeV}\\
     \textbf{distribution}
     \end{tabular}}
    &
    \multicolumn{2}{|c|}{
     \begin{tabular}{c}
     \textbf{RMS Value}\\
     \textbf{for 75GeV}\\
     \textbf{distribution}
     \end{tabular}}
    & 
    \multicolumn{2}{|c|}{
     \begin{tabular}{c}
     \textbf{RMS Value}\\
     \textbf{for 100GeV}\\
     \textbf{distribution}
     \end{tabular}}    
    \\ \cline{2-7}
      & $\Delta\eta$ RMS $$ & $\Delta\phi$ RMS &
      $\Delta\eta$ RMS $$ & $\Delta\phi$ RMS & 
      $\Delta\eta$ RMS $$ & $\Delta\phi$ RMS \\ \hline
    \begin{tabular}{l}
     Seeded\\
    \end{tabular}
    & 0.098  & 0.096  & 0.087  & 0.091  & 0.081 & 0.085 \\ \hline
    \begin{tabular}{l}
     Seedless\\
    \end{tabular}  
    & 0.129 & 0.122  & 0.095  & 0.097  & 0.085  & 0.087 \\ \hline   
  \end{tabular}
   \protect\caption{Comparison of the RMS values for the $\Delta\eta$
  and $\Delta\phi$ distributions for jet finding results from the
  seeded algorithm vs. the seedless algorithm for tracking
  plus calorimetry data.}
  \protect\label{tab:SeedVsSeedlessDir} 
\end{table}

\section{Jet Energy Correction}\label{sec:EnergyCorrection}
When the first ALICE data is available, 
jet energies will be calibrated experimentally by
measuring the $\gamma$-jet process. The energy of the observed photon will
be the same as the energy of the jet, due to energy conservation, and
therefore its energy can be used to calibrate the measured jet energy.

Presented here is a simple correction which may be applied to jet
results in the absence of the calibration to real data. This
correction takes all detector and algorithm effects on the jet energy into account
simultaneously and involves the multiplication of the reconstructed
jet energy by a constant factor $C$. 

The jet energies reconstructed by the optimised algorithms are shifted
to lower energies as shown by the reconstructed jet energy
distribution plots in section~\ref{sec:Accuracy}.
There are four main contributions to the energy shift.
Using a small cone radius ($R=0.3$) in the jet finding algorithm leads
to the exclusion of energy from jet particles which are outside the
cone but which are really part of the jet. Performing a $p_{T}$-cut of
2 GeV/$c$ on all charged tracks further excludes some jet
energy. Energy is also excluded from jet particles which are
not measurable in the detectors (e.g. $K_{L}$).
The final contribution is due to the event-by-event energy fluctuations in
the underlying event which cause a small shift to higher
energy from inaccuracies in the background energy
calculation, see Table~\ref{tab:ConePercent}.
However this shift is smaller than the combined shift of the other 
three contributions in the opposite direction. This is shown by
comparing the fraction of jet energy contained within a cone of radius
$R=0.3$ for pure PYTHIA events and combined events. The fraction is lower
for pure PYTHIA events with no background than for combined events
(see Table~\ref{tab:ConePercent}). The fractions are obtained by
fitting the reconstructed jet 
energy distributions to a gaussian distribution and taking the mean
from the fit divided by the input jet energy.

\begin{table}[!hbt]
  \center
  \begin{tabular}{|l||c|c|c|} \hline
    & \textbf{50 GeV Input} & \textbf{75 GeV Input} & \textbf{100 GeV
    Input} \\ \cline{2-4}
  \begin{tabular}{l}
    $\langle E_{Reco}\rangle$ / $E_{Input}$ \\
    (pure PYTHIA)
  \end{tabular}
  & 64.08 $\%$ & 64.89 $\%$& 65.90 $\%$ \\ \hline
  \begin{tabular}{l}
    $\langle E_{Reco}\rangle$ / $E_{Input}$ \\
    (combined events)
  \end{tabular} 
  & 67.08 $\%$ & 68.83 $\%$ & 69.33 $\%$   \\ \hline
  \end{tabular}
   \protect\caption{Percentage jet energy contained within $R=0.3$ for
  pure PYTHIA events and combined events using the seeded algorithm.}
  \protect\label{tab:ConePercent} 
\end{table}

As shown in Table.~\ref{tab:ConePercent}, the percentage of total jet
energy contained inside $R=0.3$ increases 
as a function of jet energy for pure PYTHIA events and combined events.
However, when jets are
reconstructed in experiment, it is not known \emph{a priori} what the
real jet energy was. Therefore the factor $C$, by which jet
energies are to be multiplied, needs to be independent of jet energy.

In this case, $C$ was
calculated for the case of combined events (Pb+Pb case), by weighting the fraction of
jet energy contained inside $R=0.3$ by the appropriate cross-section for
50 GeV, 75 GeV and 100 GeV jet production. The resulting value is:
$C = 1/0.6731$.     

\begin{figure}[!hbt]
          \center
          \includegraphics[scale=0.5]{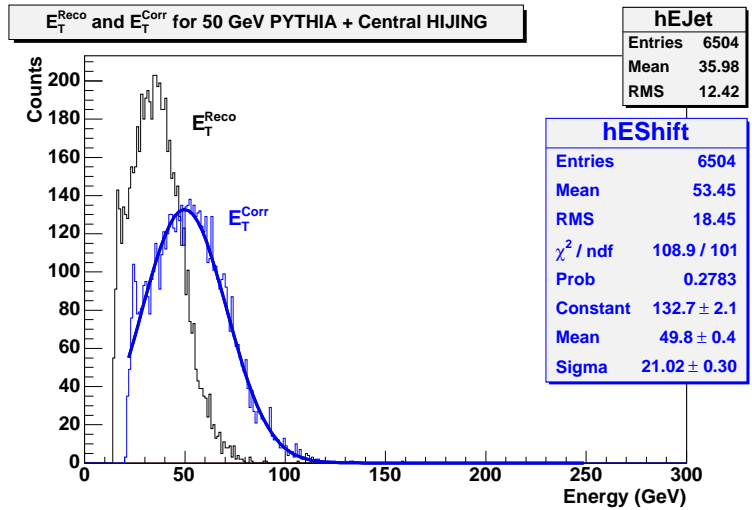}
          \protect\caption{Reconstructed ($E_{T}^{Reco}$) and
          corrected ($E_{T}^{Corr}$) jet
          energy distributions for 50 GeV jets using the method
          described in the text. }
          \protect\label{fig:EShift50}
        \end{figure}

\begin{figure}[!hbt]
          \center
          \includegraphics[scale=0.5]{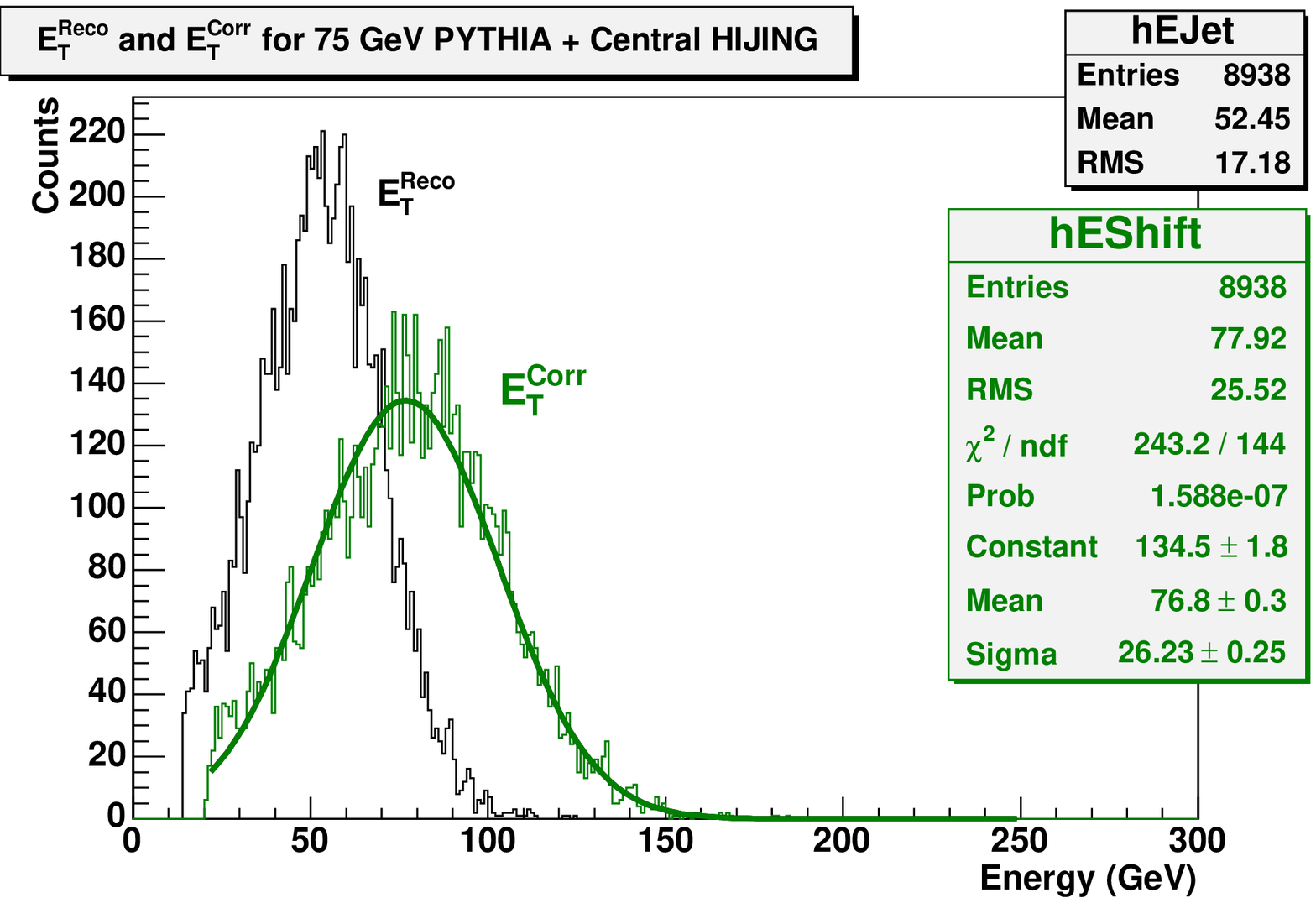}
          \protect\caption{Reconstructed ($E_{T}^{Reco}$) and
          corrected ($E_{T}^{Corr}$) jet
          energy distributions for 75 GeV jets using the method
          described in the text. }
          \protect\label{fig:EShift75}
        \end{figure}

\begin{figure}[!hbt]
          \center
          \includegraphics[scale=0.5]{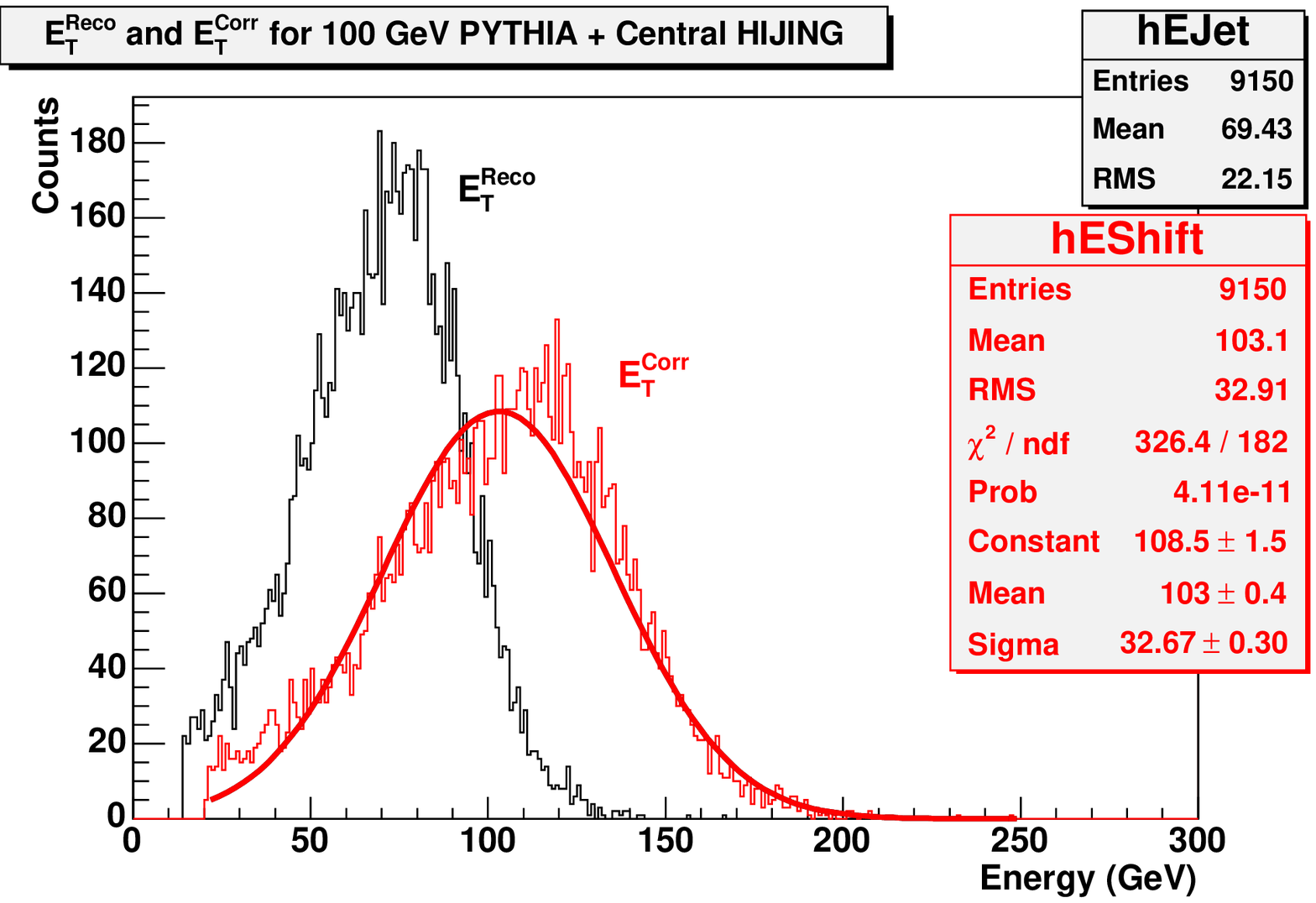}
          \protect\caption{Reconstructed ($E_{T}^{Reco}$) and
          corrected ($E_{T}^{Corr}$) jet
          energy distributions for 100 GeV jets using the method
          described in the text. }
          \protect\label{fig:EShift100}
        \end{figure}

The resulting jet energy distributions for 50 GeV, 75 GeV and 100 GeV
jets before and after performing the energy correction, are shown in
Figs.~\ref{fig:EShift50}-\ref{fig:EShift100}. (In this case, a
different sample of data was used compared to other analysis in this
chapter for technical programming reasons. However the two sets of
data are consistent.)  The
means and $\sigma$ of the fitted shifted distributions are shown in
Table.~\ref{tab:ShiftedE}. The corrected means are 
within $4\%$ of the input jet energies for the three
samples. Therefore it is possible to use one
multiplicative factor, $C$, to perform a successful jet energy
correction over a range of jet energies.

\begin{table}[!hbt]
  \center
  \begin{tabular}{|c||c|c|c|} \hline
    & \textbf{50 GeV Input} & \textbf{75 GeV Input} & \textbf{100 GeV
    Input} \\ \cline{2-4}
  \begin{tabular}{c}
    $\langle E_{Corr}\rangle$ (combined\\
    events) (GeV)
  \end{tabular}
  & 49.8 & 76.8 & 103.0 \\ \hline
  \begin{tabular}{c}
    $\sigma$ (combined\\
    events) (GeV)
  \end{tabular} 
  & 21.0 & 26.2 & 32.7   \\ \hline
  \end{tabular}
   \protect\caption{Corrected mean energies and $\sigma$ after fitting shifted
  energy distributions with a gaussian distribution.}
  \protect\label{tab:ShiftedE} 
\end{table}

\section{Reconstruction Efficiency}\label{sec:Efficiency}

\subsection{Efficiency and `Fake' Rates}
The jet reconstruction efficiency, $\epsilon$, of the algorithm can be defined as
the number of reconstructed jets divided by the number of known input jets:

\begin{equation}
  \epsilon = \frac{N^{Reco}_{Jets}}{N^{Input}_{Jets}}
\end{equation}

\noindent for a given sample of events.
If $\epsilon = 1$, it implies that all the input jets were found and
reconstructed. It is therefore the aim when optimising the
algorithm to ensure that $\epsilon$ is as close to 1 as possible.
If $\epsilon > 1$ then the jet finding algorithm found more jets than were input.

The case of $\epsilon > 1$ could 
happen for a number of reasons:
\begin{itemize}
  \item \textbf{`Underlying Event' Fluctuations }\\
    In the high multiplicity
    background environment of Central HIJING events,
    with large event-by-event fluctuations and mini-jets present, it is
    possible that the algorithm could mistake a large random
    fluctuation or mini-jet for a jet. PYTHIA events contain an
    `underlying event' which may lead to fluctuations in the
    measured energy deposition in the detectors that the algorithm
    could mistake for a real jet. These reconstructed jets will be
    classified as `fake' jets.

  \item \textbf{Unknown `Real' Jets}\\
    There is a significant
    cross-section for jet production in Central HIJING events at LHC
    energies, and therefore, there is the possibility that some of these jets
    fall within the EMCal fiducial volume and are reconstructed by the
    algorithm. There is also the possibility that the PYTHIA jet event
    contains more than two jets in it since the simulation condition
    was that \emph{at least} one jet must point in the direction of the EMCal
    (trigger jet)
    while no constraint was put on the maximum number of jets in the event. These
    extra jets could fall within the EMCal fiducial volume and
    be reconstructed by the algorithm. In this case, although the
    reconstructed jets are real, since there is no
    prior information about them as they are not purposely triggered
    on, they are treated as `fake' jets in this thesis to
    differentiate them from the known PYTHIA input
    jets. Fig.~\ref{fig:TwoJetsEMCAL} shows an event that is likely to
    be of this type.

    \begin{figure}[!hbt]
          \center
          \includegraphics[scale=0.6]{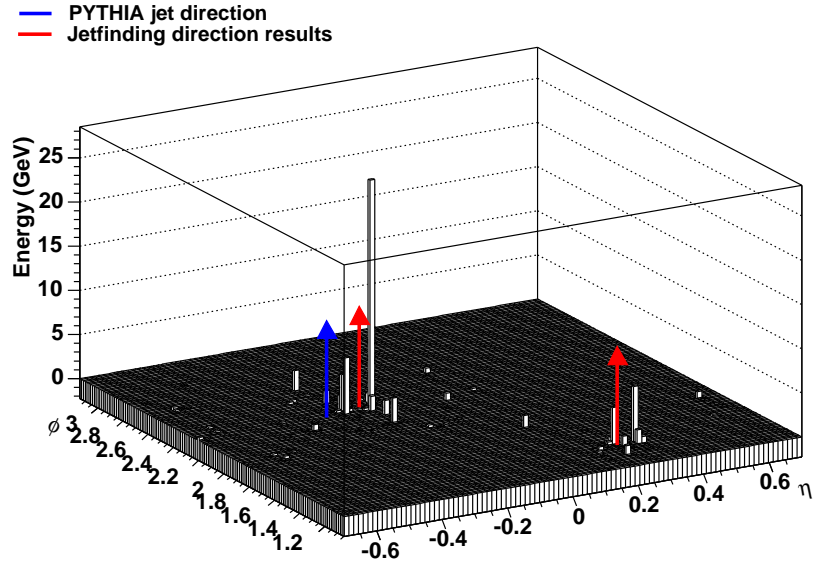}
          \protect\caption{Grid energy including tracking and
          calorimetry information for a PYTHIA event with a 100 GeV
          jet where the
          algorithm reconstructed two jets. The blue arrow indicates
          the direction of the trigger jet as calculated by PYTHIA and
          the red arrows indicate the directions of the two jets
          reconstructed by the algorithm.}
          \protect\label{fig:TwoJetsEMCAL}
        \end{figure}

    The figure shows the energy grid containing track and calorimetry
    information for a PYTHIA event, containing a 100 GeV jet, in which the seeded algorithm
    reconstructed two jets. The direction of the trigger jet as
    calculated by PYTHIA is indicated by the blue arrow ($\eta$=-0.29,
    $\phi$=2.22) and the
    directions of the two jets reconstructed by the algorithm are
    shown by the red arrows ($\eta_{1}$=-0.20, $\phi_{1}$=2.22) and
    ($\eta_{2}$=0.24, $\phi_{2}$=1.32).
    The jets reconstructed by the algorithm have    
    cone centres with an angular distance of
    $D=\sqrt{{(\Delta\eta)}^2+{(\Delta\phi)}^2} = 1.00$ away from each
    other.
    Therefore they are out of range of each other's centres to be included 
    within a cone radius of up to $R=1.0$, the typical cone jet
    definition. The angular distance of the trigger jet cone centre, as
    calculated by PYTHIA, from the centre of mass of the other energy
    deposit ($\eta_{2}$, $\phi_{2}$) is $R=1.04$. Therefore, the energy
    deposits reconstructed as two jets
    by the algorithm, are most likely real separate
    jets although the lower energy jet would be counted as a `fake'
    jet in
    this thesis. The proportion of PYTHIA events which contained two or more 
    reconstructed jets and which satisfied the following condition:
    \begin{enumerate}
    \item(2R) $D>1.0$, where $D$ is the angular distance between the two
    reconstructed jets, \label{crit1a}
    \end{enumerate}
    are shown in Table.\ref{tab:RealJets}.

    \begin{table}[!hbt]
      \center
  \begin{tabular}{|c||c|c|c|} \hline
    & \textbf{50 GeV Sample} & \textbf{75 GeV Sample} & \textbf{100 GeV
      Sample} \\ \cline{2-4}
    \begin{tabular}{c}
      Percentage of events \\ with $>1$ recon. jet \\ and
      satisfying \\ condition \ref{crit1a}(2R)
    \end{tabular}  
    & 0.02$\%$  &  0.08$\%$  &  0.17$\%$  \\ \hline
  \end{tabular}
  \protect\caption{Contribution to `fake' rate from events containing at least 2 real
     PYTHIA jets within the EMCal fiducial range.}
   \protect\label{tab:RealJets} 
    \end{table}

    It is impossible to assess the magnitude of this effect in central
    HIJING events since the relevant information is not readily
    available from the simulations.
    

  \item \textbf{Effect of Small Cone Radius}
    A small cone radius can possibly lead to `fake' jets being
    reconstructed. If a jet fragments such that its energy deposition
    is distributed in a way 
    that two energy groupings can be reconstructed with centres with an
    angular distance apart of more than the small cone radius used but
    less than $R=1.0$,
    then two jets will be reconstructed although they may be really
    part of the same jet. An example of this situation can be seen in 
    Fig.~\ref{fig:OneJetEMCAL}. 

    \begin{figure}[!hbt]
          \center
          \includegraphics[scale=0.6]{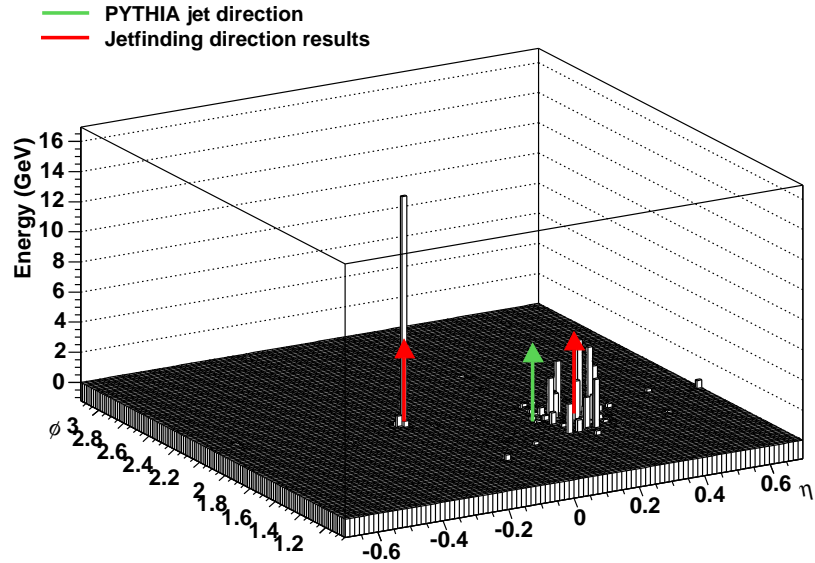}
          \protect\caption{Grid energy including tracking and
          calorimetry information for a 100 GeV PYTHIA event where the
          algorithm reconstructed two jets (different event from that shown
          in Fig.~\ref{fig:TwoJetsEMCAL}). The green arrow indicates
          the direction of the trigger jet as calculated by PYTHIA and
          the red arrows indicate the directions of the two jets
          reconstructed by the algorithm.}
          \protect\label{fig:OneJetEMCAL}
        \end{figure}
    
    The figure shows the energy grid
    containing tracking and calorimetry information for a PYTHIA event
    containing a 100 GeV jet. The green arrow indicates the 
    direction of the trigger jet as calculated by PYTHIA ($\eta$=0.17,
    $\phi$=1.80) and the red
    arrows show the directions of the two jets reconstructed by the
    seeded algorithm ($\eta_{1}$=-0.15, $\phi_{1}$=2.00) and
    ($\eta_{2}$=0.27, $\phi_{2}$=1.73). 
    In this case, the reconstructed jet cone centres were an
    angular distance of $R=\sqrt{{(\Delta\eta)}^2+{(\Delta\phi)}^2} = 0.50$ away from each
    other. If the energy weighted centroid of the two jets
    reconstructed by the algorithm is calculated, a direction
    coinciding with the jet direction calculated by PYTHIA is
    obtained. Therefore, the small cone radius of $R=0.3$ can lead to
    two jets being reconstructed where in fact there was only one. 
    The proportion of events which contained two or more 
    reconstructed jets and which satisfied the following conditions:
    \begin{enumerate}
    \item(SC) $0.3<D<1.0$, where $D$ is the angular distance between the
    two reconstructed jets \label{crit2a}
    \item(SC) $\sqrt{(\eta_{PYTHIA}-\eta_{Weight})^2 + (\phi_{PYTHIA}-\phi_{Weight})^2} < 0.01$ \label{crit2b}
    \end{enumerate}
    are shown in Table.\ref{tab:SmallCone}.

    \begin{table}[!hbt]
      \center
  \begin{tabular}{|c||c|c|c|} \hline
    & \textbf{50 GeV Sample} & \textbf{75 GeV Sample} & \textbf{100 GeV
      Sample} \\ \cline{2-4}
    \begin{tabular}{c}
      Percentage of events \\ with $>1$ recon. jet \\ and satisfying
      \\ conditions \ref{crit2a}(SC) $\&$ \ref{crit2b}(SC)
    \end{tabular}  
    &  0.02$\%$ &  0.11$\%$ &  0.35$\%$    \\ \hline
  \end{tabular}
  \protect\caption{Contribution to `fake' rate from events containing 1 real
     PYTHIA jet which the jet-finding algorithm reconstructs as 2 jets 
     due to the small cone size used ($R=0.3$).}
   \protect\label{tab:SmallCone} 
    \end{table}

  \item \textbf{Algorithm and Physics Related Effects}\\
    A combination of the physics involved and the nature of the jet
    finding algorithm may lead to the reconstruction of `fake'
    jets. It has been found that 
    cone jet finding algorithms, in which jet seeds are processed in
    decreasing order of $E_{T}$ and where a seed is removed from the list
    if it is included in a jet with a seed that is before it
    in the list, are sometimes sensitive to collinearity problems
    \cite{RunII}. For
    example, Fig.~\ref{fig:Collinear} from \cite{RunII} shows a 
    situation at the parton level where the hardest parton in the jet (left-hand picture)
    splits into two almost collinear partons (right-hand picture). In both
    cases the same jet should be reconstructed, but if the $E_{T}$ ordered seed
    algorithm is used, different jets are reconstructed as shown by the
    different cone positions in the figure. In the case of the left-hand
    picture, the central parton will be treated as a seed by the algorithm first as
    it has the highest energy and all three partons will be grouped
    together in the jet and the others removed from the seed
    list. However, for the case in the right-hand picture, 
    if the central parton splits such that one of the other two original
    partons have a higher energy than each of the split partons, then the
    algorithm could find a jet that excludes the third original parton as
    it could fall outside the chosen cone radius, $R$.
    
\end{itemize}

        \begin{figure}[!hbt]
          \center
          \includegraphics[scale=0.5]{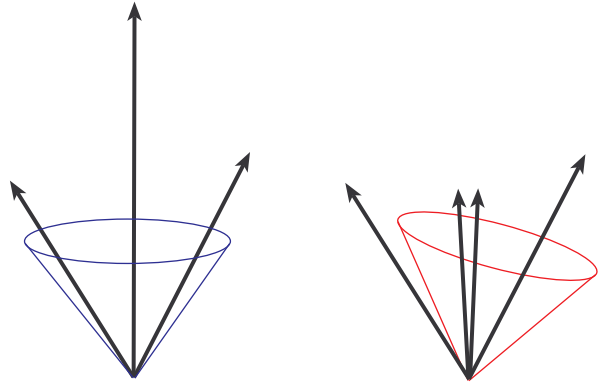}
          \protect\caption{Collinear partons (right-hand picure) resulting from the
          splitting of the central highest energy parton (left-hand
          picture) result in different jets being reconstructed by a
          jet finding algorithm with $E_{T}$ ordered seeds. The length
          of the parton arrows represent their relative $E_{T}$. From \cite{RunII}.
          }
          \protect\label{fig:Collinear}
        \end{figure}

\begin{table}[!hbt]
  \center
  \begin{tabular}{|l|c||c|c|c|c|c|}  \hline
    \multicolumn{2}{|c||}{
    \begin{tabular}{c}
       \textbf{Event Type}\\  
    \end{tabular}}
    &
    \begin{tabular}{c}
      $\epsilon$\\
    \end{tabular}
    &
    \begin{tabular}{c}
      \textbf{Events}\\
      \textbf{with 0}\\
      \textbf{reco.}\\
      \textbf{jets} ($\%$)
    \end{tabular}
    &    
    \begin{tabular}{c}
      \textbf{Events}\\
      \textbf{with 1}\\
      \textbf{reco.}\\
      \textbf{jet} ($\%$)
    \end{tabular}
    &
    \begin{tabular}{c}
      \textbf{Events}\\
      \textbf{with 2}\\
      \textbf{reco. }\\
      \textbf{jets}($\%$)       
    \end{tabular}
    &
    \begin{tabular}{c}
      \textbf{Events } \\
      \textbf{with} $>$ \\
      \textbf{2 reco.}\\
      \textbf{jets}($\%$)       
    \end{tabular}
    \\ \hline
    Seeded algorithm, & 50 GeV &0.862  & 25.5  & 63.9  & 9.69  & 0.958 \\\cline{2-7}
    combined events & 75 GeV      & 1.11 & 6.70  & 76.7  & 15.2  & 1.39 \\\cline{2-7}
    & 100 GeV                  & 1.21 & 2.82  & 75.4  & 19.8  & 1.99 \\\hline\hline
    Seeded algorithm, & 50 GeV & 0.703 & 30.1  & 69.6  & 0.345  & 0.000 \\\cline{2-7}
    PYTHIA events & 75 GeV     & 0.956  & 8.16  & 88.1  & 3.76 & 0.000 \\\cline{2-7}
    alone & 100 GeV            & 1.05  & 3.45 & 87.9  & 8.54 & 0.0745 \\\hline\hline
    Seeded algorithm, & 50 GeV & 0.735  & 33.4 & 60.2  & 6.01  & 0.414 \\\cline{2-7}
    combined events, & 75 GeV     & 0.996  & 12.0 & 77.3  & 9.68  & 0.958 \\\cline{2-7}
    tracking data only & 100 GeV & 1.09  & 6.66 & 79.0  & 13.3  & 1.10\\\hline\hline
    Seedless algorithm, & 50 GeV & 2.40 & 0.603 & 19.5  & 37.4  & 42.5 \\\cline{2-7}
    combined events & 75 GeV        & 2.46 & 0.0905 & 18.7 & 35.9  & 45.3 \\\cline{2-7}
    & 100 GeV                    & 2.50 & 0.135 & 17.5  & 36.0  & 46.4\\\hline\hline
    Seedless algorithm, & 50 GeV & 0.835  & 16.5 & 83.4  & 0.0481  & 0.000 \\\cline{2-7}
    PYTHIA events & 75 GeV       & 1.01  & 2.50 & 94.1  & 3.42  & 0.0113 \\\cline{2-7}
    alone & 100 GeV              & 1.09  & 0.713 & 89.7  & 9.49  & 0.0566 \\\hline\hline    
    \multicolumn{2}{|l||}{
    \begin{tabular}{l}
    Seeded algorithm, Central\\
    HIJING events alone
    \end{tabular}}
    & - & 84.0 & 14.7 & 1.08 & 0.215 \\\hline
    \multicolumn{2}{|l||}{
    \begin{tabular}{l}
    Seedless algorithm, Central\\
    HIJING events alone
    \end{tabular}}
    & - & 15.5 & 31.4 & 29.0 & 24.1 \\\hline    
    \end{tabular}
  \protect\caption{Efficiency results for the jet finding algorithm in
  its seeded and seedless versions for different input event
  types. Also shown is the fraction of events with no jets
  reconstructed, 1 jet only, 2 
  jets only, and greater than 2 jets reconstructed for the different
  input event types.  }
  \protect\label{tab:Efficiency} 
\end{table}

Table~\ref{tab:Efficiency} shows the efficiency of the algorithm in
its seeded and seedless versions for different event types. From
Table~\ref{tab:Efficiency}, a general trend that can be seen is that
the value of $\epsilon$ increases with increasing 
input jet energy for all event types. The definition of $\epsilon$ for
Central HIJING events has no physical meaning in this case because the
number of known input jets is zero.
Another general trend is that all the $\epsilon$ values for 100
GeV jet cases are greater than 1, implying that the algorithm
reconstructs more jets than are input for these cases. The values
of $\epsilon$ for the seeded algorithm for pure PYTHIA events with no
background, are within $5\%$ of the ideal value of 1 for 75 GeV and
100 GeV jets,
however the efficiency is smaller, $\epsilon = 0.7$, for the 50 GeV jets.

The seeded algorithm applied to combined events, with tracking plus
calorimetry data, produces values of $\epsilon$ close to unity
($\pm21\%$) although $\epsilon < 1$ for the 50 GeV input jets implies
that some real jets were excluded by the algorithm.
These efficiency values are larger than those obtained from the seeded
algorithm applied to the combined events with only tracking data.


The seedless algorithm, when applied to PYTHIA events with no
background, produced values of $\epsilon$ ranging from $19\%$ greater
(for the 50 GeV case) to $4\%$ greater (for the 100 GeV case)
than the seeded algorithm values for pure PYTHIA events. Furthermore,
the seedless algorithm found on average more than double the number of
jets than were input for all input jet energies for combined events.

The third column of Table~\ref{tab:Efficiency} shows the number of
events for each case where no jets were reconstructed i.e. the
algorithm failed to reconstruct a jet since the jet did not satisfy
the algorithm cuts. Note that the values of $\epsilon$ in column two
are calculated on an overall per \emph{input jets} basis and the values in
column three on a per \emph{event} basis. Therefore the values in columns two
and three are not expected to sum to unity.  
The percentage of events in which no jet was found decreases as a
function of increasing jet energy since the algorithm
parameters were optimised for 50 GeV jets. Therefore the higher energy
jets which result in higher tower energies above the background, are
easier to reconstruct. The jet exclusion rates are within 
$5\%$, but slightly lower than was predicted in
Tables~\ref{tab:ParamTableJets50} and \ref{tab:ParamTableJets100}.
The resulting exclusion rate obtained with the seeded algorithm on
combined events is smaller than the rate obtained for pure PYTHIA
events. This indicates that some `fake' jets are reconstructed for the
combined events since the jet signal is the same for the two cases and
the only difference is the addition of background for the combined event. The
seeded algorithm applied to combined events with only tracking data
excludes a larger fraction of real jets than the other two cases. Since
the calorimetry data is absent in this analysis while the parameters are
optimised for the use of both tracking and calorimetry data, it is expected that
the energy cuts on \emph{JetESeed} and \emph{MinJetEt} in the algorithm will
exclude more jets. 

In comparison, the seedless algorithm applied to pure PYTHIA events
reconstructs more jets than the seeded algorithm. When it is applied to
combined events, an $\epsilon$ is obtained which is double that obtained
when it is applied to pure PYTHIA events. This is due to
reconstruction of `fake' jets because of the zero cut on
\emph{JetESeed}. This is also clear from the low percentage ($<$20$\%$) of these
events where only one jet was reconstructed (column 4) compared to the
case for the seeded algorithm on combined events ($>$60$\%$).

For the case of Central HIJING events alone which serve here as an
estimate of the background, the seeded algorithm
excludes a much larger percentage of events (84$\%$) than the seedless
algorithm (16$\%$). Since, in this thesis, all reconstructed jets
which are not from the PYTHIA signal, are classified as 
`fakes', this indicates that the seeded algorithm is preferable to use
since it excludes more background events.


The seeded algorithm, for all cases except background events, reconstructs a single jet  
per event for more than $60\%$ of events. On the other hand, less than
$20\%$ of combined events are reconstructed as single jet events by the
seedless algorithm which is more than three times lower than the case
for the seeded algorithm.


The `fake' rate for the algorithm in its seeded and seedless versions
for the various input jet event types can be found from the percentage of
input events where two or more jets are reconstructed i.e. by summing
columns 5 and 6 in Table~\ref{tab:Efficiency}. The `fake' rate for the
seedless algorithm on combined events is greater than $80\%$ while the
seeded algorithm on combined events (tracks plus calorimetry and tracks
alone) produces a `fake' rate of less than 
$22\%$. For the case of PYTHIA events alone, both the seeded and
seedless algorithms produce a `fake' rate of less than $10\%$ over all
the input jet energies. For the case of Central HIJING events alone,
the `fake' rate of the algorithm can be found from the percentage of
events where \emph{at least 1} jet is
reconstructed because any jet reconstructed in a Central HIJING event
is classified as a `fake' jet here. Therefore, by summing columns 4, 5
and 6 in Table~\ref{tab:Efficiency} for the cases of Central HIJING
events alone, the `fake' rate amounts to $85\%$ for the seedless
algorithm and $16\%$ for the seeded
algorithm ($2.8\%$ higher than predicted in Table~\ref{tab:ParamTableHijing}). 
In conclusion, this investigation shows that the seeded algorithm is
better to use than the seedless version since it reconstructs far
fewer `fake' jets resulting in a purer sample of real reconstructed jets.

\subsection{`Fake' Jet Reconstruction}
\begin{figure}[!hbt]
          \center
          \includegraphics[scale=0.5]{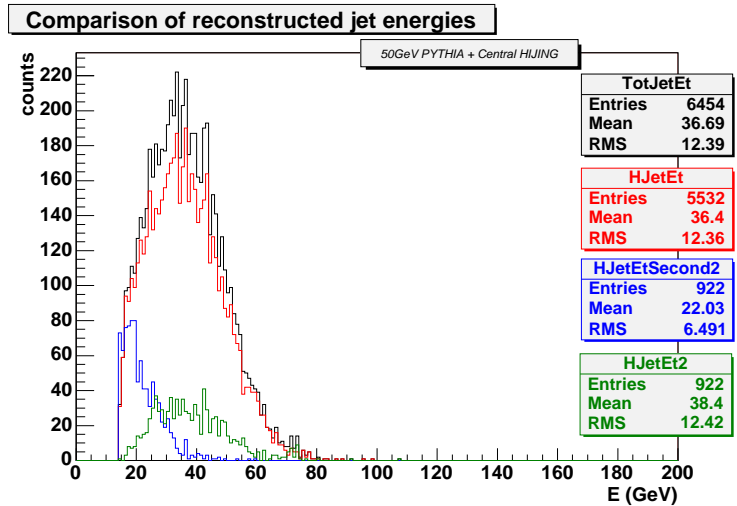}
          \protect\caption{Comparison of reconstructed jet energies
          for different cases for 50 GeV combined events and the seeded
          algorithm. The blue histogram shows the second highest
          energy jet and the green shows the highest energy jet for
          events where two or more jets were reconstructed. The red
          histogram shows the highest energy jet for events where only
          one jet was reconstructed. The black histogram shows the
          distribution for the highest energy jet for events where at
          least one jet was reconstructed (the sum of the red and
          green histograms).} 
          \protect\label{fig:Seed50SecondJet}
        \end{figure}

\begin{figure}[!hbt]
          \center
          \includegraphics[scale=0.5]{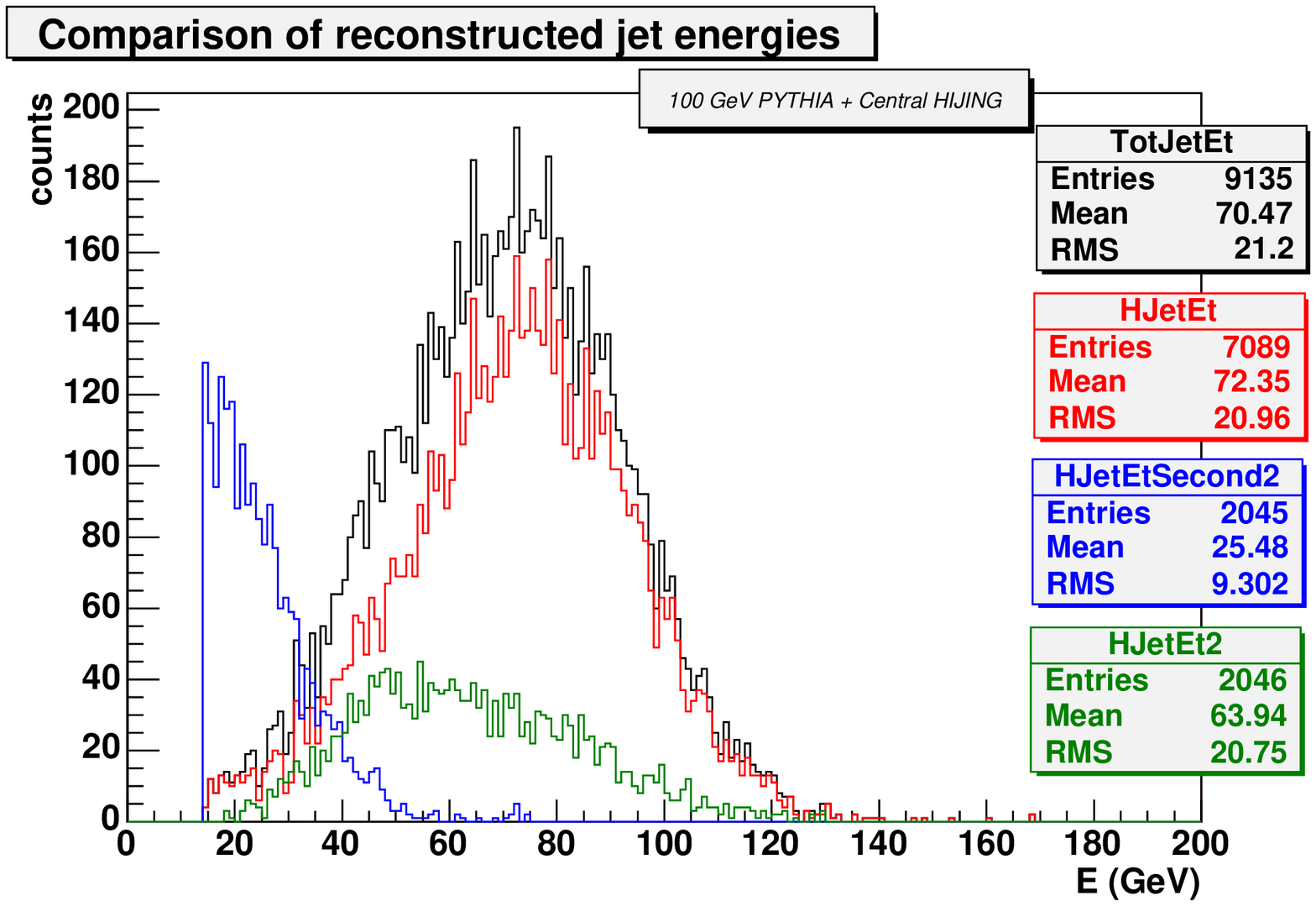}
          \protect\caption{Comparison of reconstructed jet energies
          for different cases for 100 GeV combined events and the seeded
          algorithm. The colours represent the same quantities described
          in Fig.~\ref{fig:Seed50SecondJet}. }
          \protect\label{fig:Seed100SecondJet}
        \end{figure}

\begin{figure}[!hbt]
          \center
          \includegraphics[scale=0.5]{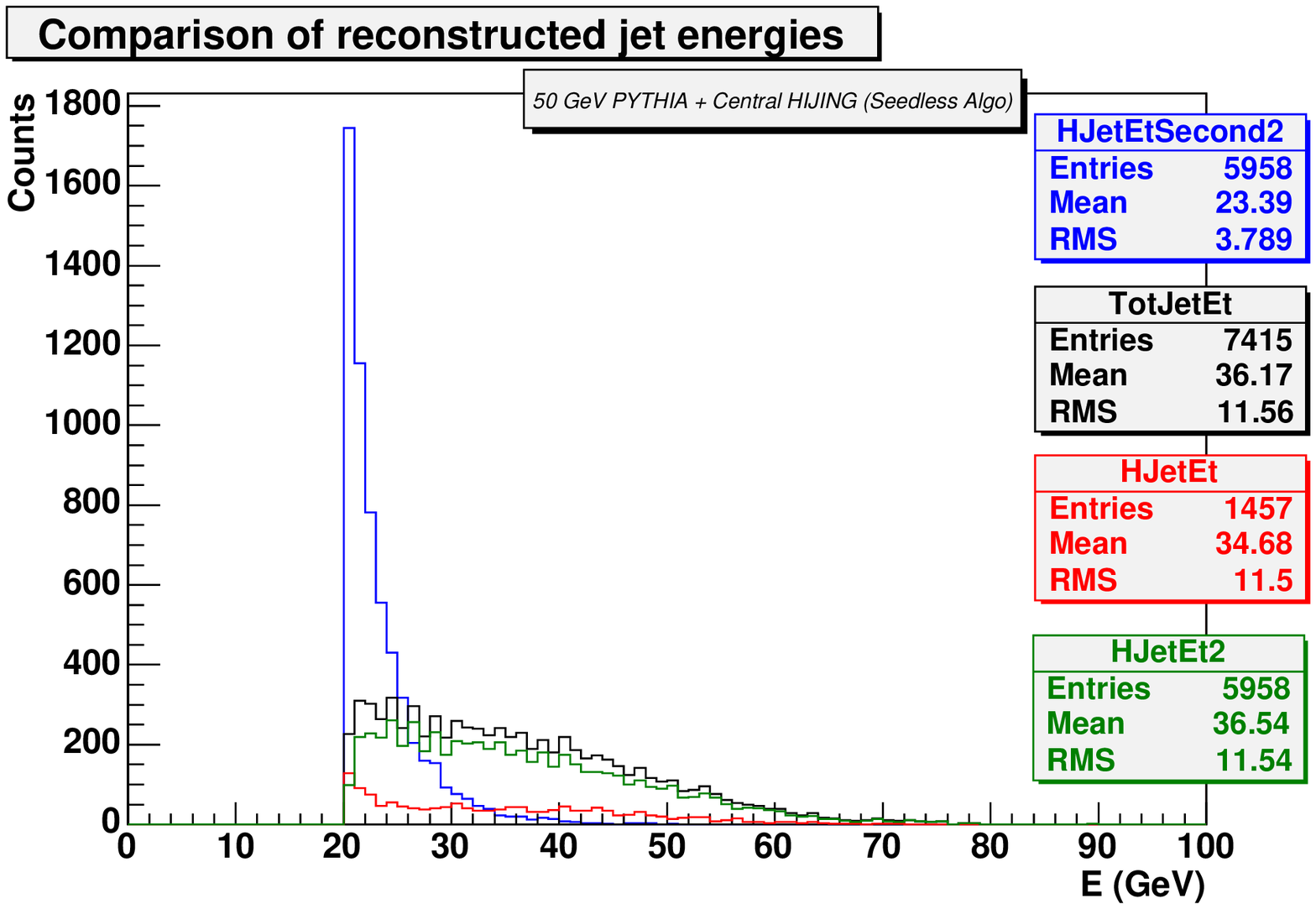}
          \protect\caption{Comparison of reconstructed jet energies
          for different cases for 50 GeV combined events and the seedless
          algorithm.The colours represent the same quantities described
          in Fig.~\ref{fig:Seed50SecondJet}.  }
          \protect\label{fig:NoSeed50SecondJet}
        \end{figure}

\begin{figure}[!hbt]
          \center
          \includegraphics[scale=0.5]{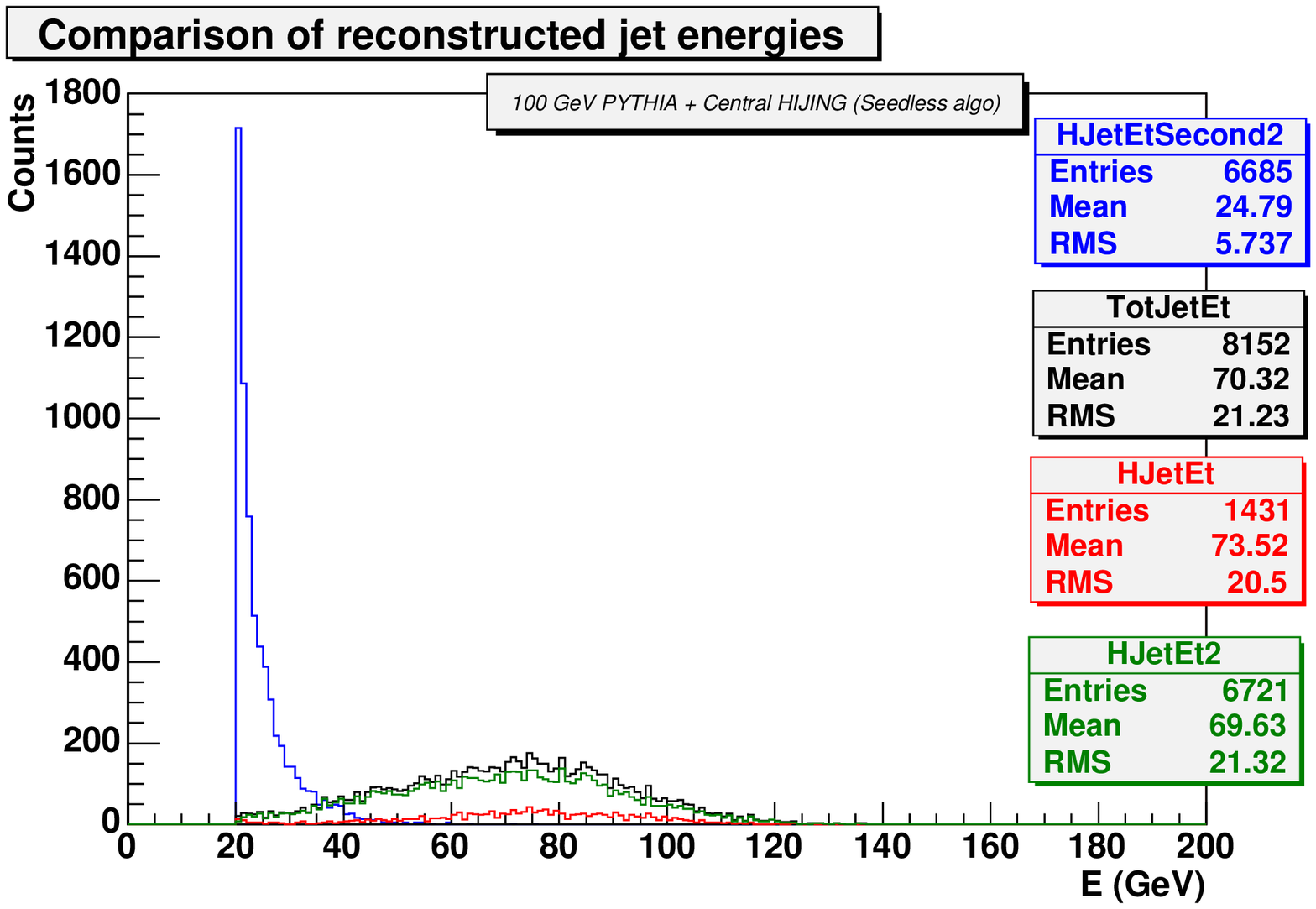}
          \protect\caption{Comparison of reconstructed jet energies
          for different cases for 100 GeV combined events and the seedless
          algorithm.The colours represent the same quantities described
          in Fig.~\ref{fig:Seed50SecondJet}. }
          \protect\label{fig:NoSeed100SecondJet}
        \end{figure}

In order to understand better the primary source of the `fake' jets that were
reconstructed, jet energy histograms were plotted for the second
highest energy jet in events where two or more jets were
reconstructed. Figs.~\ref{fig:Seed50SecondJet} and
\ref{fig:Seed100SecondJet} show a comparison of `fake' and real jets
for the seeded algorithm applied to 50 GeV combined events and 100 GeV
combined events respectively. In Figs.~\ref{fig:Seed50SecondJet} and
\ref{fig:Seed100SecondJet} the blue histogram represents the second
highest energy jet for events where two or more jets were
reconstructed. The green histogram represents the energy of the highest energy jet
per event for the same case. The red histogram represents the highest
energy jet for all events where only one jet was reconstructed and the
black histogram is the sum of the red and green histograms i.e. the
energy distribution of the highest energy jet found in all events (the
same as shown in Figs.~\ref{fig:AllJetEt50} and \ref{fig:AllJetEt100} respectively).
The same cases are plotted for the seedless algorithm applied to 50
GeV combined events and 100 GeV combined events and are shown in
Figs.~\ref{fig:NoSeed50SecondJet} and \ref{fig:NoSeed100SecondJet}
respectively.
The proportion of events where a second jet (`fake') was reconstructed is much
larger when the seedless algorithm is used as shown in
Figs.~\ref{fig:NoSeed50SecondJet} and \ref{fig:NoSeed100SecondJet}
compared to Figs.~\ref{fig:Seed50SecondJet} and
\ref{fig:Seed100SecondJet}.


\begin{figure}[!hbt]
          \center
          \includegraphics[scale=0.5]{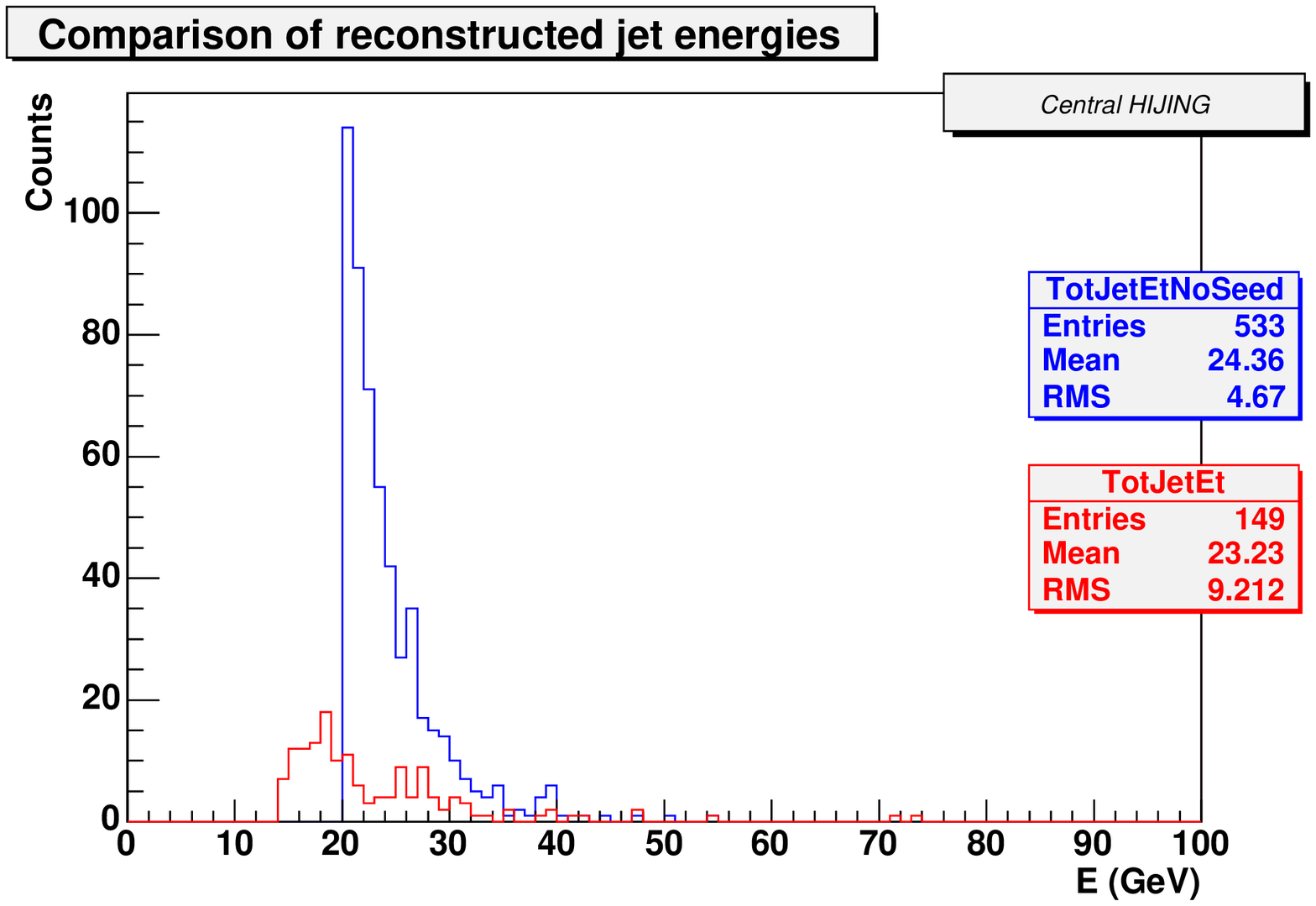}
          \protect\caption{Reconstructed jet energy distributions for
          Central HIJING events using the seeded algorithm (red) and
          the seedless algorithm (blue). }
          \protect\label{fig:HijingFakes}
        \end{figure}

For comparison the energy distributions of jets reconstructed using the
seeded algorithm (red) and the seedless algorithm (blue) in Central
HIJING events, are shown in Fig.~\ref{fig:HijingFakes}. Again more
`fakes' are reconstructed by the seedless algorithm compared to the
seeded algorithm.

The top section of Table~\ref{tab:FakeJetEnergy} shows the mean energies of
the second highest 
energy jets reconstructed in events with at least two reconstructed
jets for the cases of the seeded and seedless versions of the algorithm
and the different input event types. The bottom section of the table
shows the mean energy of the highest energy jets reconstructed using
the seeded and seedless versions of the algorithm on Central HIJING events alone.

\begin{table}[!hbt]
  \center
  \begin{tabular}{|c||c|c|c|}  \hline
    \textbf{Data Type}
    &
    \begin{tabular}{c}
      $\langle E^{2nd}_{Reco}\rangle$ \textbf{for} \\
    \textbf{50GeV input}\\
    (GeV)
    \end{tabular}
    &
    \begin{tabular}{c}
    $\langle E^{2nd}_{Reco}\rangle$ \textbf{for} \\
    \textbf{75GeV input}\\
    (GeV)
    \end{tabular}
    &
    \begin{tabular}{c}
    $\langle E^{2nd}_{Reco}\rangle$ \textbf{for} \\
    \textbf{100GeV input}\\
    (GeV)
    \end{tabular}\\ \hline
    \begin{tabular}{l}
      Seeded algorithm, \\
      combined events
    \end{tabular}
    & 22.03  & 23.18  & 25.48  \\ \hline
    \begin{tabular}{l}
      Seeded algorithm, \\
      PYTHIA events \\
      alone
    \end{tabular}
    & 16.60   & 21.22  & 25.22  \\ \hline
    \begin{tabular}{l}
      Seedless algorithm, \\
      combined events \\
    \end{tabular}
    & 23.36    & 24.21  & 24.77  \\ \hline
    \begin{tabular}{l}
      Seedless algorithm, \\
      PYTHIA events \\
      alone
    \end{tabular}
    &  20.82   & 23.69  & 26.95  \\ \hline\hline     
    \textbf{Data Type}
    &
    \multicolumn{3}{|c|}{
      $\langle E^{1st}_{Reco}\rangle$ \textbf{for all input events} (GeV) 
      }\\\hline
    \begin{tabular}{l}
      Seeded algorithm, \\
      Central HIJING\\
      events alone
    \end{tabular}
    &
    \multicolumn{3}{|c|}{
      23.23
      }\\\hline
    \begin{tabular}{l}
      Seedless algorithm, \\
      Central HIJING\\
      events alone
    \end{tabular}
    &
    \multicolumn{3}{|c|}{
      24.36
      }\\\hline
    \end{tabular}
  \protect\caption{The top section of the table shows the mean
  reconstructed energy for the second highest $E_{T}$ jets in events
  where at least two jets were reconstructed for the case of the
  seeded and seedless algorithm applied to different input event
  types. The lower section of the table shows the mean reconstructed
  energy for the highest $E_{T}$ jets for Central HIJING events alone,
  see Fig.~\ref{fig:HijingFakes}. } 
  \protect\label{tab:FakeJetEnergy} 
\end{table}

It can be argued that the primary source of `fake' jets is the
background with its large event-by-event fluctuations. The evidence to
suggest this is as follows:
For both algorithms, in the case of combined events,
the mean reconstructed jet energies for the second
highest energy jets (per event) are closer to the value of the highest
energy jets reconstructed in Central HIJING events, than is the case
for PYTHIA events alone.
For the seeded algorithm case, the reconstructed jet energies,
$\langle E^{2nd}_{Reco}\rangle$, for the 
combined events are within $10\%$ of the $\langle E^{1st}_{Reco}\rangle$ for Central
HIJING events while the reconstructed energies for PYTHIA alone differ
by up to $30\%$. A similar trend is observed for the seedless
algorithm case where for the combined events, $\langle E^{2nd}_{Reco}\rangle$ are within
$4\%$ of the $\langle E^{1st}_{Reco}\rangle$ for Central HIJING while for PYTHIA events
alone, the difference is up to $14\%$.

The `fake' rates for PYTHIA events alone using both the
seeded and seedless versions of the algorithm are very similar for
respective input jet energies (see
Table~\ref{tab:Efficiency}) but the addition of background to these
events increases the `fake' rates. The addition of background
more than doubles the `fake' rate for the case of the seeded algorithm
on combined events. For the seedless algorithm, the `fake' rate is more
than ten times higher for combined events than for PYTHIA events alone.
Therefore the primary source of `fake' jets is due to contribution from the
background or `underlying event'. 



\section{Energy Resolution}\label{sec:Resolution}
\begin{figure}[!hbt]
          \center
          \includegraphics[scale=0.6]{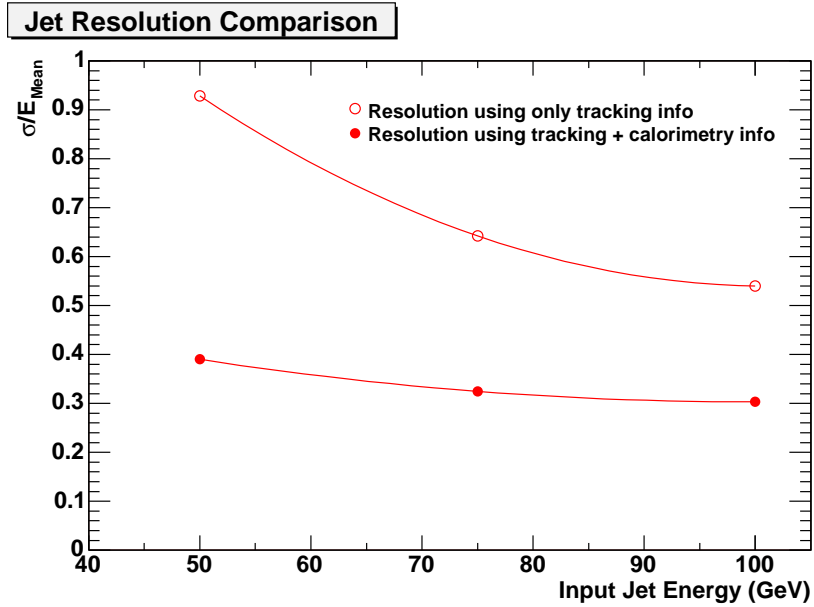}
          \protect\caption{Jet energy resolution ($\sigma/\langle
          E_{Reco}\rangle$) for combined events using the optimised seeded
          algorithm for events where only tracking information was
          used (open symbols) compared to when tracking plus
          calorimetry information was used (closed symbols). }
          \protect\label{fig:TracksHitsRes}
        \end{figure}

Fig.~\ref{fig:TracksHitsRes} compares the resulting jet energy resolutions
obtained when using the seeded version of the algorithm on combined
events with tracking 
data alone (represented by the open symbols) compared to tracking plus
calorimetry data (represented by the solid symbols). The optimised
algorithm parameters shown in Table ~\ref{tab:Param} are used.
The resolution, $\sigma/\langle E_{Fit}\rangle$, is calculated by
dividing the $\sigma$ by the
mean of the gaussian fitted reconstructed jet energy distribution.
By using a combination of tracking and calorimetry information to
reconstruct jets, an improvement of up to $54\%$ for the energy
resolution of 50 GeV jets is obtained compared to using tracking
information only. This brings the resolution of 50 GeV jets down from
$93\%$ (tracking only) to $39\%$ (tracking plus calorimetry) and of
100 GeV jets from $54\%$ (tracking only) down to $30\%$ (tracking plus
calorimetry).

\begin{figure}[!hbt]
          \center
          \includegraphics[scale=0.6]{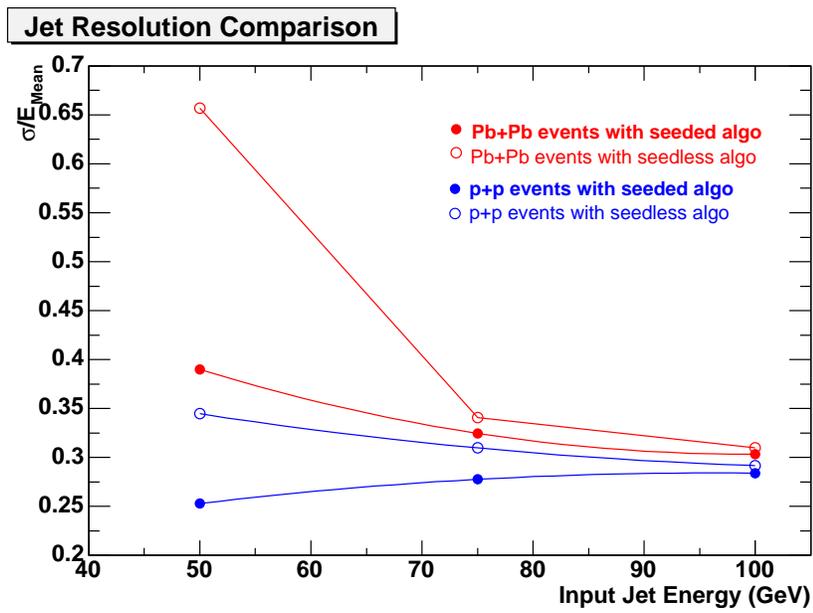}
          \protect\caption{Jet energy resolution ($\sigma/\langle
          E_{Reco}\rangle$) for combined events (red) compared to PYTHIA
          events alone (blue) using the seeded algorithm (solid
          symbols) compared to the seedless algorithm (open symbols).}
          \protect\label{fig:SeededVsSeedlessRes}
        \end{figure}

The resolutions obtained using the seeded version of the algorithm
compared to the seedless version are plotted in
Fig.~\ref{fig:SeededVsSeedlessRes}. The red lines indicate combined
events (Pb+Pb) and the blue lines indicate pure PYTHIA events (p+p). In both the
combined event and pure PYTHIA event cases, the seeded algorithm (solid
symbols) produces lower 
resolutions than the seedless algorithm (open symbols).
For the seeded algorithm (solid symbols), the resolutions obtained for
combined events approach the values obtained for the pure PYTHIA case as a
function of increasing jet energy.

\chapter{Conclusions}

This thesis demonstrates the first successful attempt to
reconstruct high-$p_{T}$ jets ($E_{Tjet}>50$ GeV) 
in high-multiplicity ($dN_{ch}/dy = 4000$) heavy-ion collisions at
$\sqrt{s_{NN}}=5.5$ TeV using a UA1-based cone algorithm modified for
this purpose
and detectors from the ALICE experiment at the LHC.
The algorithm parameters were optimised for the heavy-ion
environment and a method of calculating and 
subtracting the large background energy contribution on an
event-by-event basis was developed in order to reconstruct the jet energy.

It has been shown that due to the large event-by-event background
fluctuations, a small cone size of 
$R=0.3$ is required to optimise the measurement of the jet energy. The 
combination of information from the tracking detectors and the
electromagnetic calorimeter produces much improved jet energy resolution
results and more accurate jet direction reconstruction when compared
to using tracking information alone. The shift towards lower energy of
the reconstructed jet energies due to the small cone radius and other
algorithm effects is understood and can be corrected for using a
cross-section weighted multiplicative factor which is independent of
jet energy (in the range 50 GeV/$c$$<E_{T}<$100 GeV/$c$).

The seeded algorithm is the preferred algorithm to use since it
resulted in much lower `fake' jet rates and better jet energy resolution
than the seedless version. Artefacts of the algorithm such as the use
of a small cone radius and jet-seed $E_{T}$ ordering contribute negligibly
to the resulting `fake' jet rates.

The jet finding algorithm detailed in this thesis is now ready for
data analysis in the field of heavy-ion physics.


\appendix

\chapter{Glossary of Terms}\label{sec:Glossary}

\begin{enumerate}

  \item \textbf{Jet}\\
    A jet is a localised (in $\eta,\phi$-space) group of hadrons orginating from
    the fragmentation of a hard scattered parton.

  \item \textbf{Combined event}\\
    An event consisting of a PYTHIA jet event superimposed on a HIJING
    background event.

  \item \textbf{PYTHIA}\\
    A Monte Carlo event generator for the simulation of high energy physics events.
        
  \item \textbf{HIJING}\\
    HIJING (Heavy Ion Jet Interaction Generator) is a Monte Carlo
    event generator based on PYTHIA for simulating A+A collisions.
    
  \item \textbf{Jet-finding algorithm parameters:}\\
    \begin{enumerate}
      \item \textbf{Cone radius ($R$)}\\
        The cone radius ($R=\sqrt{(\Delta\eta)^2+(\Delta\phi)^2}$)
        defines the angle from the jet axis within which energy will
        be measured when reconstructing jets according to the cone
        algorithm prescription.
        
      \item \textbf{Track $p_{T}$-cut}\\
        A threshold momentum which determines which tracks are
        included or excluded in the analysis.
        
      \item \textbf{JetESeed}\\     
        The minimum energy a grid cell must have in order to be
        classified as a jet seed by the algorithm.
        
      \item \textbf{MinJetEt}\\
        The minimum energy contained within a cone to be classified as
        a jet by the algorithm.     
    \end{enumerate}

  \item \textbf{Seeded algorithm}\\
    The modified algorithm as presented here where the parameter $JetESeed > 0$ GeV.
    
  \item \textbf{Seedless algorithm}\\
    The modified algorithm as presented here where the parameter
    $JetESeed = 0$ GeV.

  \item \textbf{`Real' jet}\\  
    A `real' jet is classified here as the PYTHIA jet in a combined
    event. This is the jet we aim to reconstruct in the analysis.

  \item \textbf{`Fake' jet}\\  
    A `fake' jet is defined here as any jet that is reconstructed by the algorithm but
    that is not the input PYTHIA jet.

\end{enumerate}

\chapter{Reconstructed Jet Energy Fit Results}\label{sec:JetFits}

        
\begin{table}[!hbt]
  \center
  \begin{tabular}{|l|c||c|c|c|}  \hline
    \multicolumn{2}{|c||}{
    \textbf{Data Type}}
    &
    \begin{tabular}{c}
    \textbf{$\langle E_{Reco}\rangle$}\\
    (GeV)
  \end{tabular}  
    &
    \begin{tabular}{c}
    \textbf{$\sigma$}\\
    (GeV)
    \end{tabular}
    &
    \textbf{$\chi^2/ndf$} \\ \hline\hline
    Seeded algorithm, & 50 GeV & 33.54 & 14.17  & 86.61/68   \\\cline{2-5}
    combined events & 75 GeV      & 51.62 & 17.53  & 207.6/100   \\\cline{2-5}
    & 100 GeV                  & 69.33 & 22.03  & 295.8/124   \\\hline\hline
    Seeded algorithm, & 50 GeV   & 13.56 & 12.59  &  37.82/40  \\\cline{2-5}
    combined events, & 75 GeV       & 27.01 & 17.35  &  87.96/70  \\\cline{2-5}
    tracking data only & 100 GeV  & 39.25 & 21.18  & 73.52/94   \\\hline\hline
    Seeded algorithm, & 50 GeV  & 32.04 & 8.10  & 192.2/48   \\\cline{2-5}
    PYTHIA events & 75 GeV      & 48.67 & 13.51  & 252.9/82   \\\cline{2-5}
    alone & 100 GeV             & 65.90 & 18.71  & 268.9/112   \\\hline\hline
    Seedless algorithm, & 50 GeV  & 26.10 & 17.15  & 98.27/64   \\\cline{2-5}
    combined events & 75 GeV         & 51.58 & 17.57  & 140.7/92   \\\cline{2-5}
    & 100 GeV                     & 69.71 & 21.59  & 220.6/116   \\\hline\hline
    Seedless algorithm, & 50 GeV  & 28.15 & 9.70   & 95.07/42   \\\cline{2-5}
    PYTHIA events & 75 GeV        & 46.82 & 14.50  & 311.3/75   \\\cline{2-5}
    alone & 100 GeV               & 65.29 & 19.04  & 300.8/107   \\\hline 
    \end{tabular}
  \protect\caption{Reconstructed jet energy fit parameters for various
  event and algorithm types.} 
  \protect\label{tab:FitParameters} 
\end{table}





\end{document}